\pdfoutput=1
\documentclass[12pt]{article}
\usepackage[utf8]{inputenc}      
\usepackage{hyperref}
\usepackage{amsmath,amssymb,mathtools}
\usepackage[T1]{fontenc}          
\usepackage{booktabs,tabularx}
\usepackage{graphicx}
\usepackage{xspace}
\usepackage[usenames]{xcolor}\definecolor{fscolor}{RGB}{44,118,255}
\usepackage{geometry}
\usepackage{todonotes}
\usepackage{listings}
\usepackage[absolute]{textpos}
\usepackage[many]{tcolorbox}
\usepackage{xparse}
\usepackage[font=small,labelfont=bf,format=plain,margin=0.05\textwidth]{caption}
\usepackage{bbm}
\usepackage{tabularx}
\usepackage{ulem}
\usepackage{soul}
\usepackage[capitalize]{cleveref}
\usepackage{slashed}
\usepackage{subcaption}
\usepackage{pifont}
\usepackage[shortlabels]{enumitem}
\usepackage{multirow}
\usepackage[sorting=none,%
  citestyle=numeric-comp,%
  bibstyle=mynumeric,%
  giveninits=true]{biblatex}

\allowdisplaybreaks

\evensidemargin \oddsidemargin
\newcommand{\gev}{\;\text{GeV}\xspace}
\newcommand{\tev}{\;\text{TeV}\xspace}
\newcommand{\fb}{\;\text{fb}\xspace}
\newcommand{\invfb}{\;\text{fb}^{-1}\xspace}

\newcommand{\cp}{\ensuremath{\mathcal{CP}}\xspace}

\newcommand{\Is}{{\mathcal{I}_S}}
\newcommand{\If}{{\mathcal{I}_F}}
\newcommand{\Iv}{{\mathcal{I}_V}}
\newcommand{\Ms}{{\mathcal{M}_S}}
\newcommand{\Mf}{{\mathcal{M}_F}}
\newcommand{\Mv}{{\mathcal{M}_V}}

\newcommand{\Med}{{\mathcal{M}}}
\newcommand{\Inv}{{\mathcal{I}}}

\newcommand{\ETmiss}{\ensuremath{{E_T^\text{miss}}}\xspace}

\newcommand{\subfigsize}{.32}
\newcommand{\subfigsizetwo}{.3}

\NewBibliographyString{refname}
\NewBibliographyString{refsname}
\DefineBibliographyStrings{english}{%
  refname = {Ref\adddot},
  refsname = {Refs\adddot}
}
\DeclareCiteCommand{\ccite}
  {%
  \ifnum\thecitetotal=1
    \bibstring{refname}%
  \else%
    \bibstring{refsname}%
  \fi%
  \addnbspace\bibopenbracket%
  \usebibmacro{cite:init}%
   \usebibmacro{prenote}}
  {\usebibmacro{citeindex}%
   \usebibmacro{cite:comp}}
  {}
  {\usebibmacro{cite:dump}%
   \usebibmacro{postnote}%
   \bibclosebracket}
\newrobustcmd*{\Ccite}{\bibsentence\ccite}

\begin{document}

\thispagestyle{empty}
\def\thefootnote{\fnsymbol{footnote}}

\begin{flushright}
DESY-21-227 \\
EFI-12-10
\end{flushright}
\vspace{3em}
\begin{center}
{\Large\bf Simplified models for resonant neutral scalar production \\[1em]
with missing transverse energy final states}
\\
\vspace{3em}
{
Henning Bahl$^a$\footnote{email: hbahl@uchicago.edu},
Victor Martin Lozano$^b$\footnote{email: victor.lozano@desy.de},
Georg Weiglein$^{b,c}$\footnote{email: georg.weiglein@desy.de},
}\\[2em]
{\sl\small ${}^a$ University of Chicago, Department of Physics, 5720 South Ellis Avenue, Chicago, IL 60637 USA}\\[1em]
{\sl\small ${}^b$ Deutsches Elektronen-Synchrotron DESY, Notkestra{\ss}e 85, 22607 Hamburg, Germany}\\[1em]
{\sl\small ${}^c$ II.\  Institut f\"ur  Theoretische  Physik, Universit\"at  Hamburg, Luruper Chaussee 149, 22761 Hamburg, Germany}
\def\thefootnote{\arabic{footnote}}
\setcounter{page}{0}
\setcounter{footnote}{0}
\end{center}
\vspace{2ex}
\begin{abstract}
{}

Additional Higgs bosons appear in many extensions of the Standard Model (SM). While most existing searches for additional Higgs bosons concentrate on final states consisting of SM particles, final states containing beyond the SM (BSM) particles play an important role in many BSM models. In order to facilitate future searches for such final states, we develop a simplified model framework for heavy Higgs boson decays to a massive SM boson as well as one or more invisible particles. Allowing one kind of BSM mediator in each decay chain, we classify the possible decay topologies for each final state, taking into account all different possibilities for the spin of the mediator and the invisible particles. Our comparison of the kinematic distributions for each possible model realization reveals that the distributions corresponding to the different simplified model topologies are only mildly affected by the different spin hypotheses, while there is significant sensitivity for distinguishing between the different decay topologies. As a consequence, we point out that expressing the results of experimental searches in terms of the proposed simplified model topologies will allow one to constrain wide classes of different BSM models. The application of the proposed simplified model framework is explicitly demonstrated for the example of a mono-Higgs search. For each of the simplified models that are proposed in this paper we provide all necessary ingredients for performing Monte-Carlo simulations such that they can readily be applied in experimental analyses.

\end{abstract}

\newpage
\tableofcontents
\newpage
\def\thefootnote{\arabic{footnote}}


\section{Introduction}
\label{sec:intro}

The discovery of a scalar boson at the Large Hadron Collider (LHC) --- consistent with the prediction for the Higgs boson of the Standard Model (SM) within the theoretical and experimental uncertainties --- marks an important milestone for particle physics~\cite{ATLAS:2012yve,CMS:2012qbp}. It is the first spin-0 particle for which no substructure is known and which could, therefore, be a fundamental particle. While so far no conclusive hints for beyond SM (BSM) physics have been found, there is ample room for extending the scalar sector of the SM by adding additional scalar bosons.

An extended scalar sector could help to resolve several open issues of the SM like the lack of a sufficient amount of \cp violation to explain the baryon asymmetry of the Universe. Intriguingly, also an explanation for Dark Matter (DM) could be closely connected to the scalar sector. While the DM particle itself (or one of several DM particles) could be a scalar particle, also the role of the mediator between the visible and the dark sector can be played by a scalar particle.

This hypothesis can be tested in various ways. Direct or indirect DM searches concentrate on directly detecting the DM particle(s). In this work, we will, however, concentrate on LHC collider experiments, which target the mediator particle, while the DM particles cannot directly be detected but give rise to a certain amount of missing transverse energy as a trace of their presence. Since missing energy by itself is unobservable, the presence of an additional particle in the final state is required. Typical examples are a single jet, a single Higgs boson, or a single $Z$ boson.\footnote{It should be noted that these searches are also sensitive to long-lived particles which decay outside of the detector. For long-lived particles decaying inside of the detector specific searches are designed~\cite{Alimena:2019zri}.}

For interpreting the results of these ``mono-X plus missing energy'' searches, the experimental collaborations usually rely on benchmark models like the Two-Higgs-Doublet Model (THDM) with an added pseudoscalar DM portal (THDMa). In these models, an initial heavy resonance decays into a SM boson and a mediator particle, where the latter decays to invisible particles. The focus on this specific decay topology is rather restrictive. In fact,  many other well motivated BSM models with ``mono-X plus missing energy'' signatures exist in which different decay topologies (with different kinematics) are encountered.

An example for this is the decay of a heavy Higgs boson into a the lightest neutralino and the second-lightest neutralino in the Minimal Supersymmetric extension of the SM (MSSM). In the MSSM, this decay mode is of particular interest since it can efficiently be used to set constraints on the parameter region in which the lightest and second-lightest neutralino are close to each other in mass, and where direct neutralino searches are less sensitive. Accordingly, this decay mode has received considerable attention~\cite{ATLAS:2009zmv,Charlot:2006se,Bisset:2007mi,Arbey:2013jla,Craig:2015jba,Heinemeyer:2015pfa,Profumo:2017ntc,Kulkarni:2017xtf,Barman:2017swy,Bahl:2018zmf,Gori:2018pmk,Bahl:2019ago,Adhikary:2020ujn,Liu:2020muv},\footnote{Also the similar decay mode of heavy Higgs bosons into two staus resulting in a di-tau plus missing energy final state has been investigated phenomenologically~\cite{Arganda:2018hdn,Bahl:2018zmf,Arganda:2021qgi}.} but nevertheless no dedicated experimental searches for this signature have been performed at the LHC so far.

While specific well-motivated BSM models could be directly constrained experimentally, a model-independent approach allowing a re-interpretation of the results of dedicated searches for additional Higgs bosons decaying into BSM particles and of ``mono-X plus missing energy'' searches within wide classes of BSM models would clearly be more useful. The development of such an approach is the main purpose of this work. Focusing for definiteness in the present paper on mono-$Z$ and mono-Higgs plus missing transverse energy final states, we introduce simplified model topologies that describe different decay modes of the heavy neutral resonance. This approach resembles the one that has been adopted for the simplified models that are used in direct LHC searches for supersymmetric particles. As a first step, we categorize the different decay topologies. For each topology we then investigate the most relevant kinematic distributions and discuss the question to what extent the different decay topologies can be distinguished
from each other experimentally. We exemplify our approach by recasting an existing ATLAS mono-Higgs plus missing transverse energy search.

Our paper is structured as follows. In \cref{sec:simplified_model}, we detail our simplified model approach. We then use this approach to classify and kinematically compare mono-$Z$ and mono-Higgs plus missing energy final states in \cref{sec:Z_ETmiss,sec:H_ETmiss}, respectively. In \cref{sec:monoH+MET_example}, we exemplify our approach by recasting an existing ATLAS mono-Higgs plus missing transverse energy search. The conclusions can be found in \cref{sec:conclusions}. \cref{sec:model_file,sec:event_generation} give details on the employed model file and the Monte-Carlo event generation.


\section{Simplified model approach}
\label{sec:simplified_model}

Our simplified model approach represents an extension of the SM. In order to model heavy scalar resonance decays to SM particles plus missing energy, we supplement the SM particles by one additional heavy scalar $\phi$, one mediator $\Med$, and one invisible particle $\Inv$. In the present paper, we assume the heavy scalar $\phi$ to be produced resonantly via gluon fusion or bottom-associated Higgs production, which are typical production modes for BSM Higgs bosons.

We assume the mediator particle to be electrically neutral.\footnote{For the mono-$Z$ and mono-Higgs plus missing energy signatures considered in this work, an electrically charged mediator cannot contribute.} Moreover, we allow it to be either a scalar, a fermion or a vector boson. Similarly, the invisible particle, which is electrically neutral, can be a scalar, a fermion or a vector boson. This particle is assumed to be invisible for typical particle detectors. This can be the case in particular if it is stable (e.g.\ due to a protecting symmetry) or sufficiently long-lived.

In summary, our simplified model approach contains the following particles in addition to the SM particles:
\begin{itemize}
  \item heavy scalar: $\phi$,
  \item neutral mediator: $\Ms$, $\Mf$, or $\Mv$,
  \item invisible particle: $\Is$, $\If$, or $\Iv$,
\end{itemize}
where the subscript ``$S$'' is used to denote scalar particles; the subscript ``$F$'' to denote fermions; and the subscript ``$V$'' to denote vector bosons. Each simplified model topology contains only one type of the possible mediators and one type of invisible particle. While in principle it would be straightforward to include finite width effects,
we assume in this paper that all the scalar resonances and the mediators have negligible widths for the sake of clarity of the presentation. Correspondingly, we also assume in this work that all BSM particles (except of the invisible particle) have a short life time and decay promptly.

Our simplified model Lagrangian then contains --- in addition to the SM interactions --- all possible interactions between the BSM particles as well as between the BSM and SM particles. Motivated by the fact that the effects of four-point interactions in decay processes are normally suppressed due to the large amount of energy required to produce three particles at the same time (and also to reduce the complexity of our models), we include only three-point interactions. Note that this is not a restriction of our framework. It can easily be extended to include four-point interactions.

We do not explicitly fix the \cp character of the scalars as well as the chirality of the vector bosons. Instead, our model implementation allows one to set the \cp character (or the chirality) of each coupling separately. The presence of certain interactions, however, implicitly restricts the \cp character of the involved scalars. For example, the presence of a scalar--scalar--vector-boson coupling ($\Ms\Is Z$) implies that the two scalars have an opposite \cp character. Note also that due to the Lorentz symmetry all vector-boson--vector-boson--vector-boson couplings involving at least two identical vector bosons are zero.

We have implemented the model as outlined above as a \texttt{FeynRules}~\cite{Christensen:2008py,Alloul:2013bka} model. This allows one to easily derive model files for many common particle physics tools. We make use of \texttt{FeynRules} to derive a \texttt{UFO} model file~\cite{Degrande:2011ua} that is employed for generating Monte-Carlo event samples. More details on the \texttt{UFO} model file can be found in \cref{sec:model_file}. These model files are distributed as ancillary material for the present paper.

\medskip

In the subsequent Sections, we work out a complete classification of the different models and decay topologies that can be constructed using our simplified model setup. For each final-state signature (e.g.\ one $Z$ boson plus missing transverse energy), we construct all possible decay topologies of the neutral scalar decay either directly or via up to two mediators to the respective final state. For each of these decay topologies, we then formulate all possible spin configurations for the mediator and the invisible particles. In order to compare the kinematic distributions of each different topology (and spin configuration), we generate Monte-Carlo (MC) samples for each of the topologies and spin realizations (see \cref{sec:event_generation} for details on the event generation).

For our numerical analysis, we concentrate on four benchmark points (BPs),
\begin{itemize}
  \item BP1: $m_\phi = 1\tev$, $m_\Med = 400\gev$, $m_\Inv = 10\gev$,
  \item BP2: $m_\phi = 1\tev$, $m_\Med = 400\gev$, $m_\Inv = 100\gev$,
  \item BP3: $m_\phi = 1\tev$, $m_\Med = 260\gev$, $m_\Inv = 10\gev$,
  \item BP4: $m_\phi = 1.5\tev$, $m_\Med = 400\gev$, $m_\Inv = 10\gev$.
\end{itemize}
BP1 is designed to exemplify a hierarchical spectrum between the heavy scalar, the mediator, and the invisible particle. For BP2, the hierarchy between the mediator and the invisible particle is reduced by raising the mass of the invisible particle. For BP3, the mass of the mediator is lowered also resulting in a smaller hierarchy between the mediator and the invisible particle. BP4 targets the case of an even enlarged hierarchy between the heavy scalar and the other particles by raising $m_\phi$ to 1.5~TeV.

The goal of our study is, however, not only to classify all possible topologies and to discuss their kinematic differences, but also to provide a framework for the presentation of experimental searches allowing for a straightforward re-interpretation of these searches in concrete BSM models. As we will discuss in detail with a concrete example in \cref{sec:monoH+MET_example}, tabulated acceptance $\times$ efficiency values in dependence of $m_\phi$, $m_\Med$, and $m_\Inv$ allow a straightforward re-interpretation of a specific experimental search for different BSM models that realize one of the simplified model topologies.

While in many models only one simplified model topology appears, it is also possible that a BSM model gives rise to more than one simplified model topology. In this case it is still possible to apply our simplified model framework by summing the cross section $\times$ efficiency $\times$ acceptance values for each of the contributing topologies. Moreover, it is also possible that a specific topology appears multiple times in the BSM model. Also in this case the cross section $\times$ efficiency $\times$ acceptance values can be summed (see the discussion of the example in \cref{sec:monoH+MET_example}). In this procedure, possible interference effects are neglected.


\section{Simplified models for mono-\texorpdfstring{$Z$}{Z} boson + \texorpdfstring{$\ETmiss$}{ETmiss} signatures}
\label{sec:Z_ETmiss}

As first signature, we discuss the mono-$Z$ plus missing energy final state. This signature has already been searched for intensively by the experimental collaborations~\cite{Aad:2014vka,Aaboud:2016qgg,Aaboud:2017bja,Aaboud:2018xdl,Khachatryan:2015bbl,Sirunyan:2017jix,Sirunyan:2017qfc,Sirunyan:2017hci,Sirunyan:2017onm,Sirunyan:2020fwm}. The results of these searches are often interpreted in terms of $Z'$ models or the TDHM extended with a pseudoscalar dark matter portal (THDMa).


\subsection{Topologies}

In order to have a final state involving a $Z$ boson and invisible particles, the initial neutral scalar can either decay directly to a $Z$ boson and an invisible particle or the decay can be induced by up to two electrically neutral mediators.\footnote{As explained above (see \cref{sec:simplified_model}), we do not consider topologies with four-point interactions.}

\begin{figure}\centering
\begin{subfigure}[t]{\subfigsizetwo\linewidth}\centering
\includegraphics[scale=1]{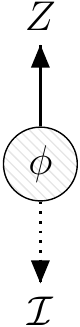}
\caption{1vs1}
\label{fig:monoZ_topologies_1vs1}
\end{subfigure}
\begin{subfigure}[t]{\subfigsizetwo\linewidth}\centering
\includegraphics[scale=1]{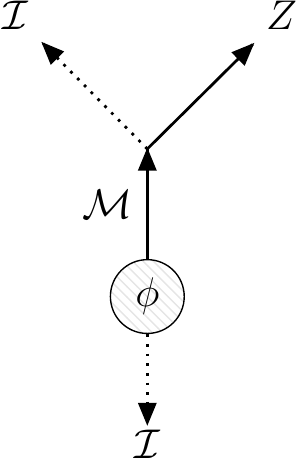}
\caption{2vs1 balanced}
\label{fig:monoZ_topologies_2vs1_balanced}
\end{subfigure}
\\[1em]
\begin{subfigure}[t]{\subfigsizetwo\linewidth}\centering
\includegraphics[scale=1]{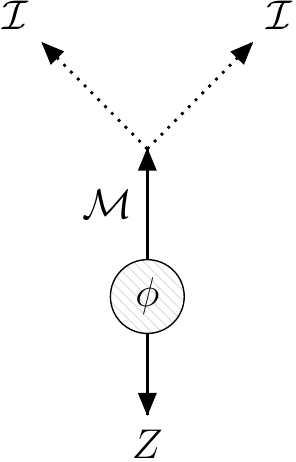}
\caption{2vs1 unbalanced}
\label{fig:monoZ_topologies_2vs1_unbalanced}
\end{subfigure}
\begin{subfigure}[t]{\subfigsizetwo\linewidth}\centering
\includegraphics[scale=1]{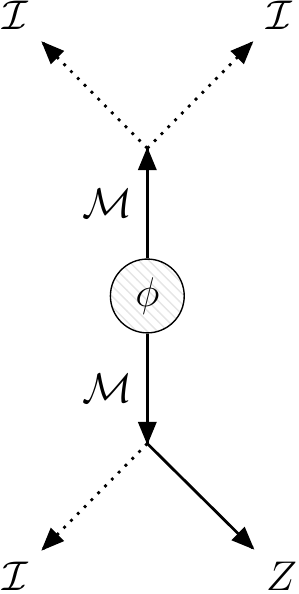}
\caption{2vs2}
\label{fig:monoZ_topologies_2vs2}
\end{subfigure}
\caption{Decay topologies of a neutral scalar boson $\phi$ decaying in its rest frame to a $Z$ boson plus $\ETmiss$.}
\label{fig:monoZ_topologies}
\end{figure}

\begin{figure}\centering
\begin{subfigure}{\subfigsizetwo\linewidth}\centering
\includegraphics[scale=1]{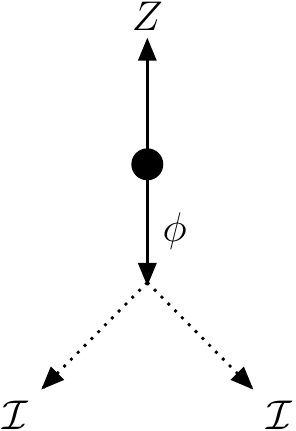}
\label{fig:monoZ_topologies_ISR1}
\end{subfigure}
\begin{subfigure}{\subfigsizetwo\linewidth}\centering
\includegraphics[scale=1]{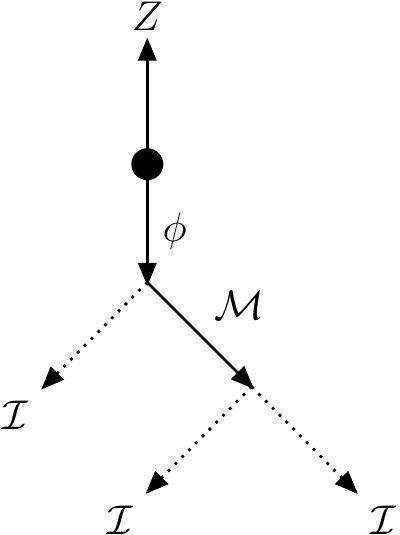}
\label{fig:monoZ_topologies_ISR2}
\end{subfigure}
\begin{subfigure}{\subfigsizetwo\linewidth}\centering
\includegraphics[scale=1]{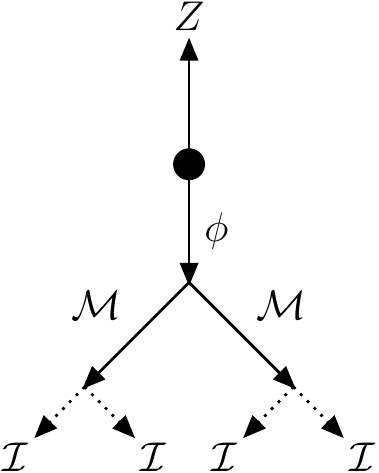}
\label{fig:monoZ_topologies_ISR3}
\end{subfigure}
\caption{Initial state radiation topologies where the $Z$ boson is radiated from the inital state and then the scalar resonance decays into invisible particles. }
\label{fig:monoZ_topologies_ISR}
\end{figure}

In general, one can distinguish five different types of final-state topologies:
\begin{enumerate}[(a)]
  \item 1vs1 unbalanced topology (see \cref{fig:monoZ_topologies_1vs1})

        The scalar resonance decays directly to an invisible particle and a $Z$ boson. The $Z$ boson recoils against the invisible particle resulting in a missing energy spectrum peaking at high \ETmiss values.

  \item 2vs1 balanced topology (see \cref{fig:monoZ_topologies_2vs1_balanced})

        The scalar resonance decays to a mediator and an invisible particle. The mediator then decays to a $Z$ boson and an invisible particle. Since the scalar resonance is produced approximately at rest, the mediator recoils against the first invisible particle resulting in a ``balanced'' missing energy spectrum.

  \item 2vs1 unbalanced topology (see \cref{fig:monoZ_topologies_2vs1_unbalanced})

        The scalar resonance decays to a mediator and a $Z$ boson. The mediator then decays to two invisible particle. For this topology, the $Z$ boson recoils against the mediator resulting in a missing energy spectrum peaking at high \ETmiss values (kinematically similar to the 1vs1 topology).

  \item 2vs2 balanced topology (see \cref{fig:monoZ_topologies_2vs2})

        The scalar resonance decays to two mediators. One of the mediators then decays to a $Z$ boson and an invisible particle. The second mediator decays to two invisible particles. This topology is kinematically similar to the 2vs1 balanced topology and features a similar \ETmiss spectrum.

  \item initial state radiation (ISR) topology (see \cref{fig:monoZ_topologies_ISR})

        The $Z$ boson is radiated from the initial state. The scalar resonance $\phi$ then decays completely to invisible particles. Note that the different decay topologies as shown in \cref{fig:monoZ_topologies_ISR} can not be distinguished experimentally.
\end{enumerate}


\subsection{Feynman diagram realizations}

\begin{figure}\centering
\begin{subfigure}{\subfigsize\linewidth}\centering
\includegraphics[scale=1]{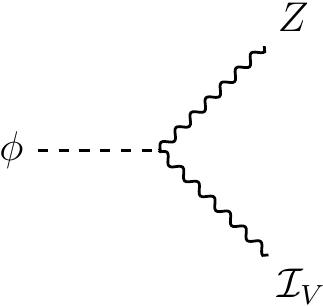}
\caption{$\Iv$}
\label{fig:monoZ_ETmiss_1vs1_Iv}
\end{subfigure}
\begin{subfigure}{\subfigsize\linewidth}\centering
\includegraphics[scale=1]{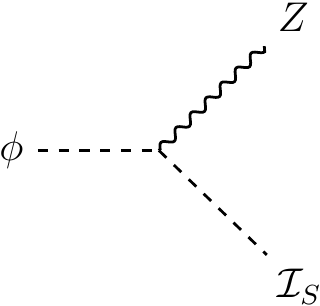}
\caption{$\Is$}
\label{fig:monoZ_ETmiss_1vs1_Is}
\end{subfigure}
\caption{Mono-$Z$ + $\ETmiss$ processes for the unbalanced 1vs1 topology.}
\label{fig:monoZ_ETmiss_1vs1}
\end{figure}

\begin{figure}\centering
\begin{subfigure}{\subfigsize\linewidth}\centering
\includegraphics[scale=1]{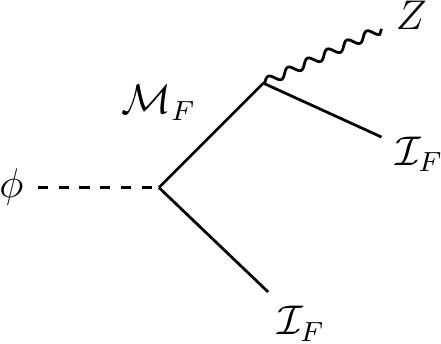}
\caption{$\Mf\If$}
\label{fig:monoZ_ETmiss_2vs1_balanced_MfIf}
\end{subfigure}
\begin{subfigure}{\subfigsize\linewidth}\centering
\includegraphics[scale=1]{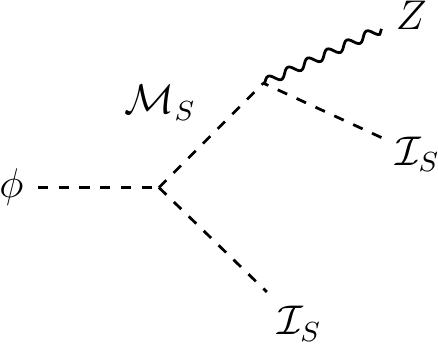}
\caption{$\Ms\Is$}
\label{fig:monoZ_ETmiss_2vs1_balanced_MsIs}
\end{subfigure}
\begin{subfigure}{\subfigsize\linewidth}\centering
\includegraphics[scale=1]{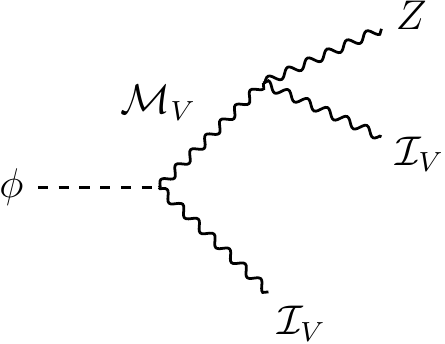}
\caption{$\Mv\Iv$}
\label{fig:monoZ_ETmiss_2vs1_balanced_MvIv}
\end{subfigure}
\begin{subfigure}{\subfigsize\linewidth}\centering
\includegraphics[scale=1]{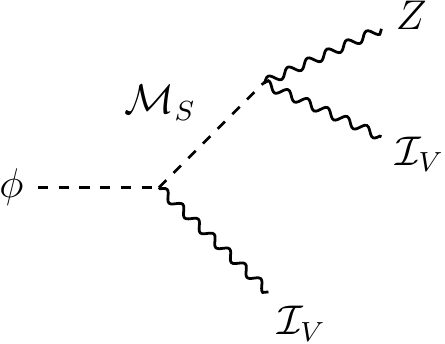}
\caption{$\Ms\Iv$}
\label{fig:monoZ_ETmiss_2vs1_balanced_MsIv}
\end{subfigure}
\begin{subfigure}{\subfigsize\linewidth}\centering
\includegraphics[scale=1]{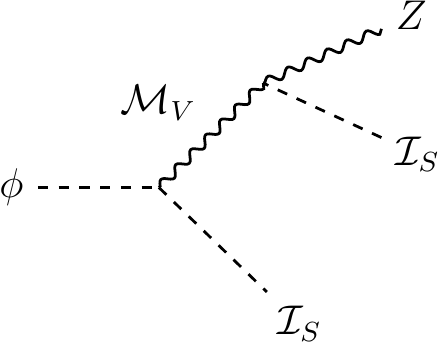}
\caption{$\Mv\Is$}
\label{fig:monoZ_ETmiss_2vs1_balanced_MvIs}
\end{subfigure}
\caption{Mono-$Z$ + $\ETmiss$ processes for the 2vs1 balanced topology.}
\label{fig:monoZ_ETmiss_2vs1_balanced}
\end{figure}

\begin{figure}\centering
\begin{subfigure}{\subfigsize\linewidth}\centering
\includegraphics[scale=1]{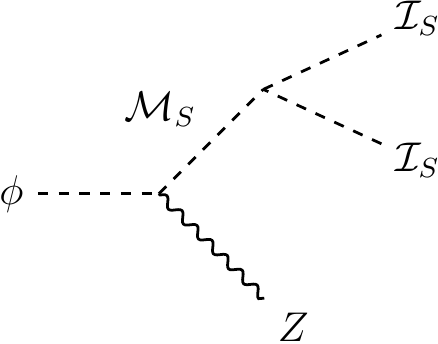}
\caption{$\Ms\Is$}
\label{fig:monoZ_ETmiss_2vs1_unbalanced_MsIs}
\end{subfigure}
\begin{subfigure}{\subfigsize\linewidth}\centering
\includegraphics[scale=1]{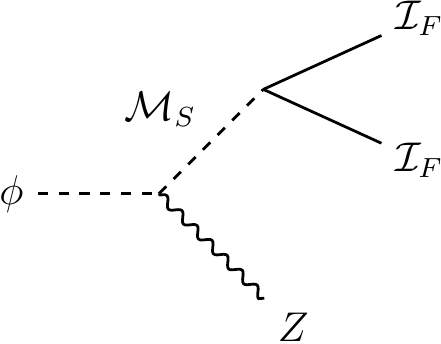}
\caption{$\Ms\If$}
\label{fig:monoZ_ETmiss_2vs1_unbalanced_MsIf}
\end{subfigure}
\\
\begin{subfigure}{\subfigsize\linewidth}\centering
\includegraphics[scale=1]{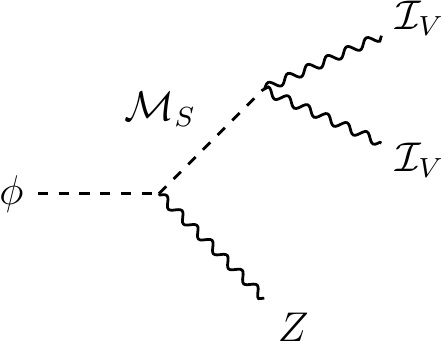}
\caption{$\Ms\Iv$}
\label{fig:monoZ_ETmiss_2vs1_unbalanced_MsIv}
\end{subfigure}
\begin{subfigure}{\subfigsize\linewidth}\centering
\includegraphics[scale=1]{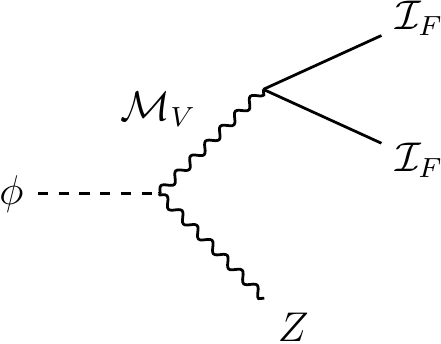}
\caption{$\Mv\If$}
\label{fig:monoZ_ETmiss_2vs1_unbalanced_MvIf}
\end{subfigure}
\caption{Mono-$Z$ + $\ETmiss$ processes for the 2vs1 unbalanced topology.}
\label{fig:monoZ_ETmiss_2vs1_unbalanced}
\end{figure}

\begin{figure}\centering
\begin{subfigure}{\subfigsize\linewidth}\centering
\includegraphics[scale=1]{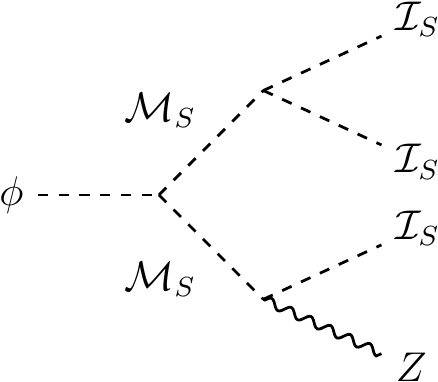}
\caption{$\Ms\Is$}
\label{fig:monoZ_ETmiss_2vs2_MsIs}
\end{subfigure}
\begin{subfigure}{\subfigsize\linewidth}\centering
\includegraphics[scale=1]{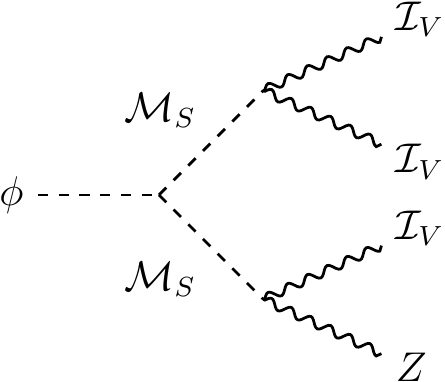}
\caption{$\Ms\Iv$}
\label{fig:monoZ_ETmiss_2vs2_MsIv}
\end{subfigure}
\caption{Mono-$Z$ + $\ETmiss$ processes for the 2vs2 topology.}
\label{fig:monoZ_ETmiss_2vs2}
\end{figure}

The Feynman diagram realizations of the different topologies (employing different spin hypothesis for the mediator and the invisible particle) are shown in \cref{fig:monoZ_ETmiss_1vs1,fig:monoZ_ETmiss_2vs1_balanced,fig:monoZ_ETmiss_2vs1_unbalanced,fig:monoZ_ETmiss_2vs2}.

For the unbalanced 1vs1 topology, two different Feynman diagram realizations exist: $\Iv$ (see \cref{fig:monoZ_ETmiss_1vs1_Iv}) and $\Is$ (see \cref{fig:monoZ_ETmiss_1vs1_Is}). This kind of topology can be present, for example, in models that contain a heavy scalar and a dark photon/$Z'$ in the case of \cref{fig:monoZ_ETmiss_1vs1_Iv} or a long-lived \cp-odd scalar, like for example in THDMa, in the case of \cref{fig:monoZ_ETmiss_1vs1_Is}.

For the balanced 2vs1 topology, five different Feynman diagram realizations exist: $\Mf\If$ (see \cref{fig:monoZ_ETmiss_2vs1_balanced_MfIf}), $\Ms\Is$ (see \cref{fig:monoZ_ETmiss_2vs1_balanced_MsIs}), $\Mv\Iv$ (see \cref{fig:monoZ_ETmiss_2vs1_balanced_MvIv}), $\Ms\Iv$ (see \cref{fig:monoZ_ETmiss_2vs1_balanced_MsIv}), and $\Mv\Is$ (see \cref{fig:monoZ_ETmiss_2vs1_balanced_MvIs}). The 2vs1 topology for a mono-$Z$ boson plus missing energy final state can be realized for example in the MSSM: a heavy Higgs boson is produced via gluon fusion or in association with bottom quarks; it then decays to the lightest (invisible particle) and the second-lightest neutralino (mediator); the second-lightest neutralino then decays to a $Z$ boson and the lightest neutralino. This cascade decay is a promising search channel at the LHC, and its cross section can exceed the cross section of direct neutralino pair production~\cite{Bahl:2018zmf,Gori:2018pmk,Bahl:2019ago,Adhikary:2020ujn,Liu:2020muv}. Another example for a concrete model realizing the $\Ms\Is$ diagram is the N2HDM, which extends the SM by a second Higgs doublet and a real singlet. Assuming that one of the doublets is inert (i.e.\ protected by a $\mathbf{Z}_2$ symmetry, see e.g.~\ccite{Engeln:2020fld}), the initial scalar resonance, which could be a dominantly singlet-like state, can decay to the heavier and lighter \cp-even states of the inert doublet. The heavier inert state then decays to a $Z$ boson and the lighter inert state which is stable due to the $\mathbf{Z}_2$ symmetry.

For the unbalanced 2vs1 topology, four different Feynman diagram realizations exist: $\Ms\Is$ (see \cref{fig:monoZ_ETmiss_2vs1_unbalanced_MsIs}), $\Ms\If$ (see \cref{fig:monoZ_ETmiss_2vs1_unbalanced_MsIf}), $\Ms\Iv$ (see \cref{fig:monoZ_ETmiss_2vs1_unbalanced_MsIv}), and $\Mv\If$ (see \cref{fig:monoZ_ETmiss_2vs1_unbalanced_MvIf}). One possible realization of the unbalanced 2vs1 topology can occur in the Two-Higgs-Doublet model with an additional \cp-odd singlet, dubbed as THDMa, which serves as a DM portal~\cite{Ipek:2014gua,No:2015xqa,Goncalves:2016iyg,Bauer:2017ota,Tunney:2017yfp}. This model is also frequently used as a benchmark model in experimental analyses, see for instance Ref.~\cite{ATLAS:2021gcn}. As discussed e.g.\ in \ccite{Tunney:2017yfp}, searches for a $\bar b b Z(\to\ell\ell) + \ETmiss$ final state have the potential to probe still unconstrained regions of the parameter space compatible with an explanation of the observed Galactic Centre gamma ray excess~\cite{TheFermi-LAT:2015kwa}. Another examplary model realizing the unbalanced 2vs1 topology is the MSSM: a heavy \cp-even Higgs boson decays to a \cp-odd Higgs boson and a $Z$ boson. The \cp-odd Higgs boson subsequently decays to two neutralinos.

For the 2vs2 topology, only two different Feynman diagrammatic realization exist: $\Ms\Is$ (see \cref{fig:monoZ_ETmiss_2vs2_MsIs}) and $\Ms\Iv$ (see \cref{fig:monoZ_ETmiss_2vs2_MsIv}). This topology can be present, for example, in the singlet-extended THDM (NTHDM) or the Next-To-Minimal-Supersymmetric extension of the SM (NMSSM), where the mediator $\Ms$ can be a singlet Higgs and $\Is$ a light pseudoscalar.

For the ISR topology, several Feynman diagrammatic realizations are possible. Experimentally, these are, however, not distinguishable. Therefore, we do not explicitly show Feynman diagrams here.


\subsection{Kinematic analysis}

For the kinematic analysis of the mono-$Z$ plus missing energy final state, we concentrate on the $Z$ boson decay to two muons. This decay mode should provide the best experimental resolution. Moreover, we assume that the scalar resonance is produced via bottom-associated Higgs production. We expect our result to be qualitatively similar for the other $Z$ boson decay modes (including off-shell decays) and for Higgs production via gluon fusion.

Since there is only one visible particle in the final state (apart from the usual QCD radiation), the number of relevant observables is low. As a consequence of the scalar resonance being produced approximately at rest, the momentum of the $Z$ boson has no preferred direction. The transverse momentum of the $Z$ boson, however, contains valuable information about the underlying process.\footnote{A full multivariate analysis could be able to extract additional information. Carrying out such an analysis is, however, beyond the scope of the current work.}

\medskip

\begin{figure}\centering
\begin{subfigure}{.48\linewidth}\centering
\includegraphics[width=\linewidth]{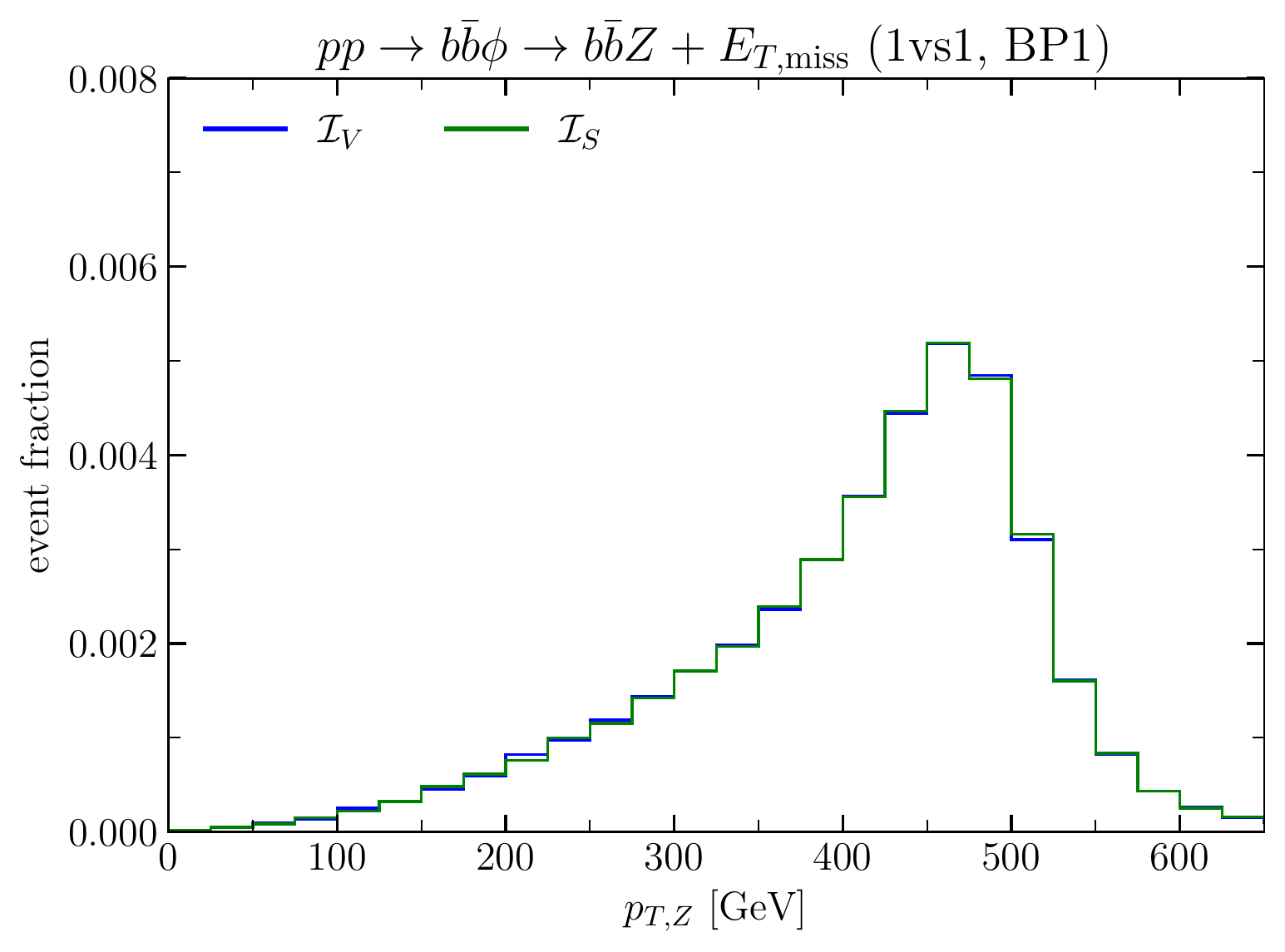}
\end{subfigure}
\begin{subfigure}{.48\linewidth}\centering
\includegraphics[width=\linewidth]{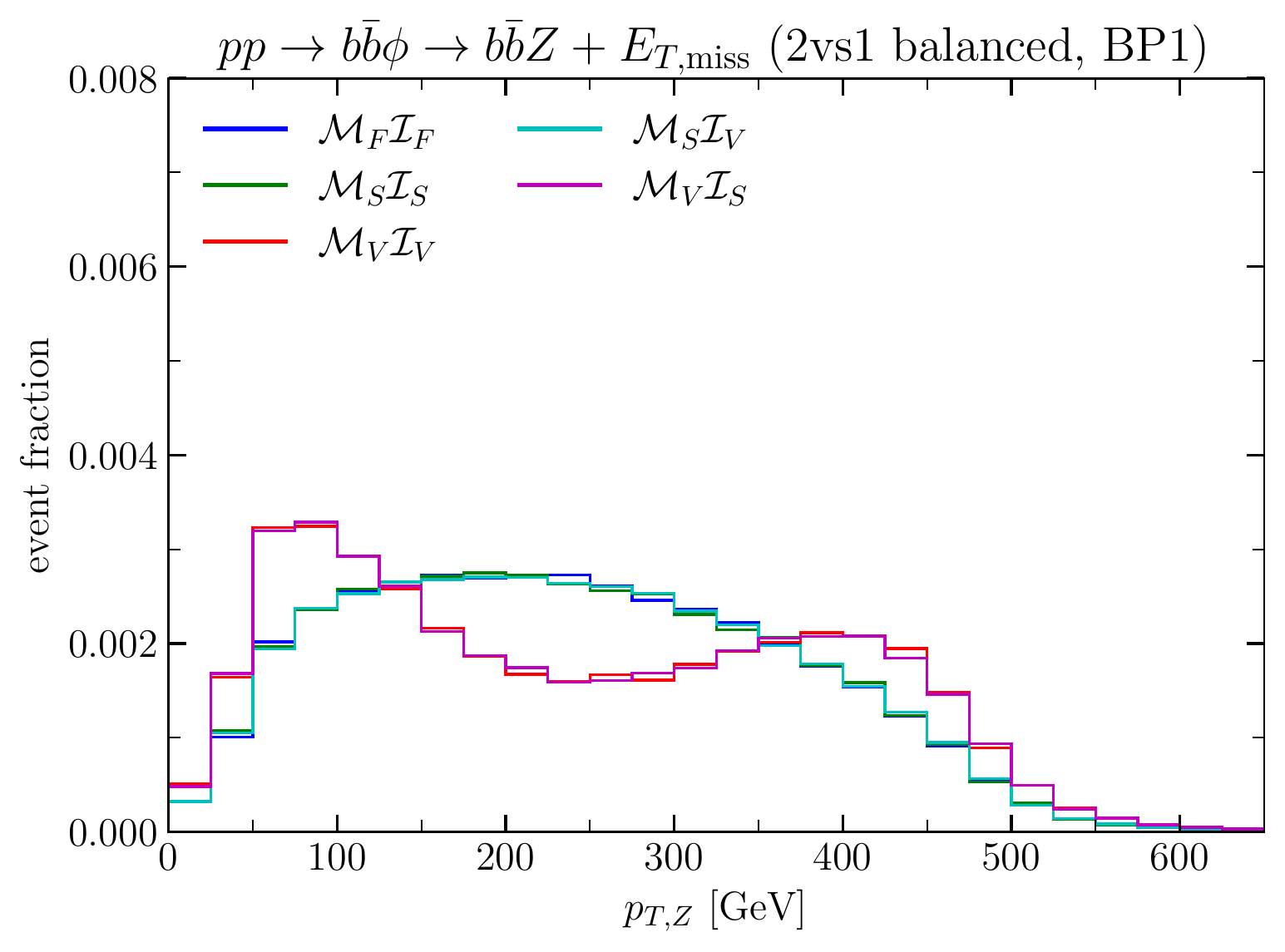}
\end{subfigure}
\begin{subfigure}{.48\linewidth}\centering
\includegraphics[width=\linewidth]{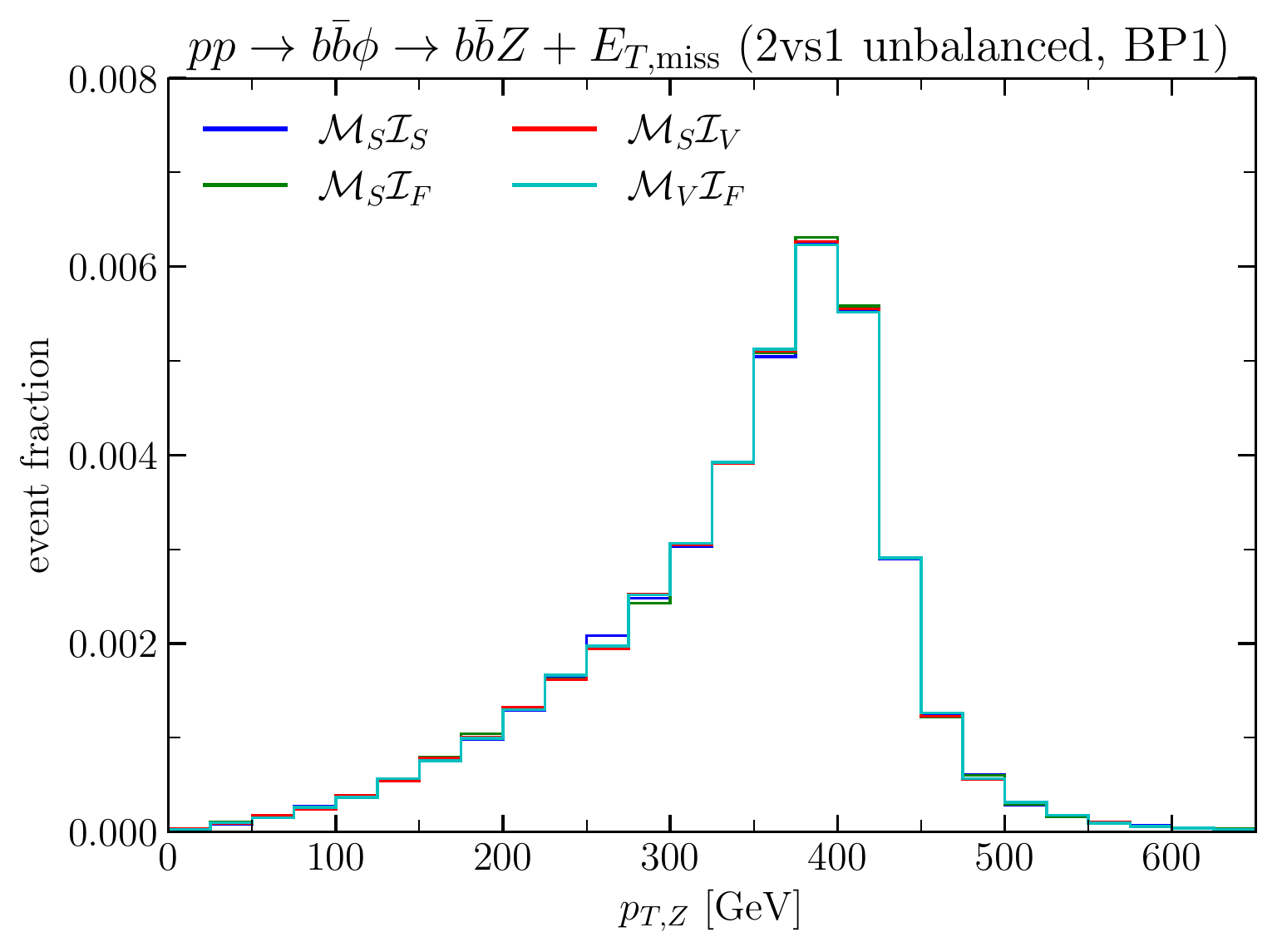}
\end{subfigure}
\begin{subfigure}{.48\linewidth}\centering
\includegraphics[width=\linewidth]{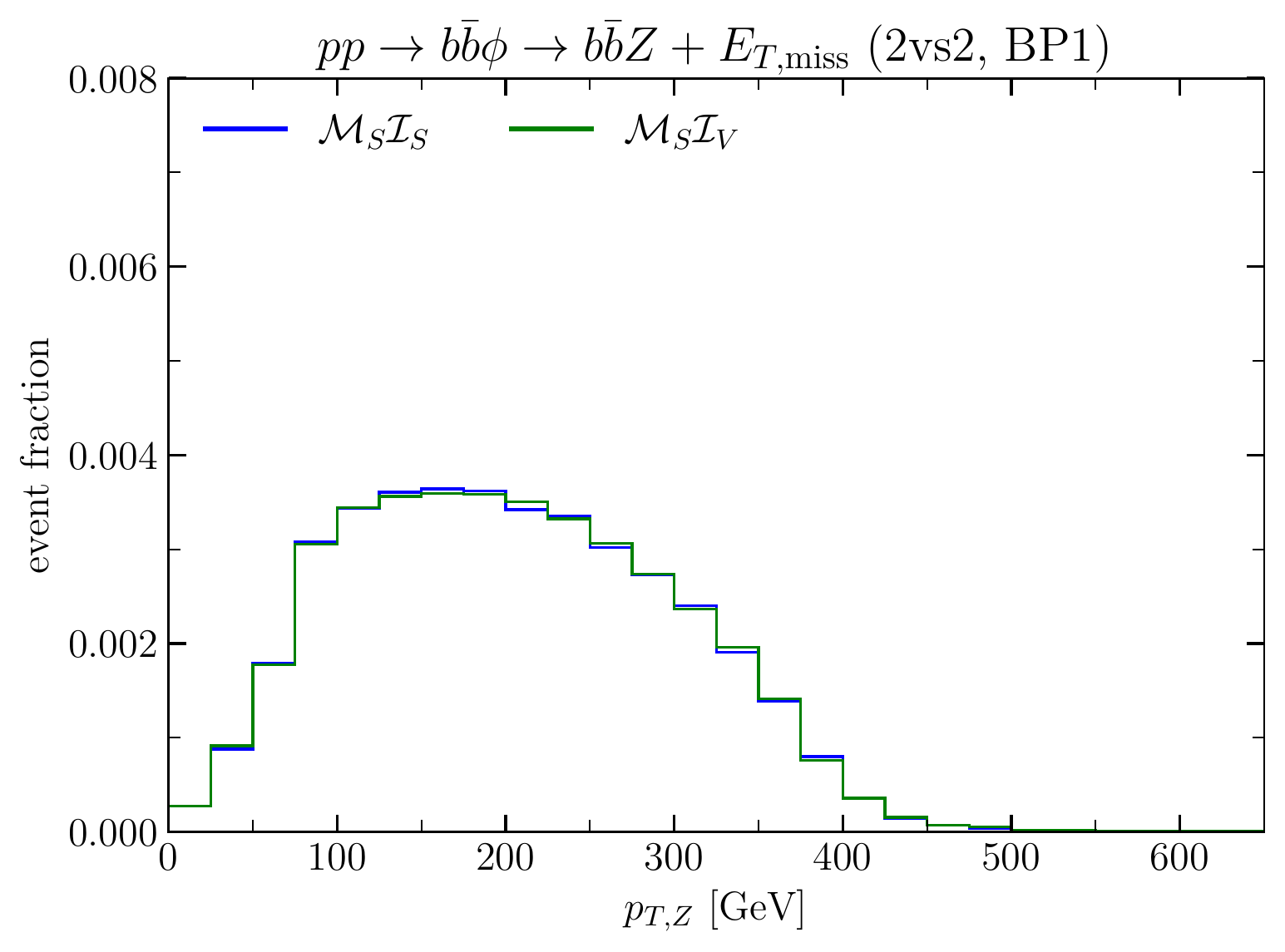}
\end{subfigure}
\begin{subfigure}{.48\linewidth}\centering
\includegraphics[width=\linewidth]{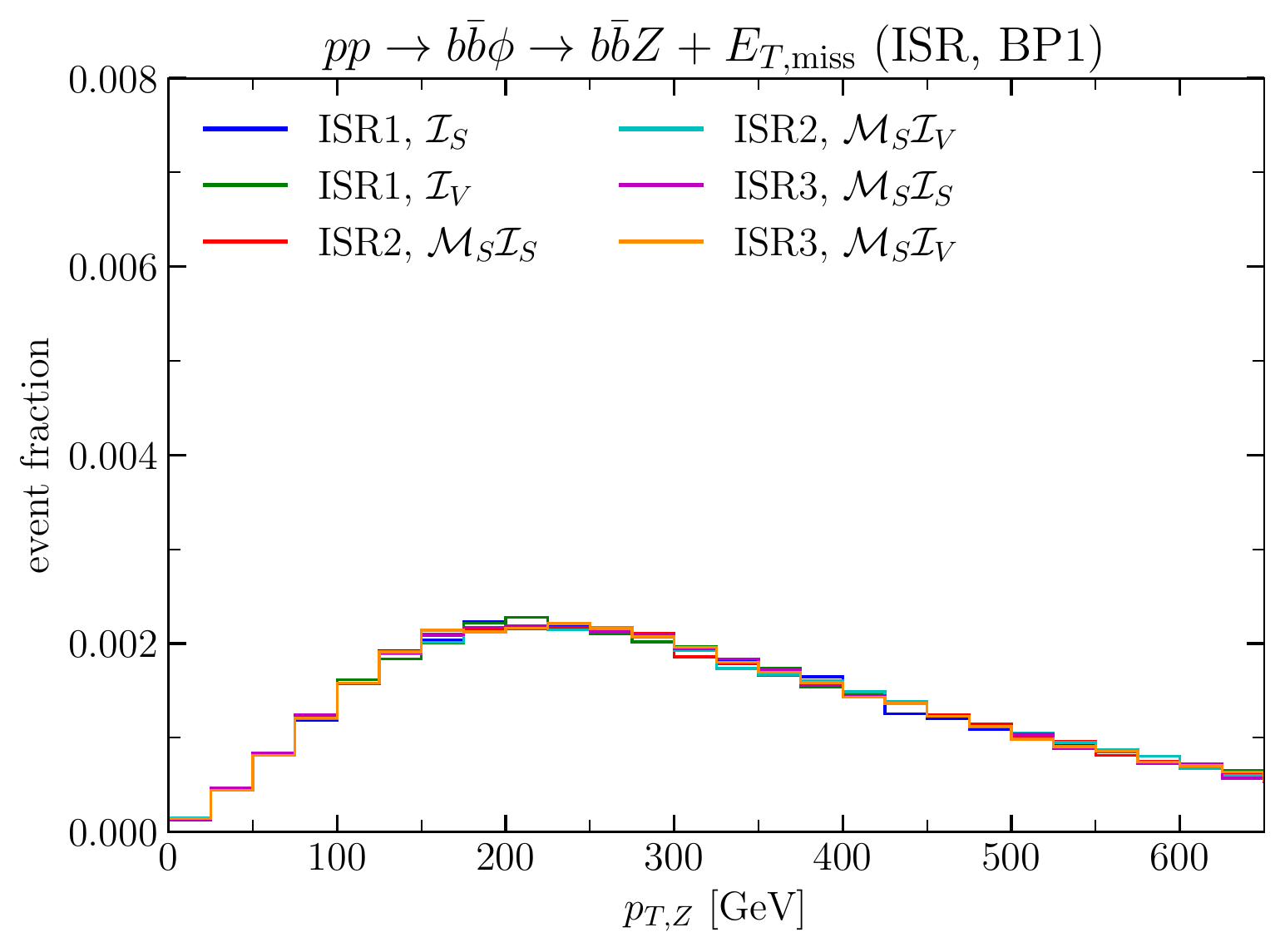}
\end{subfigure}
\caption{Mono-$Z + \ETmiss$ final state: transverse momentum distribution of the di-muon system for BP1. The upper left panel displays the results for the 1vs1 balanced topology; the upper right panel for the 2vs1 balanced topology; the middle left panel for the 2vs1 unbalanced topology; the middle right panel for of the 2vs2 topology; and the lower panel for the ISR topology.}
\label{fig:monoZ_top_comp}
\end{figure}

In \cref{fig:monoZ_top_comp}, we compare the transverse momentum distributions\footnote{It is important to note that all shown distributions are obtained after applying showering and detector effects to the parton level events. For more details about the event production chain see \cref{sec:event_generation}.} of the $Z$ boson (labelled with $p_{T,\mu\mu}$) for BP1. All distributions are generated assuming bottom-associated $\phi$ production. As mentioned above, we expect very similar results for $\phi$ production via gluon fusion.

In the upper left panel, the distributions for the 1vs1 topology are shown. The distributions peak at $\sim 450\gev$. If no detector effects are taken into account, the distribution has a sharp endpoint at $\sim 495\gev$. The $\Iv$ and $\Is$ spin realizations are not distinguishable experimentally for this distribution.

In the upper right panel of \cref{fig:monoZ_top_comp}, we display the distributions for the 2vs1 balanced topology. Since the $Z$ boson is not recoiling against a single invisible particle, the transverse momentum spectrum has no clear peak and is spread out up to values of $\sim 550\gev$. While the $\Mf\If$, $\Ms\Iv$, and $\Ms\Is$ spin hypotheses are hardly distinguishable featuring a broad peak at around $250\gev$, the spin hypotheses with a vector mediator --- $\Mv\If$ and $\Mv\Is$ --- give rise to a kinematic spectrum featuring two peaks at $\sim 80\gev$ and $\sim 400\gev$. This difference in the kinematic spectra arises as a consequence of the Lorentz structure of the scalar--vector--vector (or scalar--scalar--vector) coupling between the scalar resonance, the mediator, and the invisible particle. In order to answer the question whether this difference between the vector- and the non-vector-mediator spin hypotheses is experimentally distinguishable, a dedicated study implementing a realistic analysis flow would be required.

The kinematic distributions of the 2vs1 unbalanced topology --- as shown in the middle left panel of \cref{fig:monoZ_top_comp} --- are very similar to the distributions of the 1vs1 topology. They feature a clear peak at $\sim 400\gev$. The different spin hypothesis are not distinguishable experimentally.

For the 2vs2 topology --- as shown in the middle right panel of \cref{fig:monoZ_top_comp} --- the kinematic distributions are similar to the 2vs1 balanced topology. As only scalar mediators are possible, the two different spin hypotheses are not distinguishable. The distributions have a broad peak at $\sim 150\gev$. In comparison to the 2vs1 balanced topology, the distributions have a slightly lower endpoint at $\sim 450\gev$ (vs.\ $\sim 550\gev$ for the 2vs1 unbalanced topology).

As a final topology, we take into account the ISR topology shown in the lower panel of \cref{fig:monoZ_top_comp}. As already mentioned above, the different spin hypotheses are indistinguishable. The $p_{T,\mu\mu}$ distribution is very broad reaching its maximum at $\sim 200\gev$.

\medskip

To sum up, we find that the results for the different simplified model topologies are robust with respect to variations of the spins of the mediator and the invisible particles. Our results indicate that the different spin hypotheses will not be experimentally distinguishable, with the possible exception of the 2vs1 balanced topology for which the distribution in case of a vector mediator shows different patterns than for the other spin hypotheses.

\medskip

\begin{figure}\centering
\begin{subfigure}{.48\linewidth}\centering
\includegraphics[width=\linewidth]{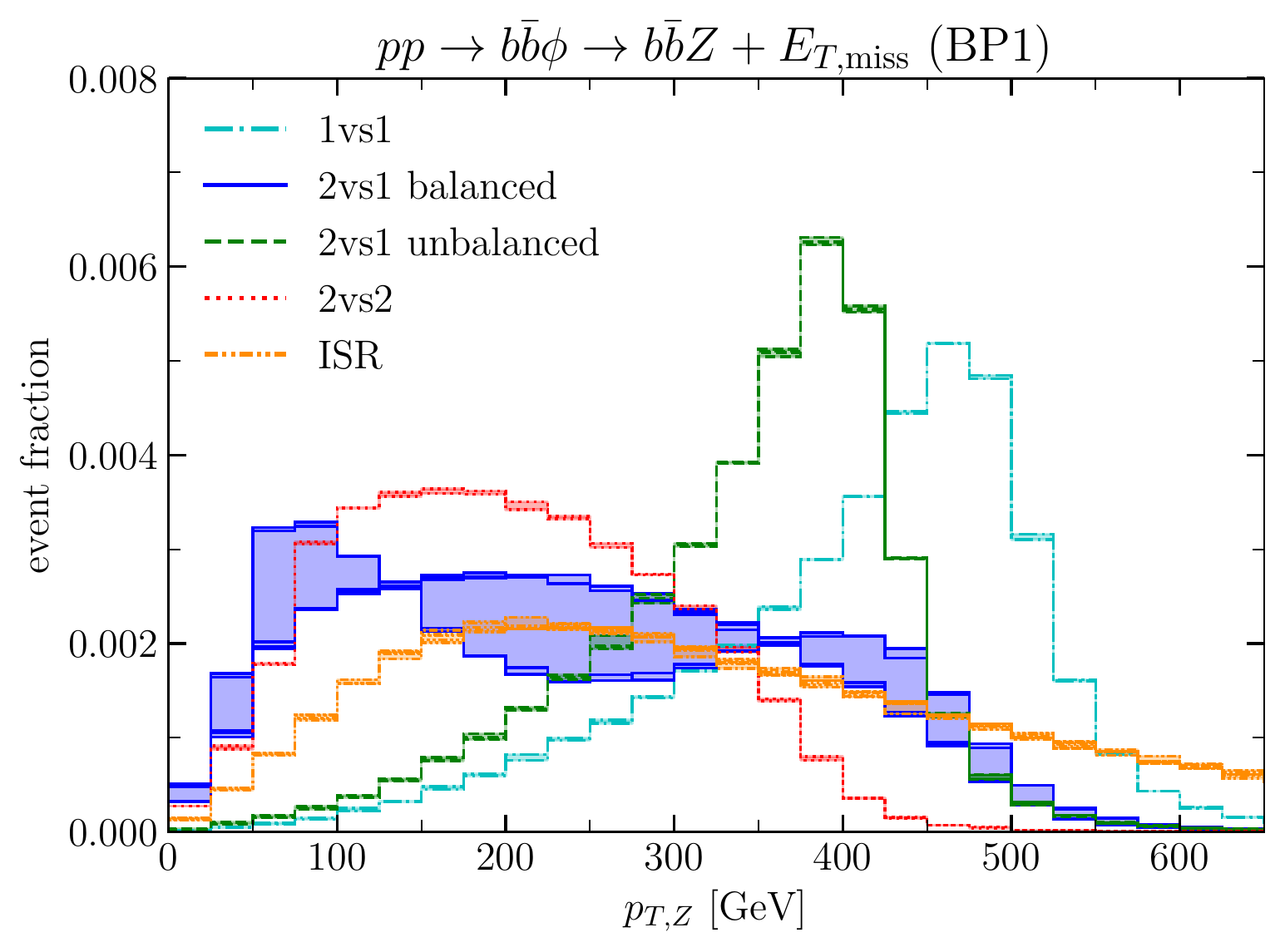}
\end{subfigure}
\begin{subfigure}{.48\linewidth}\centering
\includegraphics[width=\linewidth]{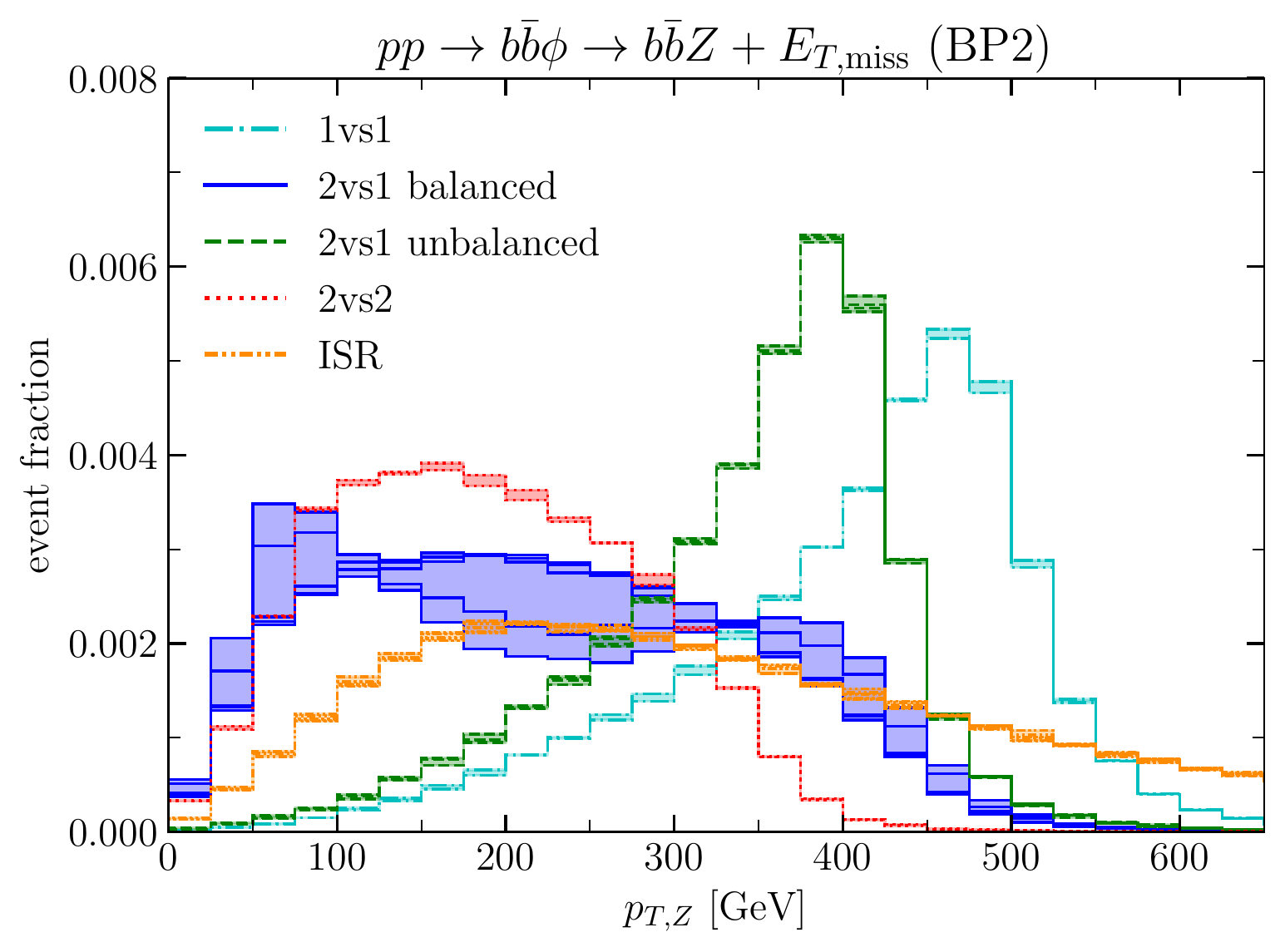}
\end{subfigure}
\begin{subfigure}{.48\linewidth}\centering
\includegraphics[width=\linewidth]{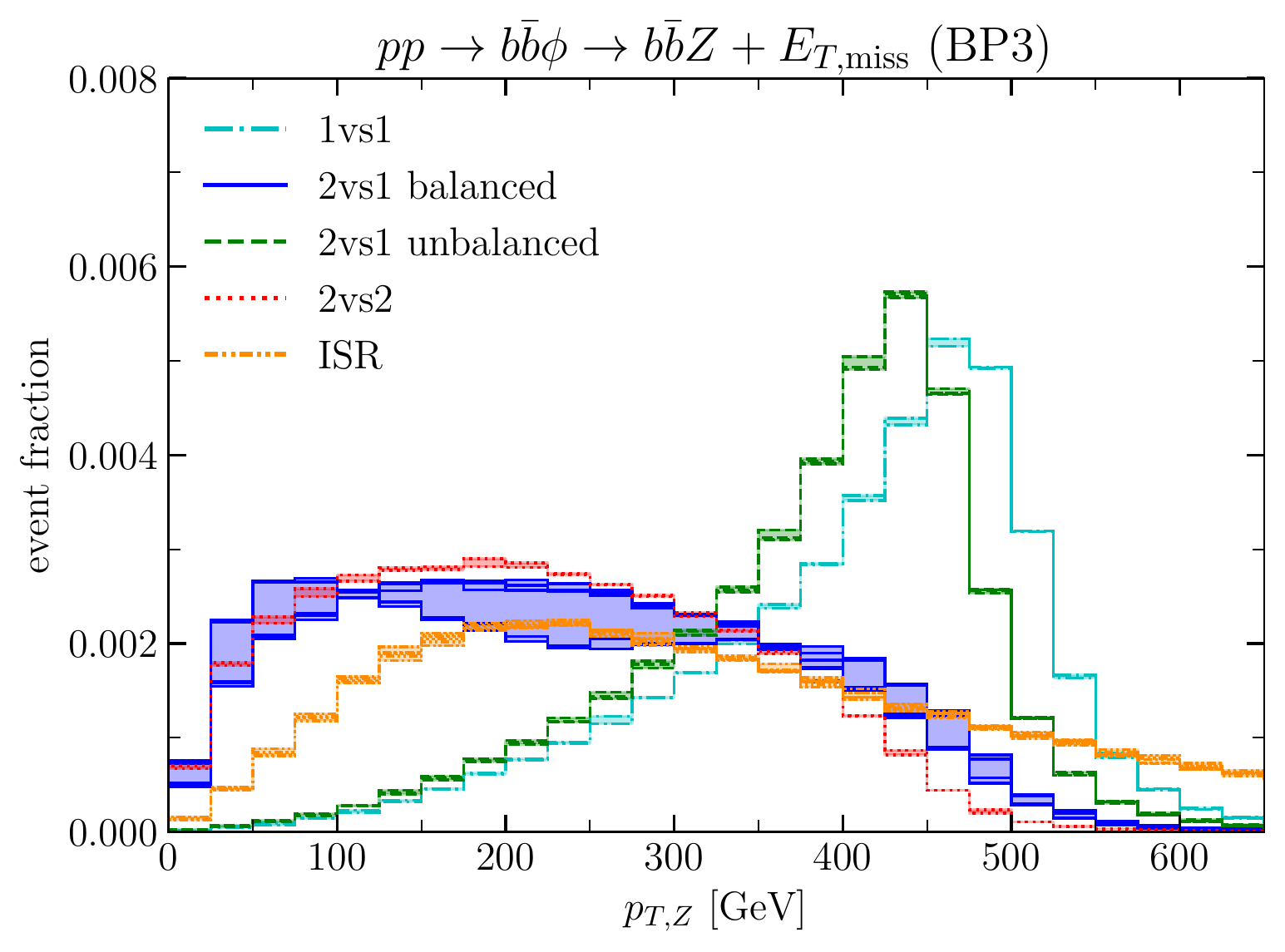}
\end{subfigure}
\begin{subfigure}{.48\linewidth}\centering
\includegraphics[width=\linewidth]{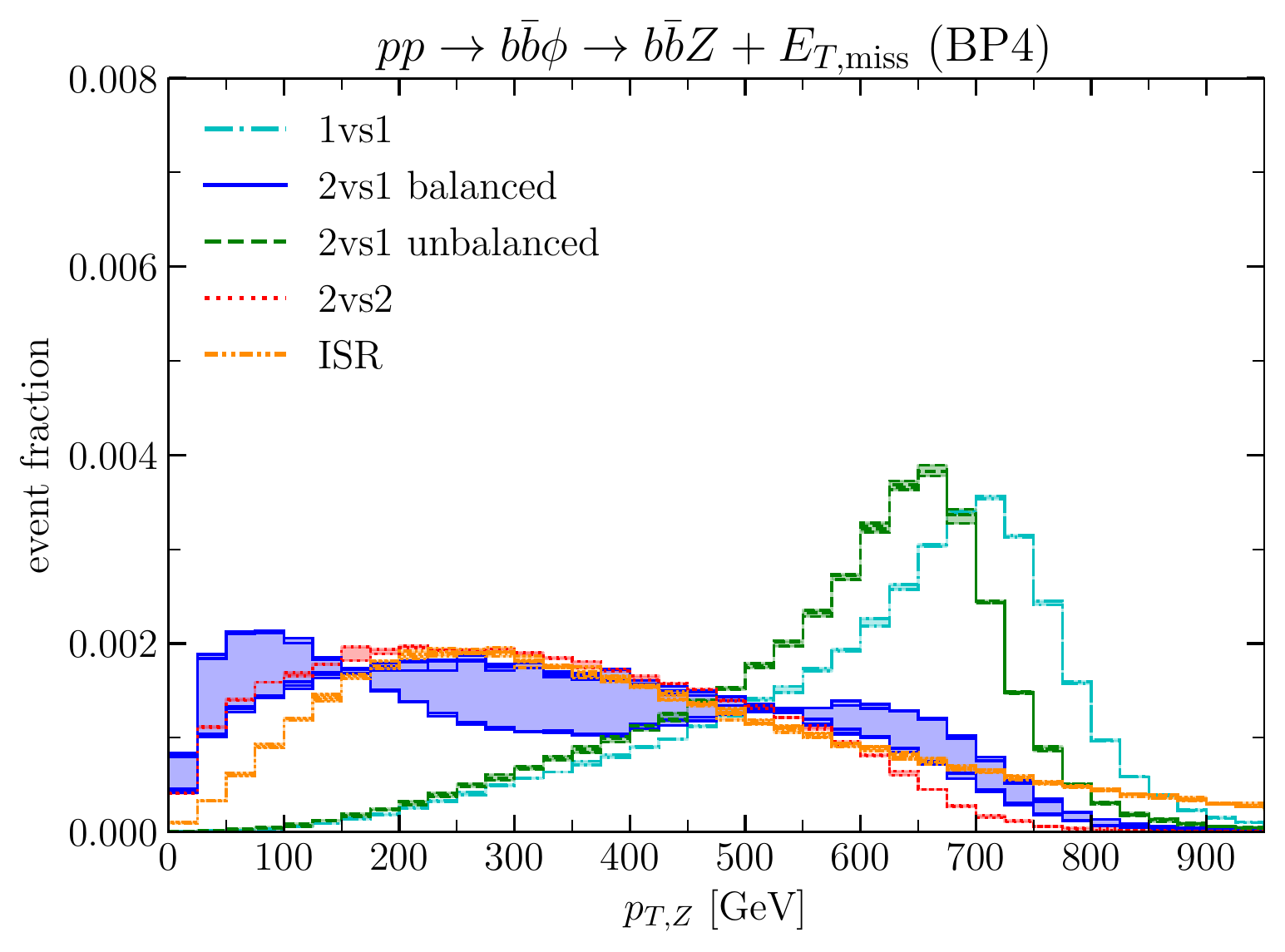}
\end{subfigure}
\caption{Mono-$Z + \ETmiss$ final state: transverse momentum distribution of the di-muon system evaluated for BP1 (upper left panel), BP2 (upper right panel), BP3 (lower left panel), and BP4 (lower right panel). All different curves of the 1vs1 topology are contained in the turquoise band; the curves of the 2vs1 balanced topology in the dark blue band; the curves of the 2vs1 unbalanced topology in the green band; and, the curves of the 2vs2 topology in the red band. The ISR topology curve is shown in orange.}
\label{fig:monoZ_BP_comp}
\end{figure}

After comparing the topologies (and the corresponding spin hypotheses), we investigate the dependence on the masses of the mediator and the invisible particle in \cref{fig:monoZ_BP_comp}. In this Figure, the spin hypotheses for a given topology are shown as thin lines: turquoise dot-dashed for the 1vs1 topology, blue solid for the 2vs1 balanced topology, green dashed for the 2vs1 unbalanced topology, red dotted for the 2vs2 topology, and orange dot-dot-dashed for the ISR topology. For each bin, the area between the minimum and maximum value for a given topology
obtained from varying the different spin hypotheses
is filled in the respective color.

The upper left plot of \cref{fig:monoZ_BP_comp} shows the $p_{T,\mu\mu}$ distribution for BP1. The curves in this plot correspond to the curves in \cref{fig:monoZ_top_comp}. In the upper right plot of \cref{fig:monoZ_BP_comp}, the distributions are shown for BP2, for which $m_\Inv = 100\gev$ in contrast to $m_\Inv = 10\gev$ for BP1. The increased mass of the invisible particle hardly affects the kinematic distributions.

Stronger effects are visible for BP3 (see lower left plot of \cref{fig:monoZ_BP_comp}). Here, the lowered mediator mass ($m_\Med = 260\gev$ in comparison to $m_\Med = 400\gev$ for BP1 and BP2) shifts the distributions of the 2vs1 unbalanced and the 2vs2 toplogies to slightly higher $p_{T,\mu\mu}$ values. While the overall shape of the 2vs1 balanced distribution is very similar to BP1 and BP2, the effect of varying the spin hypotheses is diminished. The distributions of the 1vs1 and the ISR topologies are not dependent on the mediator mass.

For BP4 (see lower right plot of \cref{fig:monoZ_BP_comp}), featuring an increased scalar resonance mass in comparison to the other benchmark points, the distributions are in general shifted to higher $p_{T,\mu\mu}$ values. The peaks of the 1vs1 and 2vs1 unbalanced topologies are at $\sim 700\gev$, whereas the other topologies reach a broad maximum at $\sim 200\gev$.

\medskip

As expected, we observe for the various benchmark points that the missing energy distributions of the unbalanced topologies (1vs1 and 2vs1 unbalanced) peak at high values, whereas the balanced topologies (2vs1 balanced, 2vs2, and ISR) are broader and also have a significant event fraction at low \ETmiss values. Most existing experimental searches apply a lower cut on \ETmiss of $\sim 80 -100\gev$~\cite{Aaboud:2017bja,Khachatryan:2015bbl,Sirunyan:2017qfc,Sirunyan:2017onm,Sirunyan:2020fwm,ATLAS:2021gcn}. While these searches are expected to have the highest sensitivity to the unbalanced topologies, we also expect them to show a significant sensitivity to the balanced topologies. This is in contrast to the experimental searches performed in \ccite{Aad:2014vka,Aaboud:2016qgg,Aaboud:2018xdl,Sirunyan:2017jix,Sirunyan:2017hci,} demanding $\ETmiss \gtrsim 150-250\gev$, which we expect to have a much lower sensitivity to the balanced topologies. We also want to point out that the existing searches concentrate on heavy resonances produced via gluon fusion or via light quarks in the initial state. In order to suppress background events, $b$ jets are often vetoed. As a consequence, these searches are not sensitive to heavy scalar resonances produced in association with bottom quarks. This production channel is, however, one of the dominant channels in many extensions of the SM Higgs sector (e.g.\ for the TDHM type-II at high $\tan\beta$ values).


\section{Simplified models for mono-Higgs boson + \texorpdfstring{$\ETmiss$}{ETmiss} signatures}
\label{sec:H_ETmiss}

As a second experimental signature, we investigate a mono-Higgs plus missing energy final state. Also for this signature, a lot of experimental searches already exist~\cite{Aad:2015dva,Aaboud:2016obm,Aaboud:2017yqz,Aaboud:2017uak,ATLAS:2018bvd,ATLAS:2020jpb,ATLAS:2021qwk,Sirunyan:2017hnk,Sirunyan:2017hnk,Sirunyan:2018fpy,Sirunyan:2018fpy,Sirunyan:2018gdw,Sirunyan:2019zav}. These searches are typically interpreted in $Z'$ models, the THDM extended with an additional gauged $U(1)$ symmetry, or in the THDMa.


\subsection{Topologies}

As for the mono-$Z$ topologies, the mono-Higgs plus missing energy final state can be produced by the decay of the initial neutral scalar either directly into a Higgs boson and an invisible particle or mediated by up to two electrically neutral mediators.\footnote{As mentioned above (see \cref{sec:simplified_model}), we do not consider topologies with four-point interactions.}

\begin{figure}\centering
\begin{subfigure}[t]{\subfigsizetwo\linewidth}\centering
\includegraphics[scale=1]{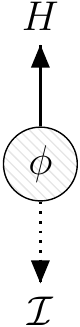}
\caption{1vs1}
\label{fig:monoH_topologies_1vs1}
\end{subfigure}
\begin{subfigure}[t]{\subfigsizetwo\linewidth}\centering
\includegraphics[scale=1]{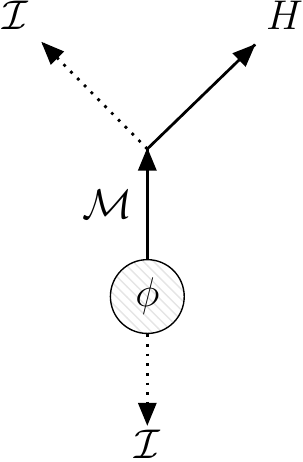}
\caption{2vs1 balanced}
\label{fig:monoH_topologies_2vs1_balanced}
\end{subfigure}
\\[1em]
\begin{subfigure}[t]{\subfigsizetwo\linewidth}\centering
\includegraphics[scale=1]{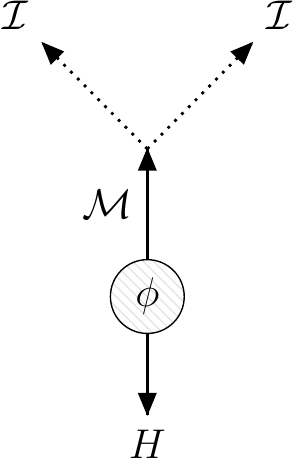}
\caption{2vs1 unbalanced}
\label{fig:monoH_topologies_2vs1_unbalanced}
\end{subfigure}
\begin{subfigure}[t]{\subfigsizetwo\linewidth}\centering
\includegraphics[scale=1]{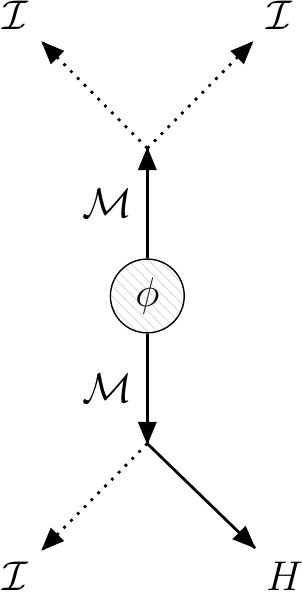}
\caption{2vs2}
\label{fig:monoH_topologies_2vs2}
\end{subfigure}
\caption{Sketch of an asymmetric neutral scalar boson $\phi$ decay in its rest frame leading to a mono-Higgs ($H$) plus $\ETmiss$ collider signature.}
\end{figure}

\begin{enumerate}[(a)]
  \item 1vs1 unbalanced topology (see \cref{fig:monoH_topologies_2vs1_balanced})

        The scalar resonance decays directly to an invisible particle and a Higgs boson. The Higgs boson recoils against the invisible particle resulting in a missing energy spectrum peaking at high \ETmiss values.

  \item 2vs1 balanced topology (see \cref{fig:monoH_topologies_2vs1_balanced})

        The scalar resonance decays to a mediator and an invisible particle. The mediator then decays to a Higgs boson and an invisible particle. Since the scalar resonance is produced approximately at rest, the mediator recoils against the first invisible particle resulting in a ``balanced'' missing energy spectrum.

  \item 2vs1 unbalanced topology (see \cref{fig:monoH_topologies_2vs1_unbalanced})

        The scalar resonance decays to a mediator and a Higgs boson. The mediator then decays to two invisible particles. For this topology, the Higgs boson recoils against the mediator resulting in a missing energy spectrum peaking at high \ETmiss values (kinematically similar to the 1vs1 topology).

  \item 2vs2 balanced topology (see \cref{fig:monoH_topologies_2vs2})

        The scalar resonance decays to two mediators. One of the mediators then decays to a Higgs boson and an invisible particle. The second mediator decays to two invisible particles. This topology is kinematically similar to the 2vs1 balanced topology and features a similar \ETmiss spectrum.
\end{enumerate}

Note that we do not include an initial-state radiation topology for the mono-Higgs plus \ETmiss final state. While it is in principle easily possible to take into account this topology, the associated cross section is negligibly small for most BSM models.


\subsection{Feynman diagram realizations}

\begin{figure}\centering
\begin{subfigure}{\subfigsize\linewidth}\centering
\includegraphics[scale=1]{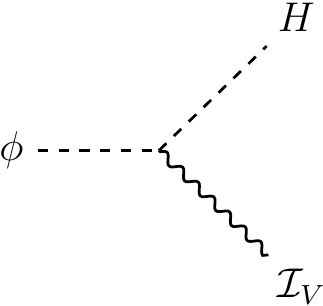}
\caption{$\Iv$}
\label{fig:monoH_ETmiss_1vs1_Iv}
\end{subfigure}
\begin{subfigure}{\subfigsize\linewidth}\centering
\includegraphics[scale=1]{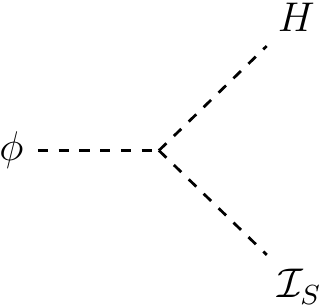}
\caption{$\Is$}
\label{fig:monoH_ETmiss_1vs1_Is}
\end{subfigure}
\caption{Mono-$H$ + $\ETmiss$ processes (1vs1 topology).}
\label{fig:monoH_ETmiss_1vs1}
\end{figure}

\begin{figure}\centering
\begin{subfigure}{\subfigsize\linewidth}\centering
\includegraphics[scale=1]{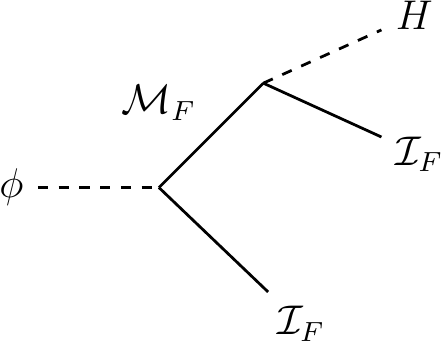}
\caption{$\Mf\If$}
\label{fig:monoH_ETmiss_2vs1_balanced_MfIf}
\end{subfigure}
\begin{subfigure}{\subfigsize\linewidth}\centering
\includegraphics[scale=1]{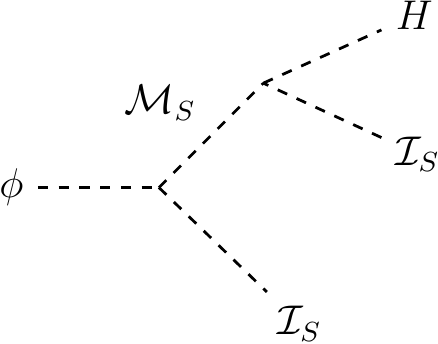}
\caption{$\Ms\Is$}
\label{fig:monoH_ETmiss_2vs1_balanced_MsIs}
\end{subfigure}
\begin{subfigure}{\subfigsize\linewidth}\centering
\includegraphics[scale=1]{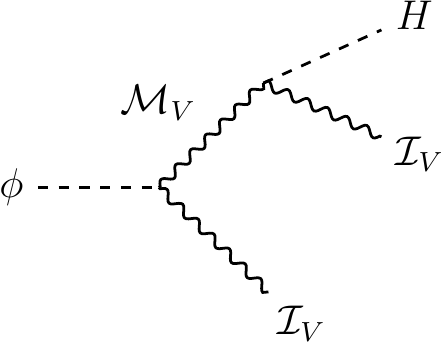}
\caption{$\Mv\Iv$}
\label{fig:monoH_ETmiss_2vs1_balanced_MvIv}
\end{subfigure}
\begin{subfigure}{\subfigsize\linewidth}\centering
\includegraphics[scale=1]{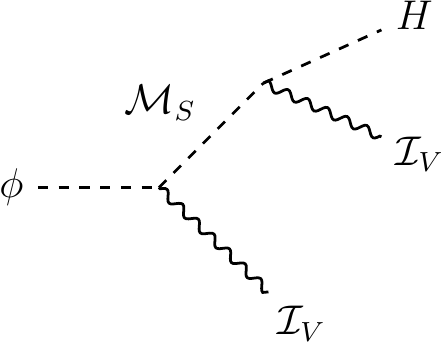}
\caption{$\Ms\Iv$}
\label{fig:monoH_ETmiss_2vs1_balanced_MsIv}
\end{subfigure}
\begin{subfigure}{\subfigsize\linewidth}\centering
\includegraphics[scale=1]{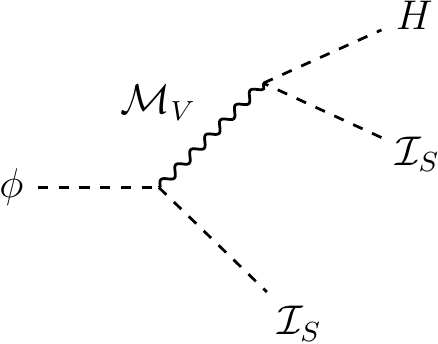}
\caption{$\Mv\Is$}
\label{fig:monoH_ETmiss_2vs1_balanced_MvIs}
\end{subfigure}
\caption{Mono-Higgs ($H$) + $\ETmiss$ processes (balanced, 2vs1 topology).}
\label{fig:monoH_ETmiss_2vs1_balanced}
\end{figure}

\begin{figure}\centering
\begin{subfigure}{\subfigsize\linewidth}\centering
\includegraphics[scale=1]{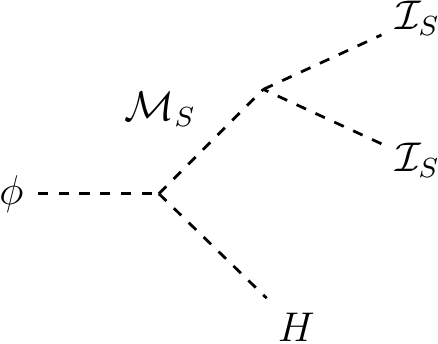}
\caption{$\Ms\Is$}
\label{fig:monoH_ETmiss_2vs1_unbalanced_MsIs}
\end{subfigure}
\begin{subfigure}{\subfigsize\linewidth}\centering
\includegraphics[scale=1]{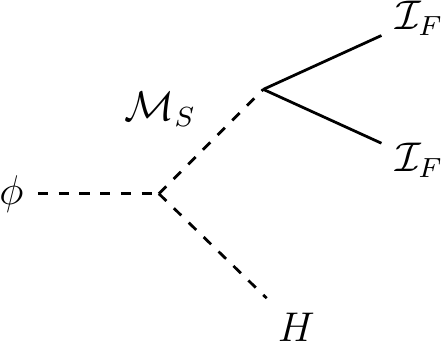}
\caption{$\Ms\If$}
\label{fig:monoH_ETmiss_2vs1_unbalanced_MsIf}
\end{subfigure}
\\
\begin{subfigure}{\subfigsize\linewidth}\centering
\includegraphics[scale=1]{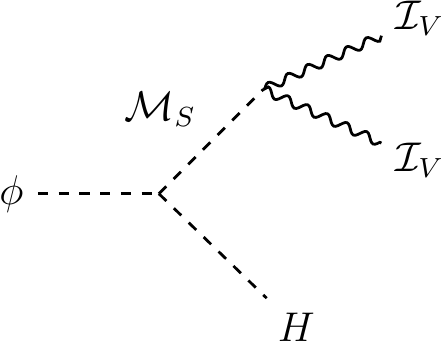}
\caption{$\Ms\Iv$}
\label{fig:monoH_ETmiss_2vs1_unbalanced_MsIv}
\end{subfigure}
\begin{subfigure}{\subfigsize\linewidth}\centering
\includegraphics[scale=1]{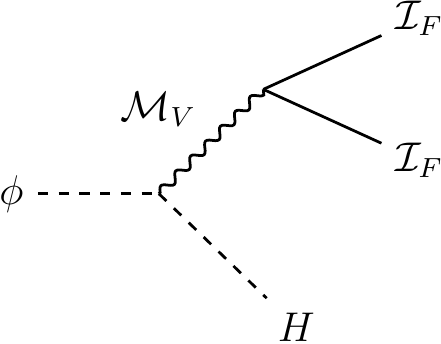}
\caption{$\Mv\If$}
\label{fig:monoH_ETmiss_2vs1_unbalanced_MvIf}
\end{subfigure}
\caption{Mono-Higgs ($H$) + $\ETmiss$ processes (unbalanced, 2vs1 topology).}
\label{fig:monoH_ETmiss_2vs1_unbalanced}
\end{figure}

\begin{figure}\centering
\begin{subfigure}{\subfigsize\linewidth}\centering
\includegraphics[scale=1]{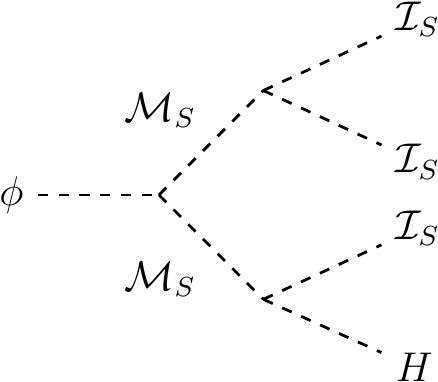}
\caption{$\Ms\Is$}
\label{fig:monoH_ETmiss_2vs2_MsIs}
\end{subfigure}
\begin{subfigure}{\subfigsize\linewidth}\centering
\includegraphics[scale=1]{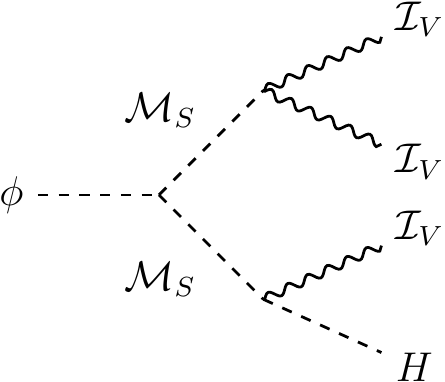}
\caption{$\Ms\Iv$}
\label{fig:monoH_ETmiss_2vs2_MsIv}
\end{subfigure}
\caption{Mono-Higgs ($H$) + $\ETmiss$ processes (2vs2 topology).}
\label{fig:monoH_ETmiss_2vs2}
\end{figure}

We show the Feynman diagram realizations of the different mono-Higgs topologies in \cref{fig:monoH_ETmiss_1vs1,fig:monoH_ETmiss_2vs1_balanced,fig:monoH_ETmiss_2vs1_unbalanced,fig:monoH_ETmiss_2vs2}.

Two different Feynman diagram realizations exist for the 1vs1 topology: $\Iv$ (see \cref{fig:monoH_ETmiss_1vs1_Iv}) and $\Is$ (see \cref{fig:monoH_ETmiss_1vs1_Is}).

Five different Feynman diagram realizations can be constructed for the 2vs1 balanced topology: $\Mf\If$ (see \cref{fig:monoH_ETmiss_2vs1_balanced_MfIf}), $\Ms\Is$ (see \cref{fig:monoH_ETmiss_2vs1_balanced_MsIs}), $\Mv\Iv$ (see \cref{fig:monoH_ETmiss_2vs1_balanced_MvIv}), $\Ms\Iv$ (see \cref{fig:monoH_ETmiss_2vs1_balanced_MsIv}), and $\Mv\Is$ (see \cref{fig:monoH_ETmiss_2vs1_balanced_MvIs}). The $\Mf\If$ realization can be encountered e.g.\ in the MSSM: a heavy Higgs boson is produced via gluon fusion or bottom-associated Higgs production; it then decays into the lightest neutralino and a heavier neutralino, which then decays to a SM-like Higgs boson and the lightest neutralino. As for the mono-$Z$ final state, the mono-$H$ final state with the $\Ms\Is$ realization can be constructed in the N2HDM if one of the Higgs doublet is inert.

The 2vs1 unbalanced topology can be realized by five different Feynman diagrams: $\Ms\Is$ (see \cref{fig:monoH_ETmiss_2vs1_unbalanced_MsIs}), $\Ms\If$ (see \cref{fig:monoH_ETmiss_2vs1_unbalanced_MsIf}), $\Ms\Iv$ (see \cref{fig:monoH_ETmiss_2vs1_unbalanced_MsIv}), and $\Mv\If$ (see \cref{fig:monoH_ETmiss_2vs1_unbalanced_MvIf}). The $\Ms\If$ topology can be realized e.g.\ in the THDMa or the MSSM, with the lightest neutralino playing the role of $\If$. Another example for the $\Ms\Iv$ realization is the SM extended by a complex scalar singlet which is subject to an additional gauged $U(1)$ symmetry, whose gauge boson plays the role of the invisible particle.

Only two Feynman diagram realizations exist for the 2vs2 topology: $\Ms\Is$ (see \cref{fig:monoH_ETmiss_2vs2_MsIs}) and $\Ms\Iv$ (see \cref{fig:monoH_ETmiss_2vs2_MsIv}). This topology can be realized in extensions of the THDM.


\subsection{Kinematic analysis}

In order to study the kinematics of the mono-Higgs plus missing energy final state, we assume the SM-like Higgs boson to decay into two photons. This decay channel has a high mass resolution and
low background. We assume the initial scalar resonance to be produced in association with bottom quarks. Apart from the mass resolution, we expect very similar results for other Higgs decay channels (including off-shell decays) and production via gluon fusion. As for the mono-$Z$ final state, the transverse momentum of the di-gamma system contains valuable information about the underlying process.

\medskip

\begin{figure}\centering
\begin{subfigure}{.48\linewidth}\centering
\includegraphics[width=\linewidth]{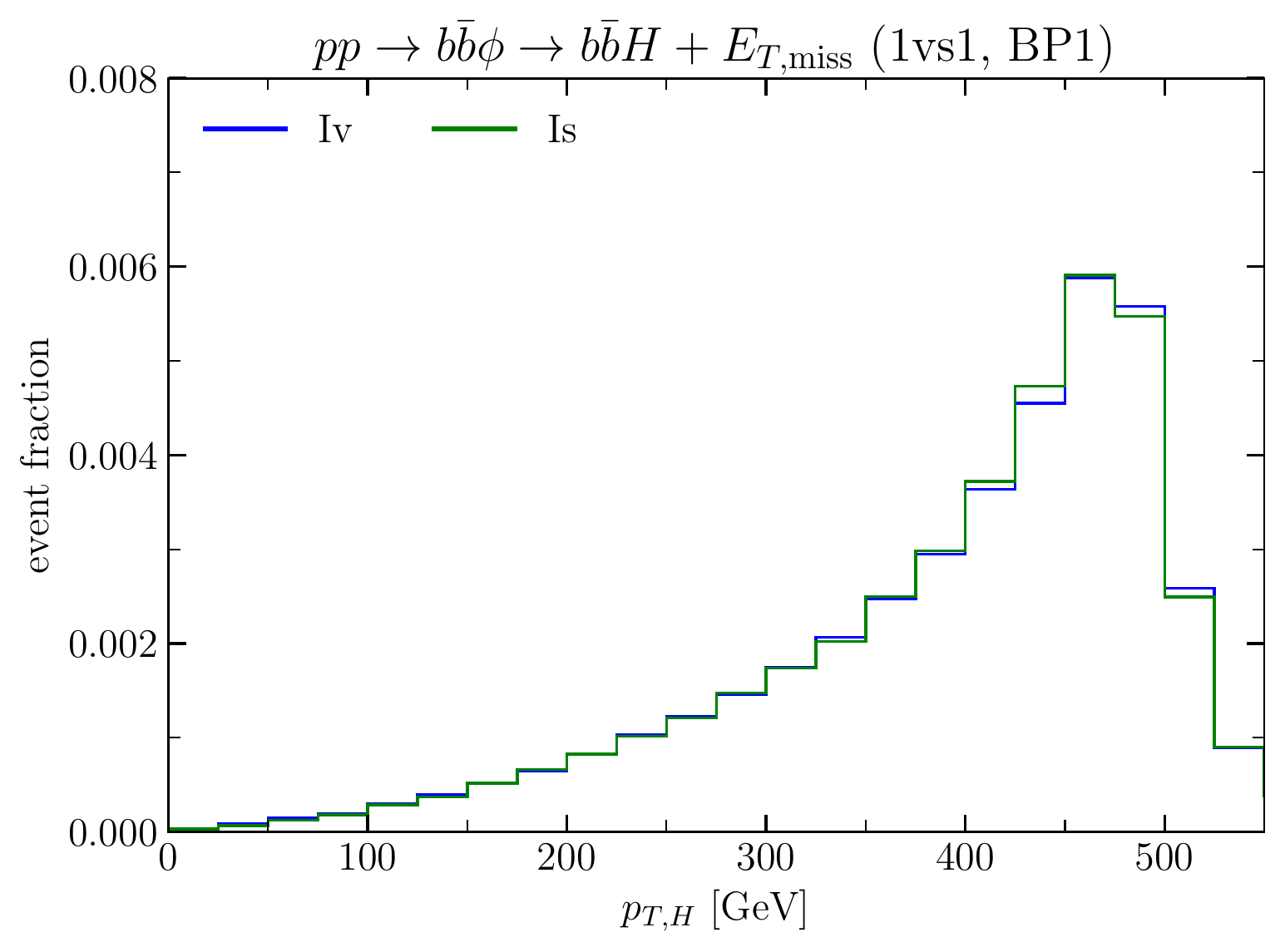}
\end{subfigure}
\begin{subfigure}{.48\linewidth}\centering
\includegraphics[width=\linewidth]{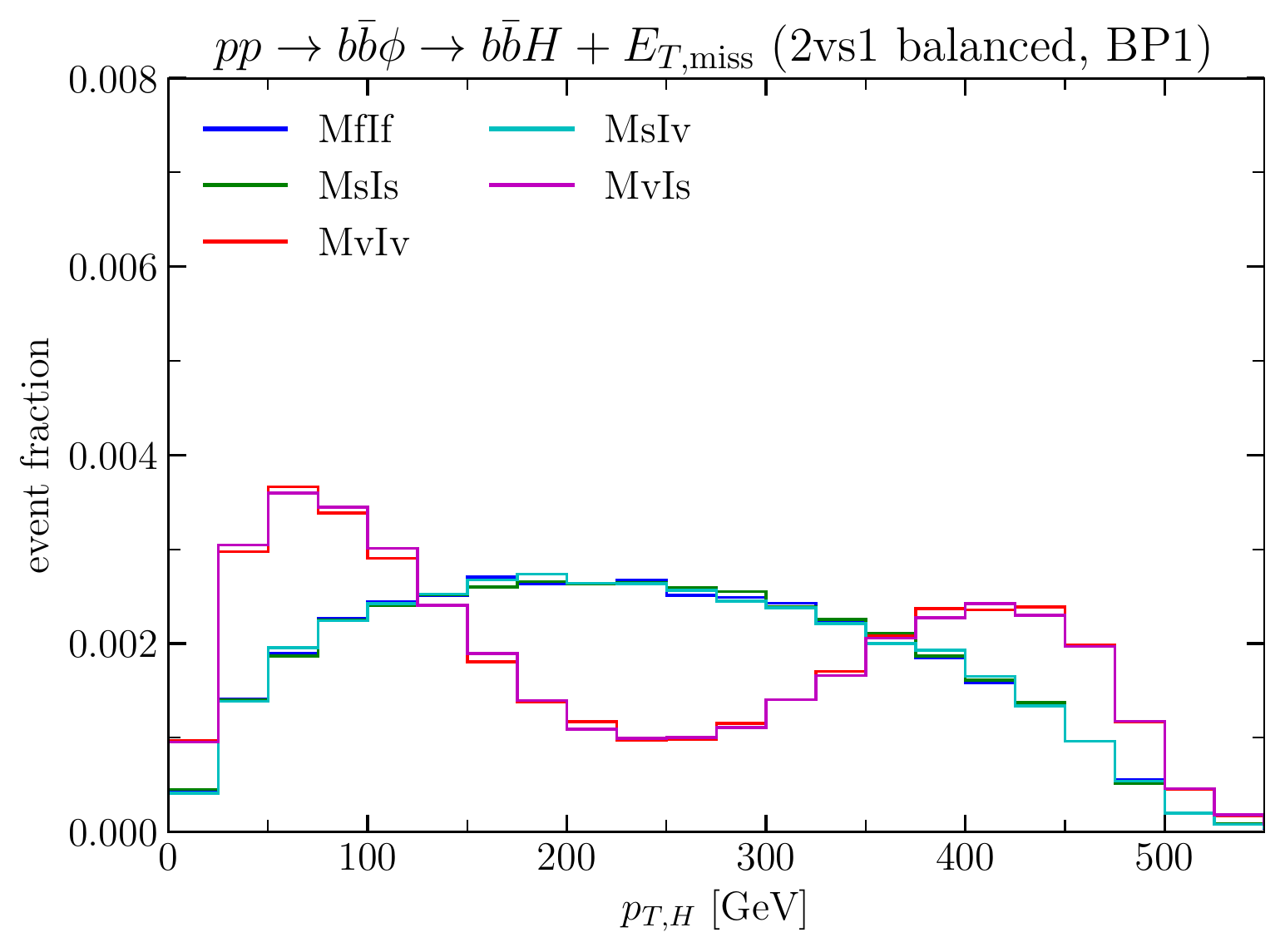}
\end{subfigure}
\begin{subfigure}{.48\linewidth}\centering
\includegraphics[width=\linewidth]{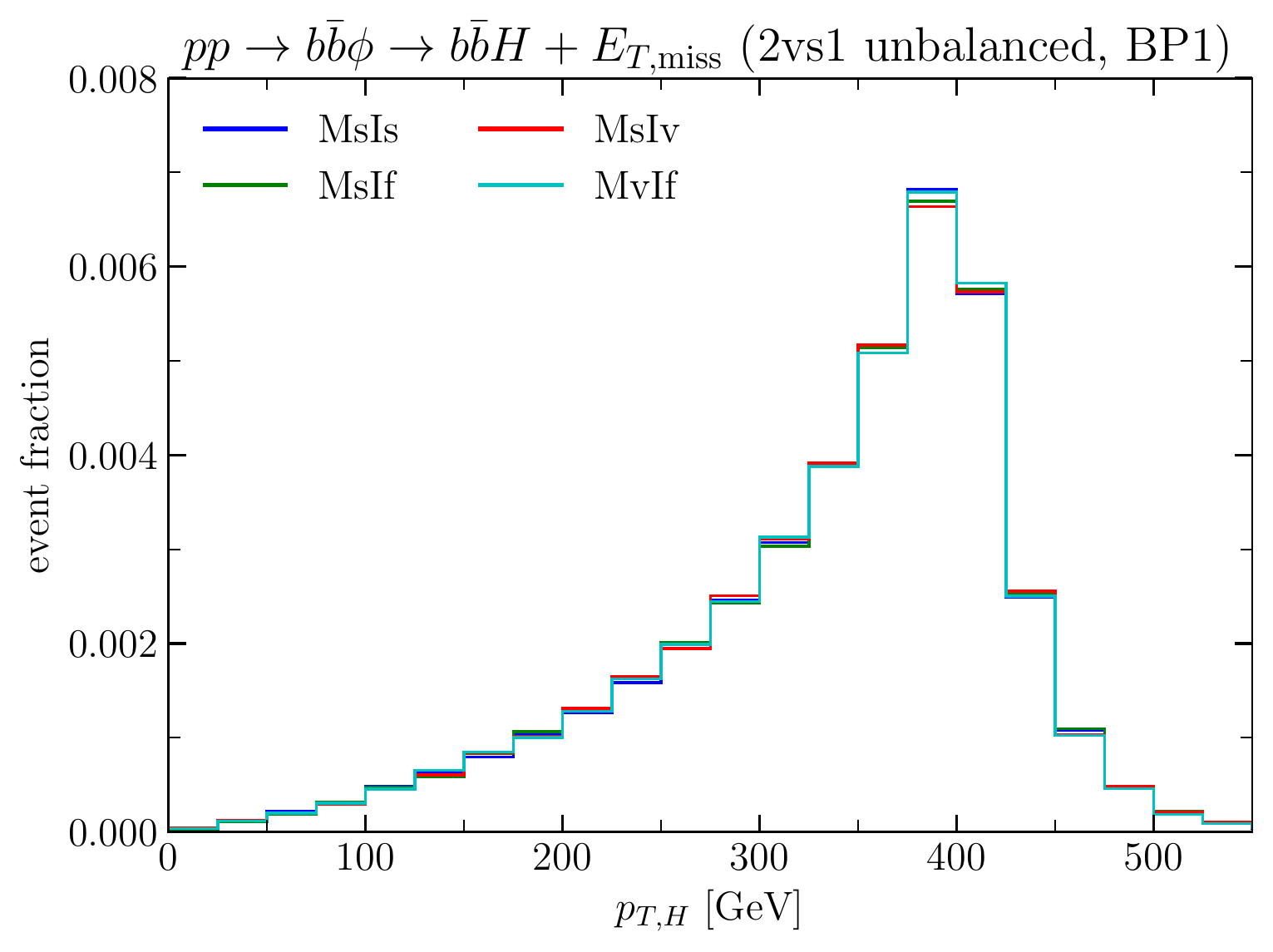}
\end{subfigure}
\begin{subfigure}{.48\linewidth}\centering
\includegraphics[width=\linewidth]{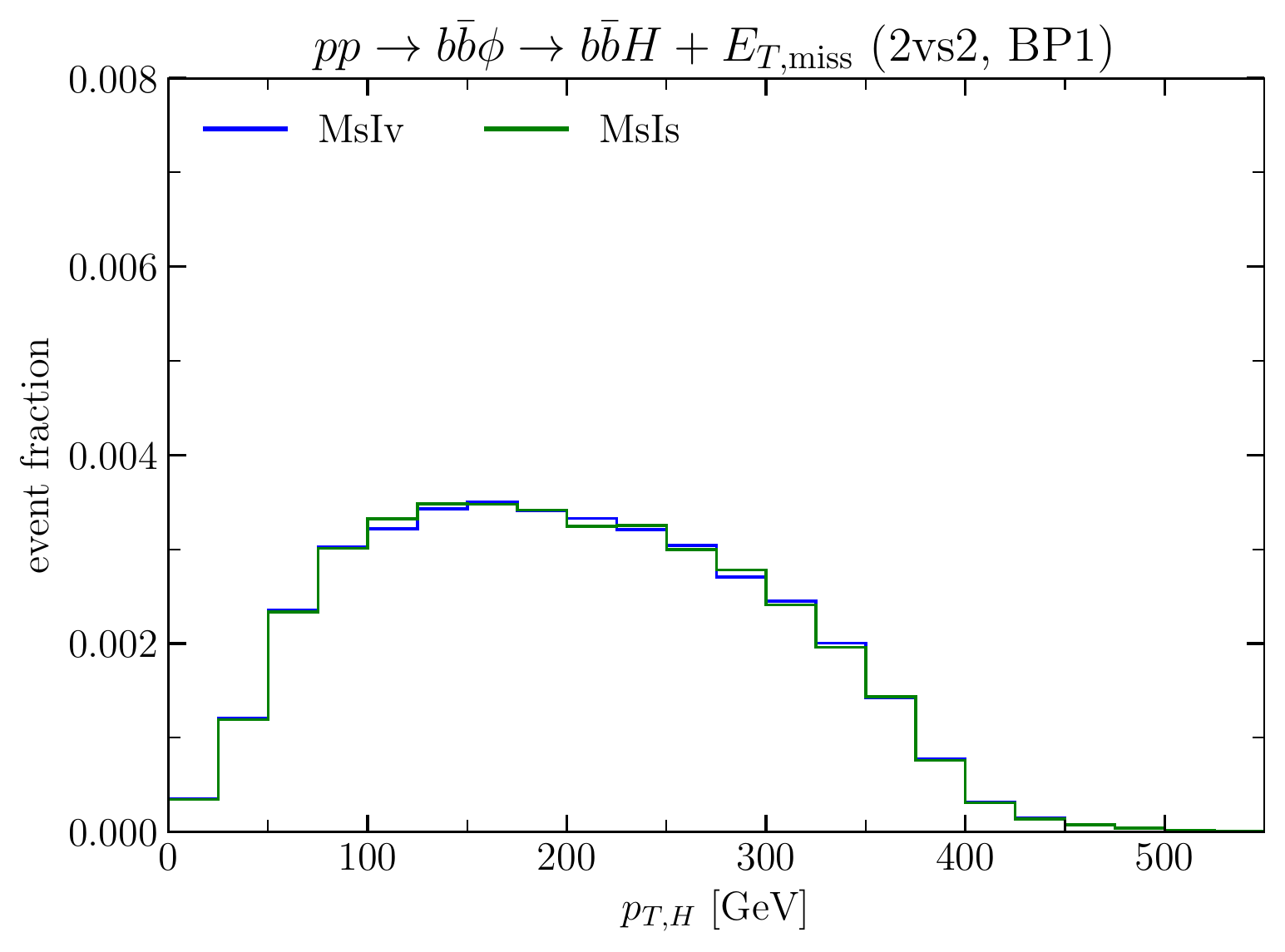}
\end{subfigure}
\caption{Mono-$H + \ETmiss$ final state: transverse momentum distribution of the di-gamma system for BP1. The upper left panel displays the results for the 1vs1 balanced topology; the upper right panel for the 2vs1 balanced topology; the lower left panel for the 2vs1 unbalanced topology; and the lower right panel for the 2vs2 topology.}
\label{fig:monoH_top_comp}
\end{figure}

The transverse momentum distribution of the di-gamma system, labelled with $p_{T,\gamma\gamma}$, is shown for BP1 in \cref{fig:monoH_top_comp}. As in \cref{sec:Z_ETmiss}, all distributions are generated for the scalar resonance being produced in association with bottom quarks. We expect very similar results for $\phi$ production via gluon fusion.

In the upper left panel of \cref{fig:monoH_top_comp}, we show the $p_{T,\gamma\gamma}$
distribution for the 1vs1 topology. The distributions peak at $\sim 480\gev$, falling off quickly for higher $p_{T,\gamma\gamma}$ values. The $\Iv$ and $\Is$ distributions are not distinguishable experimentally for this distribution.

The spin realizations for the 2vs1 balanced topology, shown in the upper right panel of \cref{fig:monoH_top_comp}, could, however, potentially be distinguishable experimentally. Whereas the distributions of the $\Mv\Iv$ and $\Mv\Is$ spin realizations show a double peak structure (with peaks at $\sim 80\gev$ and $\sim 420\gev$) due to the momentum dependence of the scalar--vector couplings, the other distributions feature only one broad peak at $\sim 200\gev$.

The 2vs1 unbalanced configuration presented in the lower left panel of \cref{fig:monoH_top_comp}, however, shows no significant difference among the different spin realizations. All the distributions show a clear peak at $p_{T,\gamma\gamma}\sim 380\gev$.

As a final case the 2vs2 topology is shown in the lower right panel of \cref{fig:monoH_top_comp}. We can see that the two spin configurations $\Ms\Is$ and $\Ms\Iv$ are practically identical, so that the invisible particle nature has no impact in the distribution.

As we saw above for the case of the mono-$Z$ signal the different spin hypotheses are not
expected to be experimentally
distinguishable, with the possible exception of the 2vs1 balanced topology in the case of vector mediators that shows a different distribution for the $p_{T,\gamma\gamma}$ variable.

\begin{figure}\centering
\begin{subfigure}{.48\linewidth}\centering
\includegraphics[width=\linewidth]{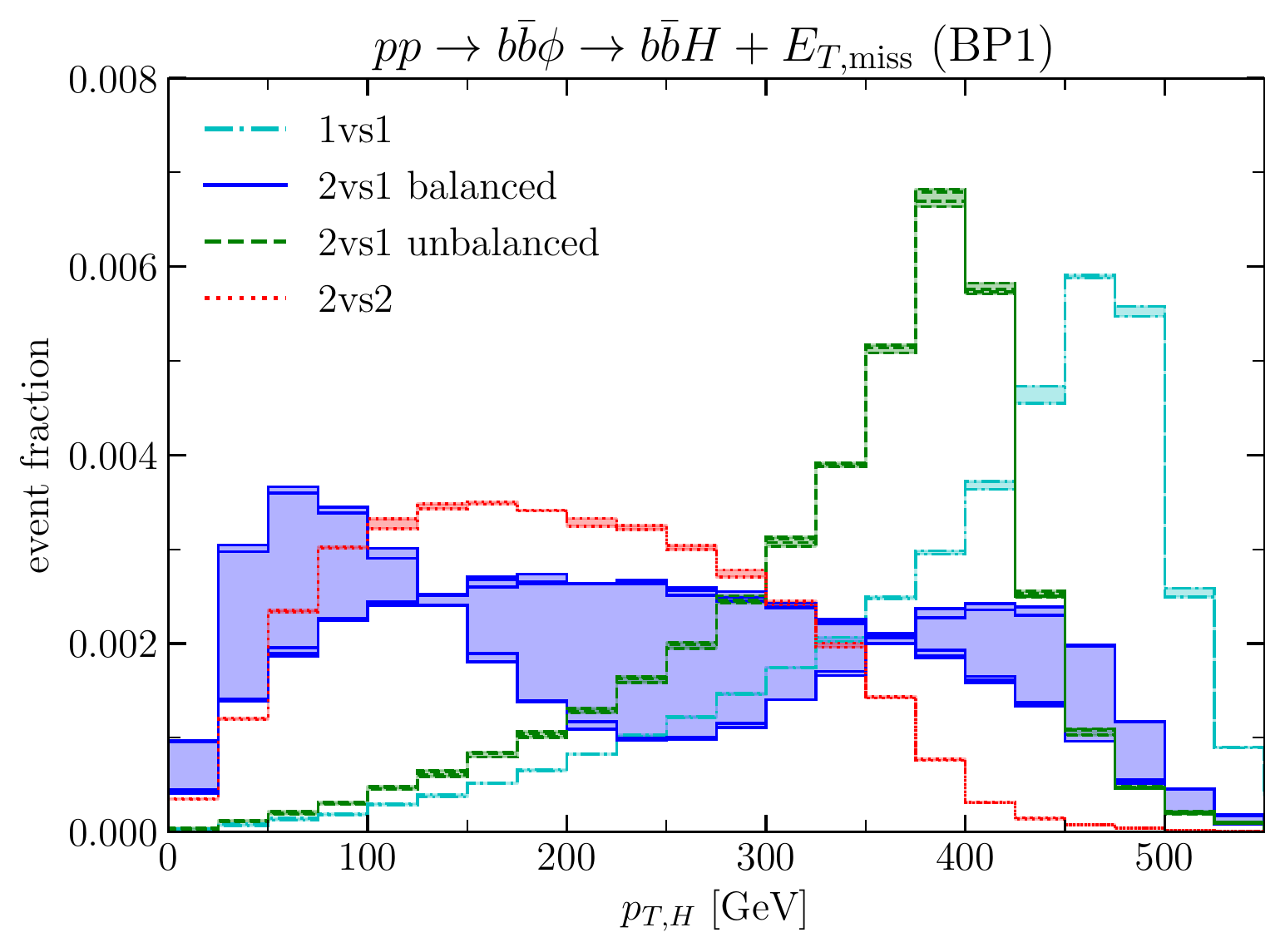}
\end{subfigure}
\begin{subfigure}{.48\linewidth}\centering
\includegraphics[width=\linewidth]{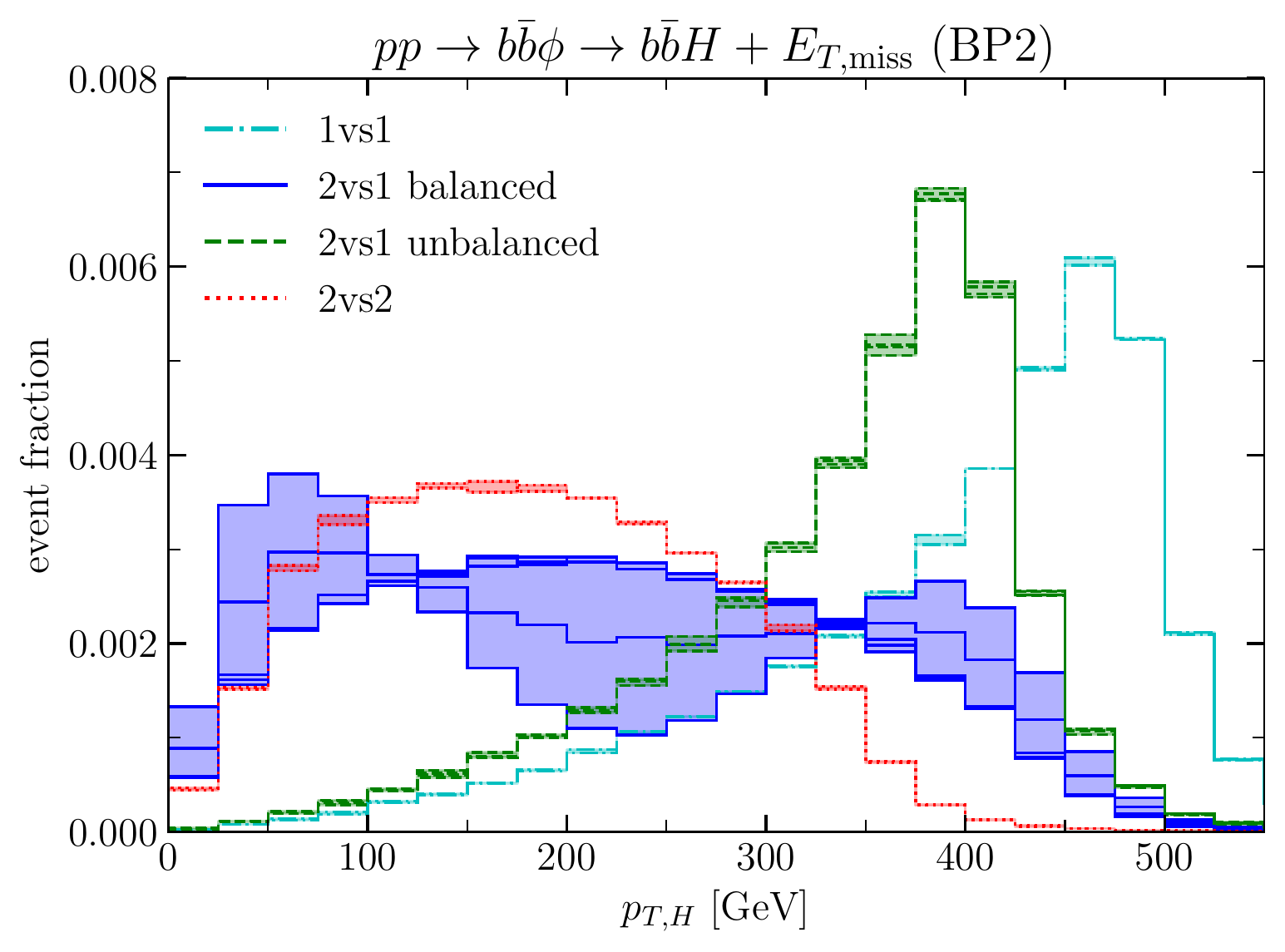}
\end{subfigure}
\begin{subfigure}{.48\linewidth}\centering
\includegraphics[width=\linewidth]{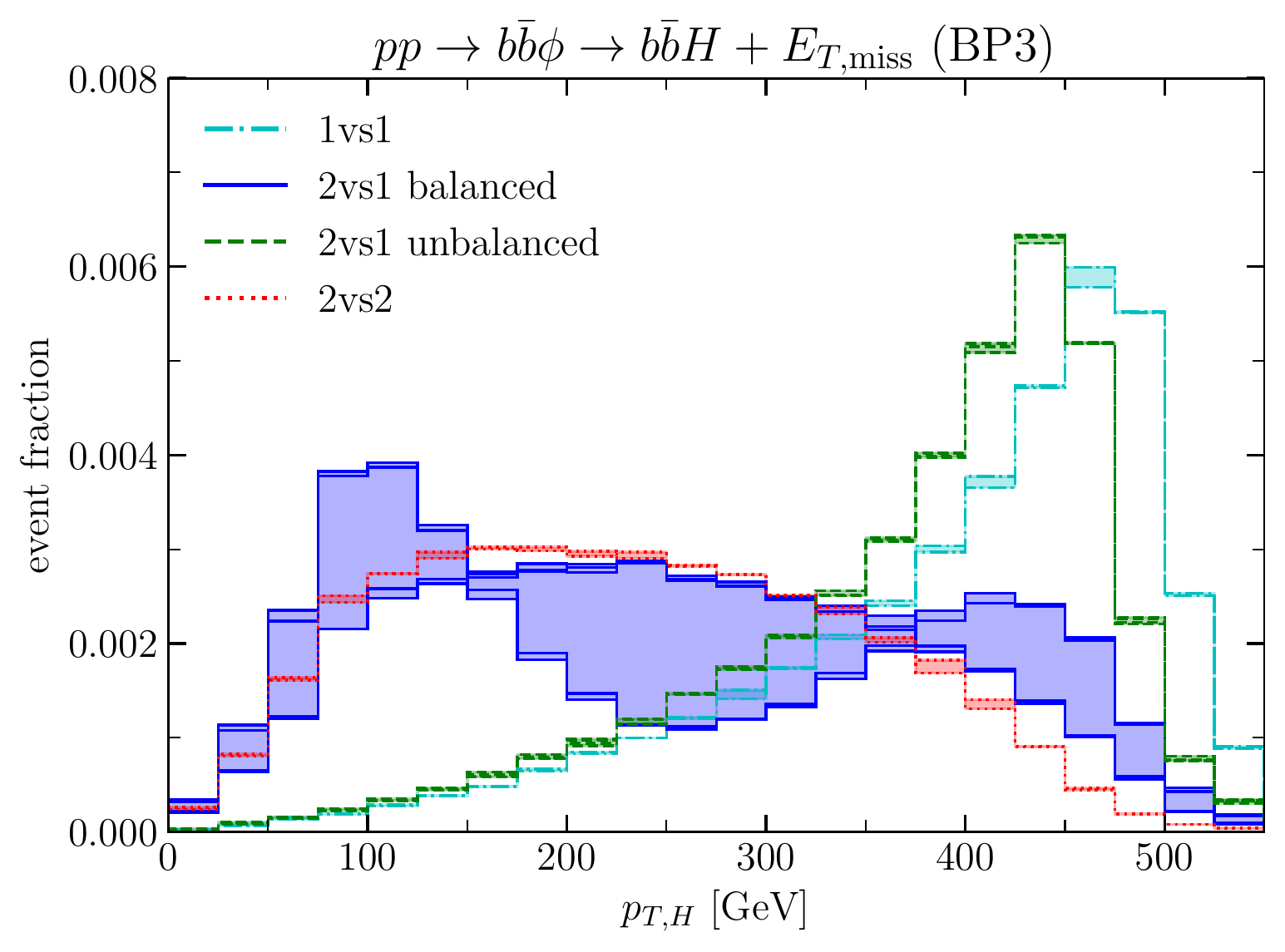}
\end{subfigure}
\begin{subfigure}{.48\linewidth}\centering
\includegraphics[width=\linewidth]{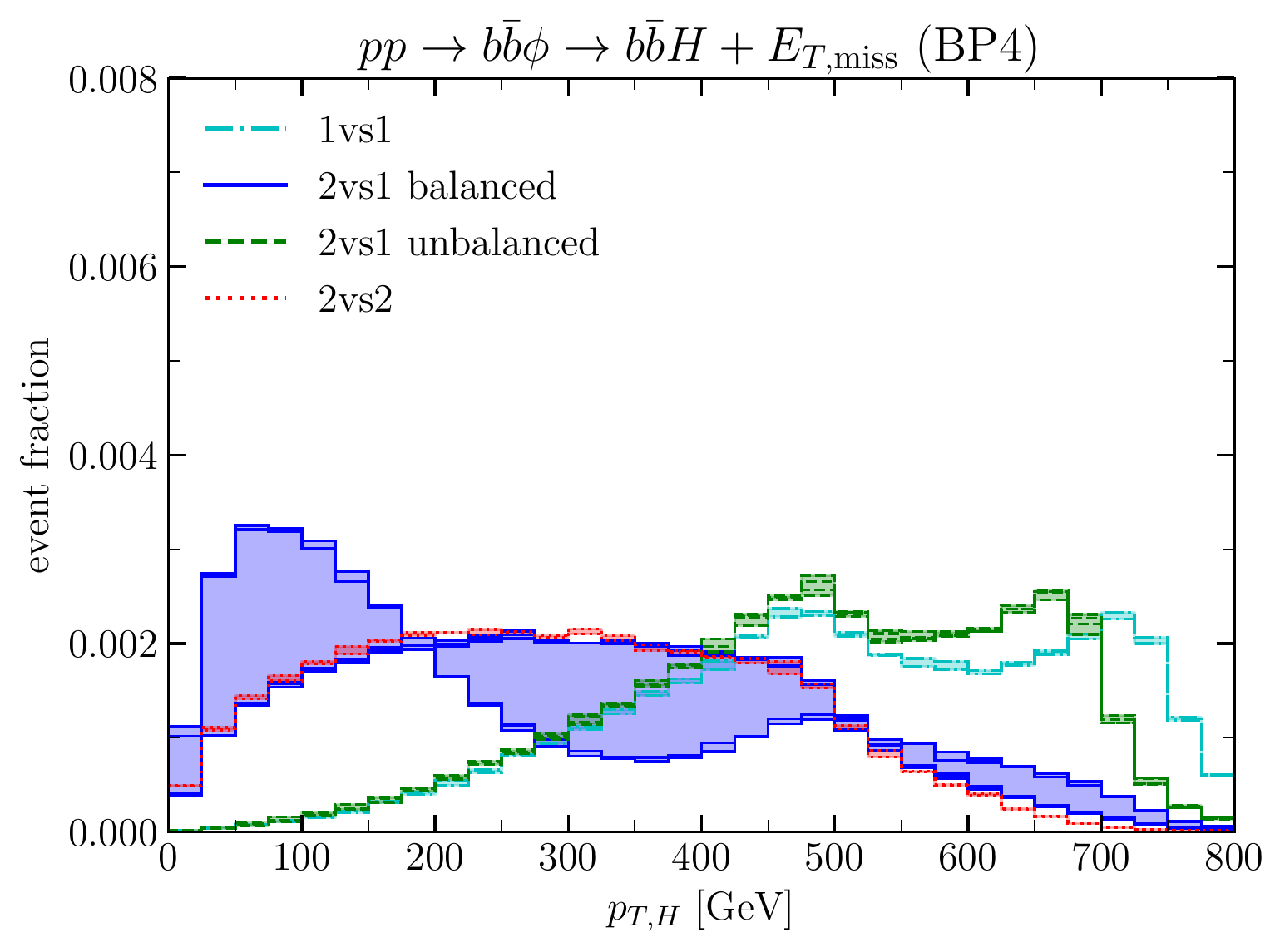}
\end{subfigure}
\caption{Mono-$H + \ETmiss$ final state: transverse momentum distribution of the di-gamma system evaluated for BP1 (upper left panel), BP2 (upper right panel), BP3 (lower left panel), and BP4 (lower right panel). All different curves of the 1vs1 topology are contained in the turquoise band; the curves of the 2vs1 balanced topology in the dark blue band; the curves of the 2vs1 unbalanced topology in the green band; and the curves of the 2vs2 topology in the red band.}
\label{fig:monoH_BP_comp}
\end{figure}

After comparing the topologies and spin hypotheses, we assess the dependence on the masses of the involved particles by evaluating the distributions for the four different benchmark points (see \cref{fig:monoH_BP_comp}). The description of the Figure is analogous to \cref{fig:monoZ_BP_comp} for the mono-$Z$ signature. Comparing the $p_{T,\gamma\gamma}$ distributions for BP1 and BP2 --- upper left and right panels of \cref{fig:monoH_BP_comp}, respectively --- no differences are visible (with the exception of the 2vs1 balanced $\Mv\Iv$ topology). This clearly shows that the kinematical distributions are insensitive to the invisible particle mass as we also observed for the mono-$Z$ signature. The case of BP3 --- lower left panel of \cref{fig:monoH_BP_comp} --- is different, since the mass of the mediator is smaller than for the cases BP1 and BP2. In this case, we can see a shift in the peak of the distributions in comparison to BP1 and BP2. For example for the 2vs1 balanced topology (with a vector mediator), we observe the first peak to be shifted to $\sim 100\gev$. For the 2vs1 unbalanced topology, the peak is shifted to $\sim 440\gev$. Consequently, the distributions of the 1vs1 and 2vs1 unbalanced topologies almost overlap. For the 2vs2 topology, we can see that the distribution reaches larger values of $p_{T,\gamma\gamma}$ than for BP1 and BP2. Examining BP4 for which the mass of $m_\phi$ is larger (see lower right panel of \cref{fig:monoH_BP_comp}), we see a clear shift in the distributions of all the topologies in comparison to the other benchmark points. First of all the increase of mass of the heavy scalar makes the distributions reach larger values. The 2vs1 balanced topology gives rise to a peak at $\sim 60\gev$ while the 1vs1 and the 2vs1 unbalanced topologies feature a first peak at $\sim 500\gev$ and then a second one at $\sim 660\gev$ for the 2vs1 balanced topology and at $\sim 750\gev$ for the 1vs1 topology.

We can see for the mono-$H$ signature that the 1vs1 and 2vs1 unbalanced distributions of $p_{T, H}$ peak at higher values. The balanced topologies (2vs1 balanced and 2vs2) show broader peaks at lower values of the distribution. Several experimental searches apply a relatively large cut value of $\ETmiss > 150\gev$~\cite{Aad:2015dva,Aaboud:2016obm,Aaboud:2017yqz,Aaboud:2017uak,ATLAS:2018bvd,ATLAS:2020jpb,ATLAS:2021qwk,Sirunyan:2018fpy,Sirunyan:2018fpy,Sirunyan:2019zav}. While these searches could be sensitive to the unbalanced topologies, they are relatively insensitive to the balanced topologies. However, there are also some searches~\cite{Sirunyan:2017hnk,Sirunyan:2018gdw} with cuts at lower values ($\ETmiss > 90$--$105\gev$) that can be more sensitive to the balanced topologies by capturing those events lying in the low-energy area of the distribution.


\section{Exemplary application: ATLAS mono-Higgs + \texorpdfstring{$\ETmiss$}{ETmiss} search}
\label{sec:monoH+MET_example}

In this Section, we show how the simplified models worked out in the previous Sections can be used in practice. For our exemplary study, we focus on the mono-Higgs plus \ETmiss search conducted in \ccite{ATLAS:2017uis} (using $36.1\invfb$). We have chosen this search, besides the fact that the search signature is covered by the simplified models worked out in \cref{sec:H_ETmiss}, since the analysis performed in \ccite{ATLAS:2017uis} is conveniently already available in the public analysis database of the \texttt{MadAnalysis~5} framework~\cite{Dumont:2014tja,Conte:2012fm,Conte:2014zja,Conte:2018vmg,DVN/KAAYFM_2021}.

We want to stress here that recasting an existing analysis is not optimal. Instead of extracting the information we discuss below by recasting an existing analysis, this information could most easily and most precisely be provided by the experimental collaborations as part of their analyses, for instance as supplementary material. If the information that is needed for the simplified model interpretation is directly provided as part of the experimental analysis, a re-implementation or recasting of a previously done analysis, which is always associated with approximations, is avoided. Instead, the experimental collaborations can employ the full tool chain of their actual analysis without relying on fast detector simulation tools like \texttt{Delphes}. Accordingly, the main purpose of the recasting we perform below is to demonstrate the use of our simplified model framework --- as a proposal for how this framework could be applied by the experimental groups.


\subsection{Acceptance \texorpdfstring{$\times$}{x} efficiency maps}
\label{sec:monoH+MET_example_eff_maps}

First, we discuss the efficiency of the mono-Higgs plus \ETmiss search for the different simplified model topologies worked out in \cref{sec:H_ETmiss}. The search conducted in \ccite{ATLAS:2017uis} was optimized for (and interpreted in) a Two-Higgs-Doublet Model with an additional $U(1)$ gauge symmetry. This model gives rise to a final state corresponding to the 2vs1 unbalanced topology within our simplified model framework. It should be noted in this context that the heavy resonance is assumed to be a vector boson in \ccite{ATLAS:2017uis} instead of a scalar as in this work. As we will illustrate below, the analysis of \ccite{ATLAS:2017uis} can nevertheless be re-interpreted in the proposed simplified model framework.

In order to assess the sensitivity also to the other topologies, we perform a scan in the $(m_\phi,m_\Med,m_\Inv)$ parameter space generating MC events for each parameter point. For each MC sample, we then calculate the kinematic acceptance times detector efficiency using the \texttt{MadAnalysis} recasting functionality~\cite{Dumont:2014tja,Conte:2018vmg}.

\begin{figure}\centering
\begin{subfigure}{.49\linewidth}\centering
\includegraphics[width=\textwidth]{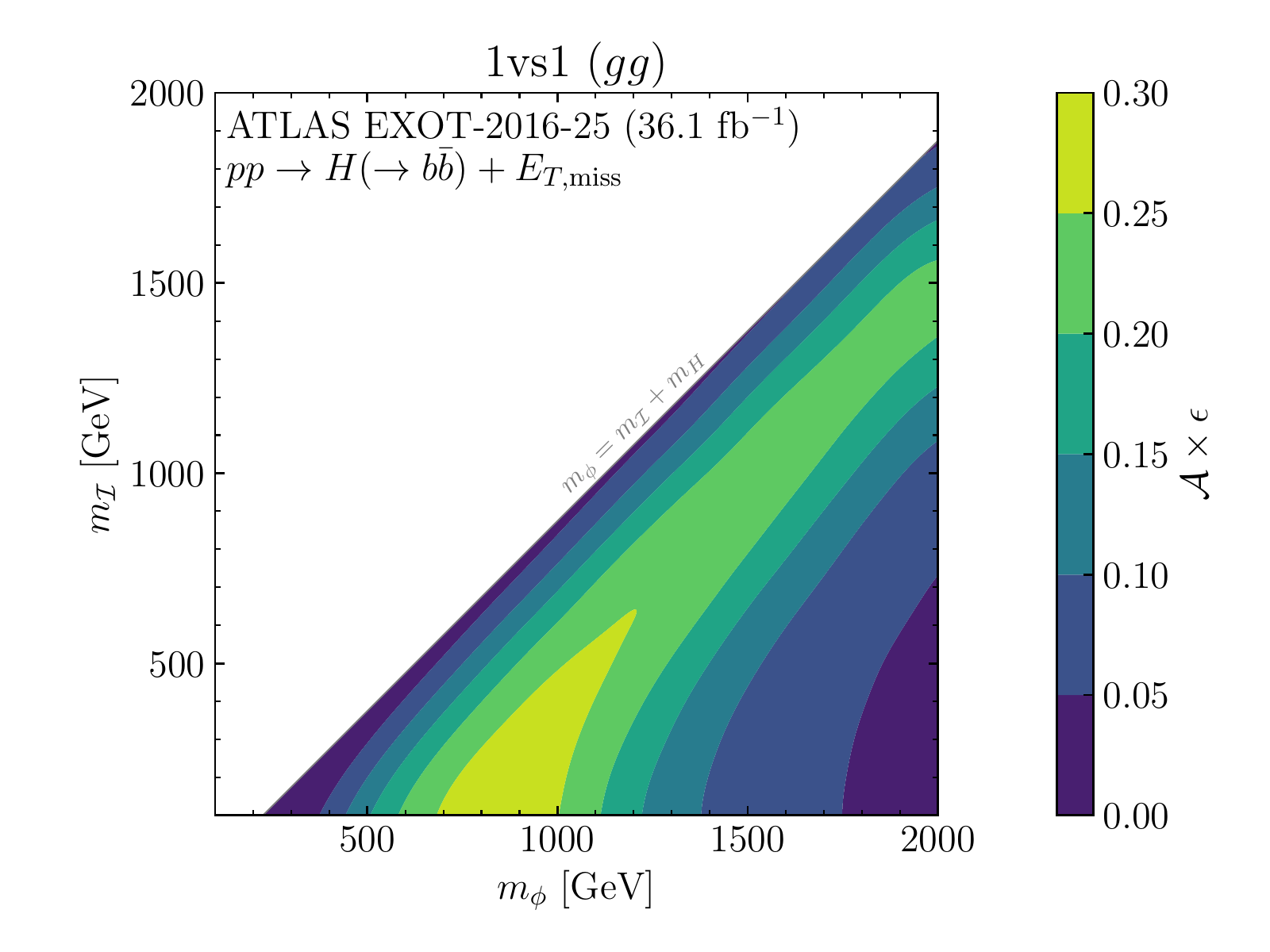}
\end{subfigure}
\begin{subfigure}{.49\linewidth}\centering
\includegraphics[width=\textwidth]{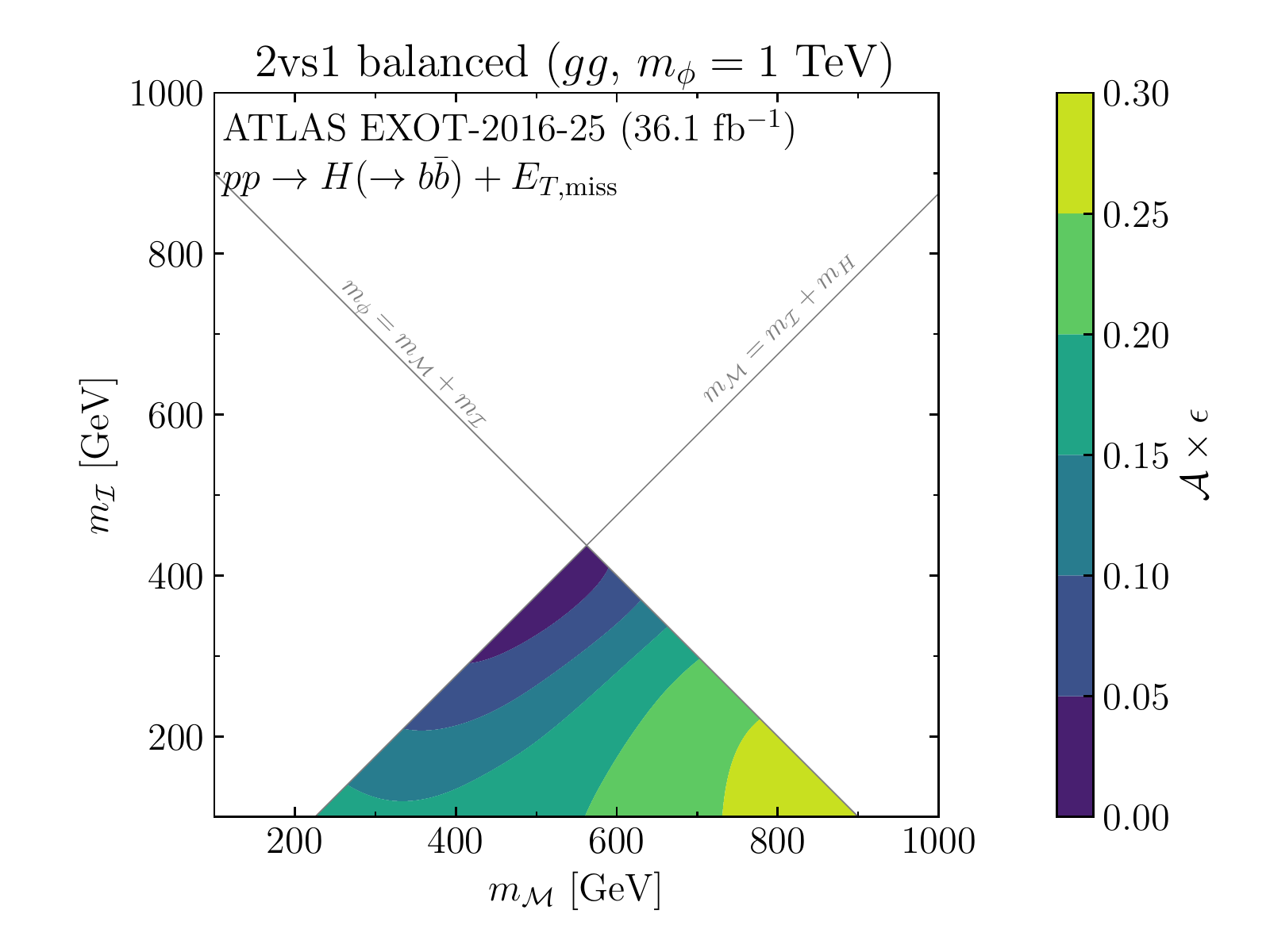}
\end{subfigure}
\begin{subfigure}{.49\linewidth}\centering
\includegraphics[width=\textwidth]{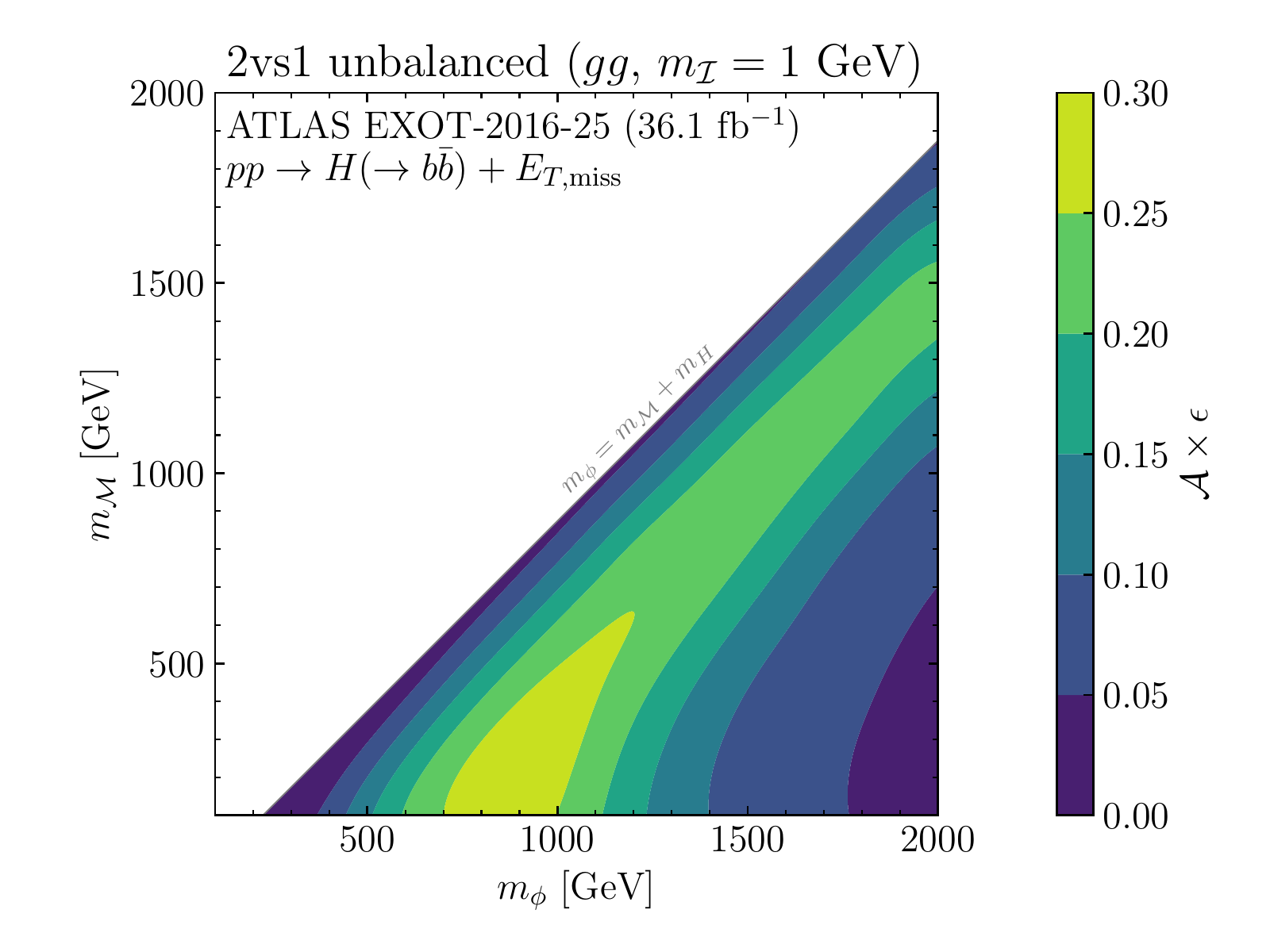}
\end{subfigure}
\begin{subfigure}{.49\linewidth}\centering
\includegraphics[width=\textwidth]{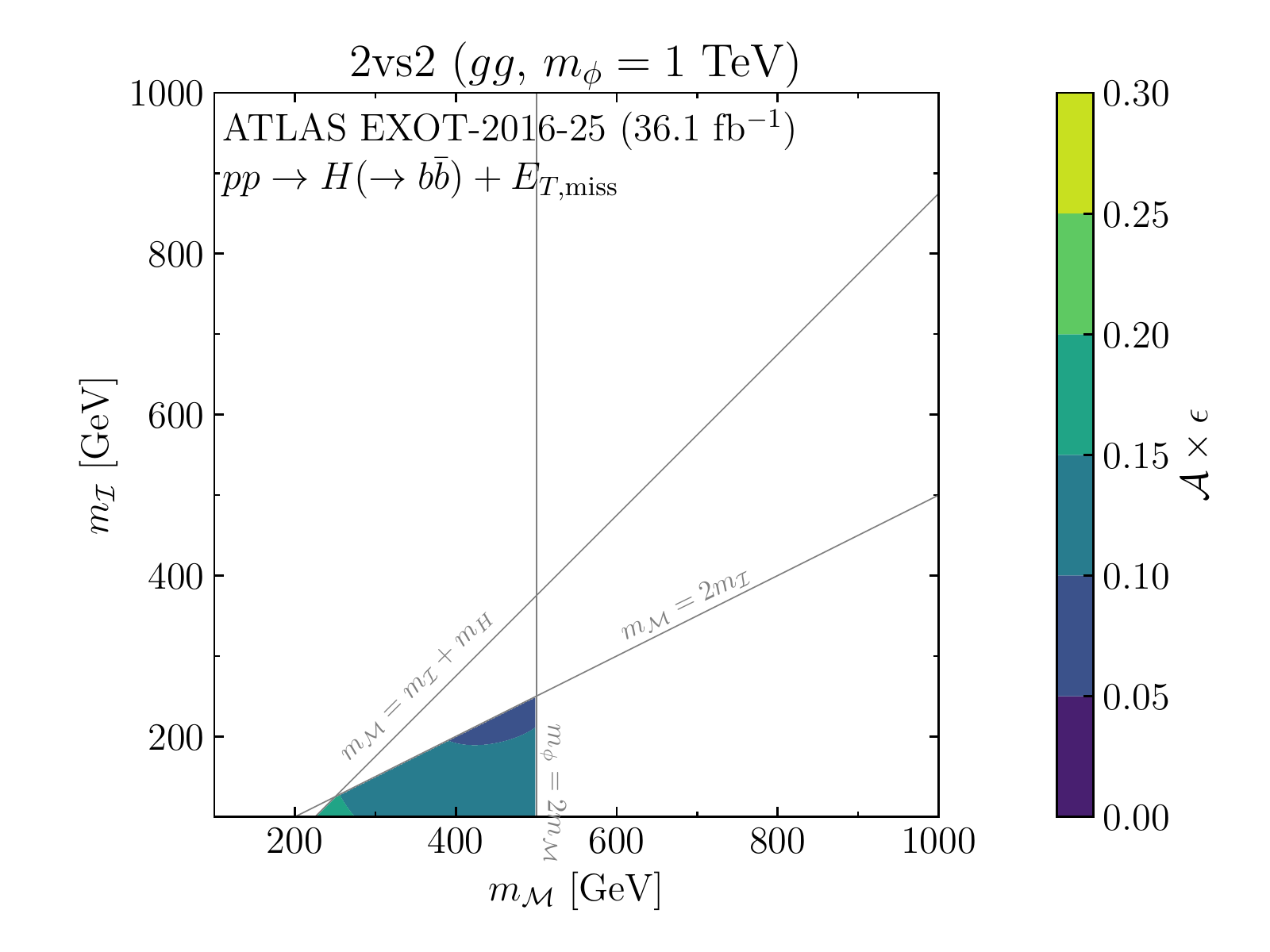}
\end{subfigure}
\caption{Acceptance $\times$ efficiency maps for the $gg$-initiated mono-Higgs plus \ETmiss simplified model topologies derived by recasting the ATLAS mono-Higgs plus \ETmiss search of \ccite{ATLAS:2017uis}. The results are shown in the $(m_\phi,m_\Inv)$ parameter plane for the 1vs1 topology (upper left panel), in the $(m_\Med,m_\Inv)$ parameter plane for the 2vs1 balanced topology (upper right panel), in the
$(m_\phi,m_\Med)$ parameter plane for the 2vs1 unbalanced topology (lower right panel), and in the
$(m_\Med,m_\Inv)$ parameter plane for the 2vs2 topology (lower right panel). The kinematic constraints for each topology are depicted by gray lines.}
\label{fig:monoH+MET_example_eff_maps}
\end{figure}

The resulting efficiency maps for gluon-initiated $\phi$ production are presented in \cref{fig:monoH+MET_example_eff_maps}. In the upper left panel, the acceptance $\times$ efficiency map for the 1vs1 topology is shown. In the upper right panel, it is shown for the 2vs1 balanced topology; in the lower left panel for the 2vs1 unbalanced topology; and in the lower right panel, for the 2vs2 topology. For the 1vs1 topology, the parameter scan is restricted to the two parameters $m_\phi$ and $m_\Inv$ (as no mediator appears in the 1vs1 topology), which are varied in the interval $[100,2000]\gev$. For the 2vs1 unbalanced topology, we set $m_\Inv = 1\gev$ and scan $m_\phi$ and $m_\Med$ in the interval $[100,2000]\gev$. For the other topologies, we fix $m_\phi = 1\tev$ and vary $m_\Med$ and $m_\Inv$ in the interval $[100,1000]\gev$. For all plots, the kinematic constraints are depicted by gray lines.\footnote{As we assume here for simplicity that all involved particles have a negligible decay width, no off-shell effects are incorporated, and the kinematic constraints result in sharp boundaries.}

For the 1vs1 topology, we find the highest values of acceptance $\times$ efficiency ($\sim 0.28$) for $m_\phi\sim 900\gev$ and $m_\Inv \sim 200$ $\gev$. The acceptance $\times$ efficiency values decrease if the kinematic limit of $m_\phi = m_\Inv + m_H$ is approached. In this limit, the invisible particle and the Higgs boson are produced at rest (in the $\phi$ rest frame) implying that almost no missing transverse energy is recorded. Since the analysis of \ccite{ATLAS:2017uis} requires a minimum amount of missing transverse energy ($\ETmiss \ge 150\gev$), this directly results in a lower acceptance. The acceptance also decreases towards the lower right corner of the parameter plane (high $m_\phi$, low $m_\Inv$). This is mainly a consequence of the \texttt{MadAnalysis} implementation of \ccite{ATLAS:2017uis}. While \ccite{ATLAS:2017uis} defines two signal regions, namely the resolved signal region, in which the two $b$ jets can be disentangled, and the merged signal region, in which the two $b$ jets are merged into a single large-radius jet, only the resolved signal region is available within \texttt{MadAnalysis}. For the resolved signal region, $\ETmiss \le 500\gev$ is required. For high $m_\phi$ and low $m_\Inv$, the amount of missing transverse energy surpasses this threshold.

The highest acceptance $\times$ efficiency ($\sim 0.27$) for the 2vs1 balanced topology is reached for the highest values of $m_\Med$ and the lowest values for $m_\Inv$, i.e.\ $m_\Med \sim 900 \gev$ and $m_\Inv \sim 100\gev$ in the depicted mass range. The low acceptance values for $m_\phi \sim 2 m_\Inv + m_H$ are again a consequence of the Higgs boson and the invisible particles being produced approximately at rest. For the lower right corner of the shown parameter plane, the invisible particle produced by the initial $\phi$ decay is approximately produced at rest. The decay of the additionally produced mediator then mimics the situation of the 1vs1 topology, where the lower right corner of the upper right panel of \cref{fig:monoH+MET_example_eff_maps} corresponds to the $m_\phi = 1\tev$ and $m_\Inv = 100\gev$ parameter point in the upper left panel of \cref{fig:monoH+MET_example_eff_maps}.

For the 2vs1 unbalanced topology, we obtain the highest acceptance $\times$ efficiency values ($\sim 0.29$) for $m_\phi\sim 900\gev$ and $m_\Med\sim 100\gev$. Since the mediator decays to two invisible particles, the mass of the invisible particle does not influence the visible kinematics (as long as $m_\Med > 2 m_\Inv$). Therefore, the 2vs1 unbalanced topology behaves like the 1vs1 topology with the mediator particle in the 2vs1 unbalanced topology playing the role of the invisible particle in the 1vs1 topology. Only for small mediator masses, small differences are visible in comparison to the 1vs1 topology.

For the 2vs2 topology, the highest values of acceptance $\times$ efficiency ($\sim 0.16$) are obtained for $m_\Med \sim 250\gev$ and $m_\Inv \sim 100\gev$. This topology behaves very similar to the 2vs1 balanced topology. The kinematically accessible parameter region is, however, reduced in comparison to the 2vs1 balanced topology.

In summary, while the maximum acceptance $\times$ efficiency values are reached in different regions of the parameter space for the different topologies, the maximum values are of similar magnitude. This clearly shows that the analysis of \ccite{ATLAS:2017uis} has also a significant sensitivity for other topologies than the 2vs1 unbalanced topology, for which the analysis was originally designed. Since the patterns of the acceptance $\times$ efficiency maps that we have found for the different topologies are a direct consequence of the missing transverse energy requirements of the experimental search, we expect similar results for other mono-Higgs (or mono-$Z$) plus \ETmiss searches.

\begin{figure}\centering
\begin{subfigure}{.49\linewidth}\centering
\includegraphics[width=\textwidth]{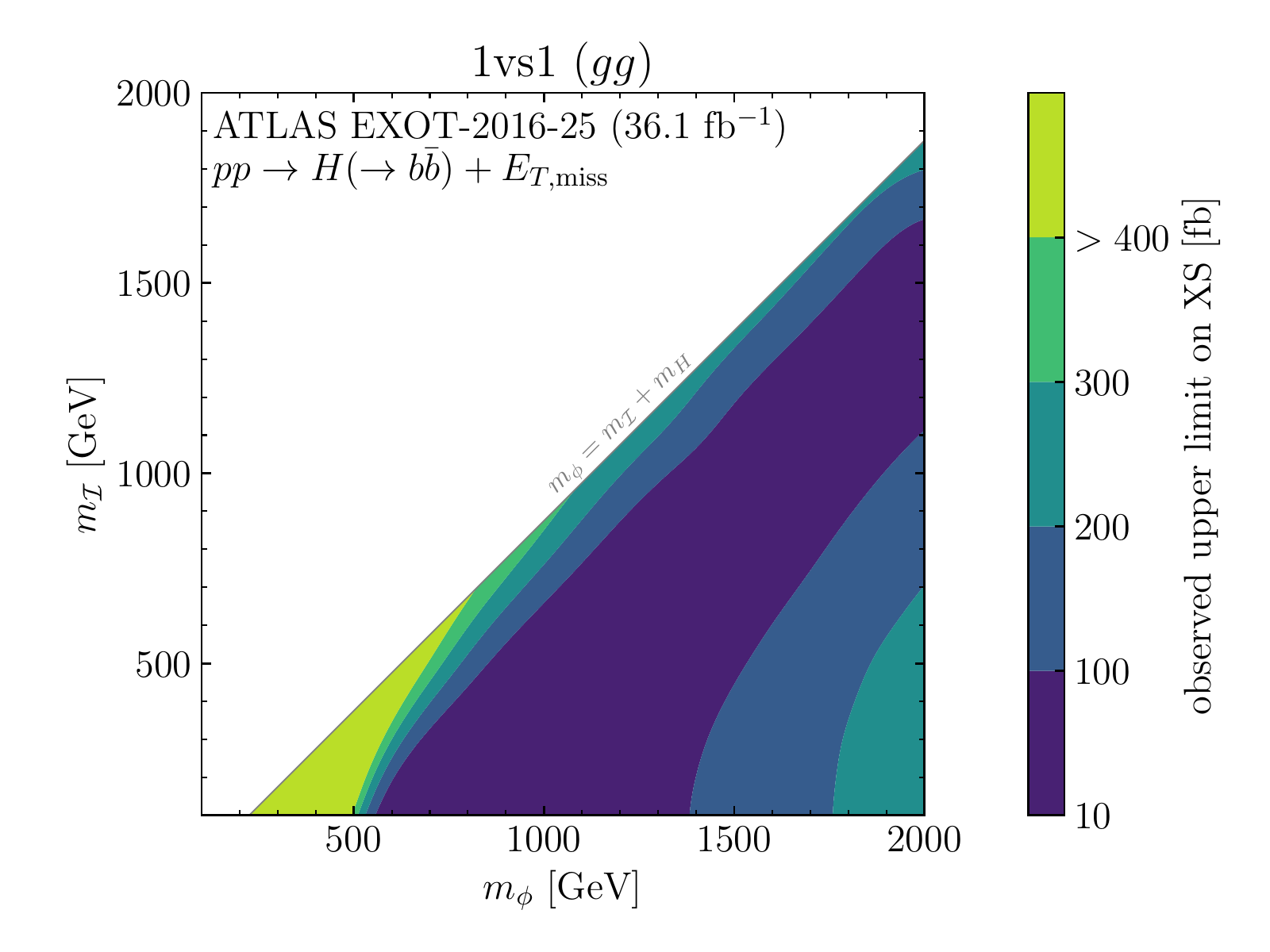}
\end{subfigure}
\begin{subfigure}{.49\linewidth}\centering
\includegraphics[width=\textwidth]{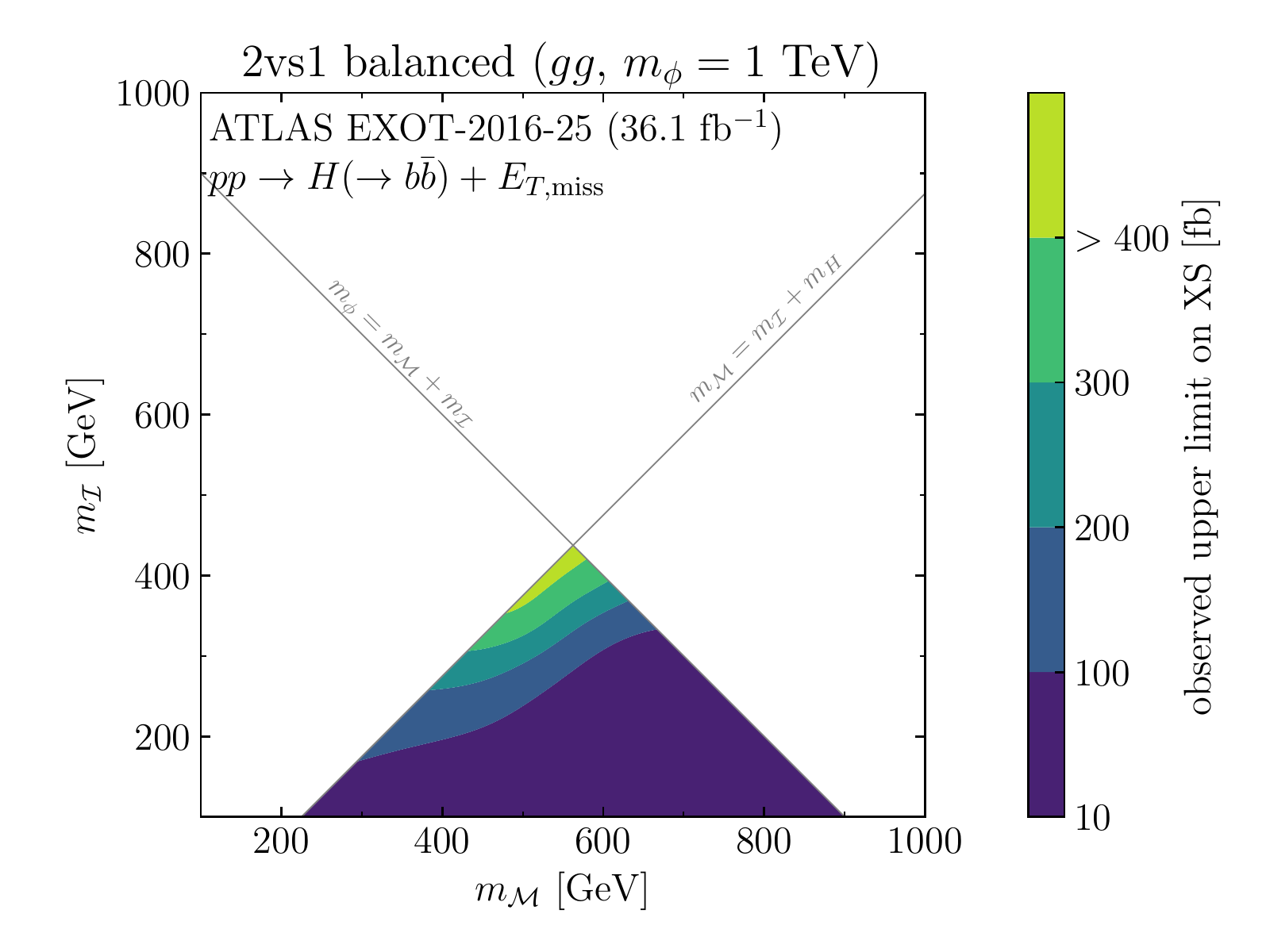}
\end{subfigure}
\begin{subfigure}{.49\linewidth}\centering
\includegraphics[width=\textwidth]{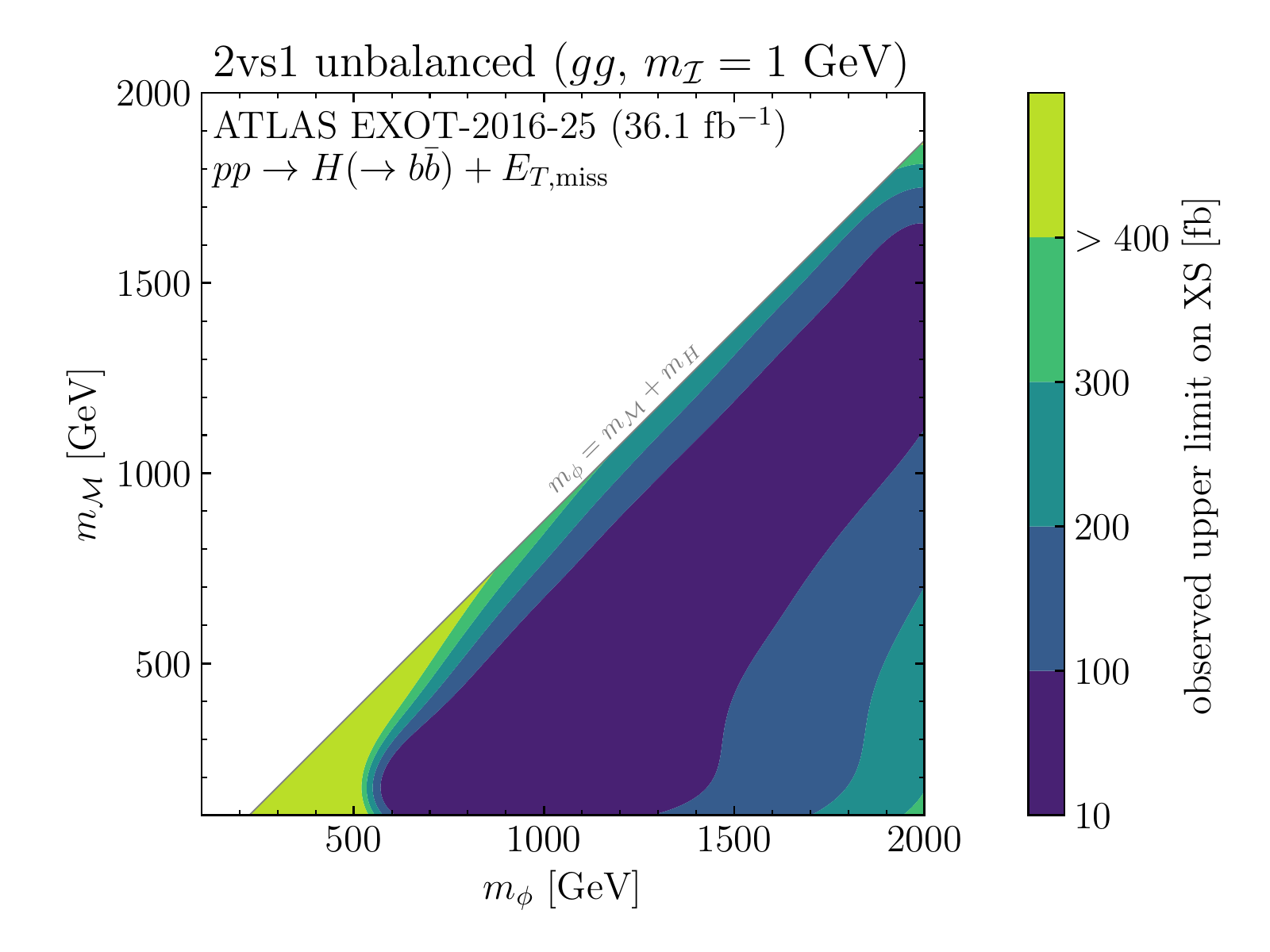}
\end{subfigure}
\begin{subfigure}{.49\linewidth}\centering
\includegraphics[width=\textwidth]{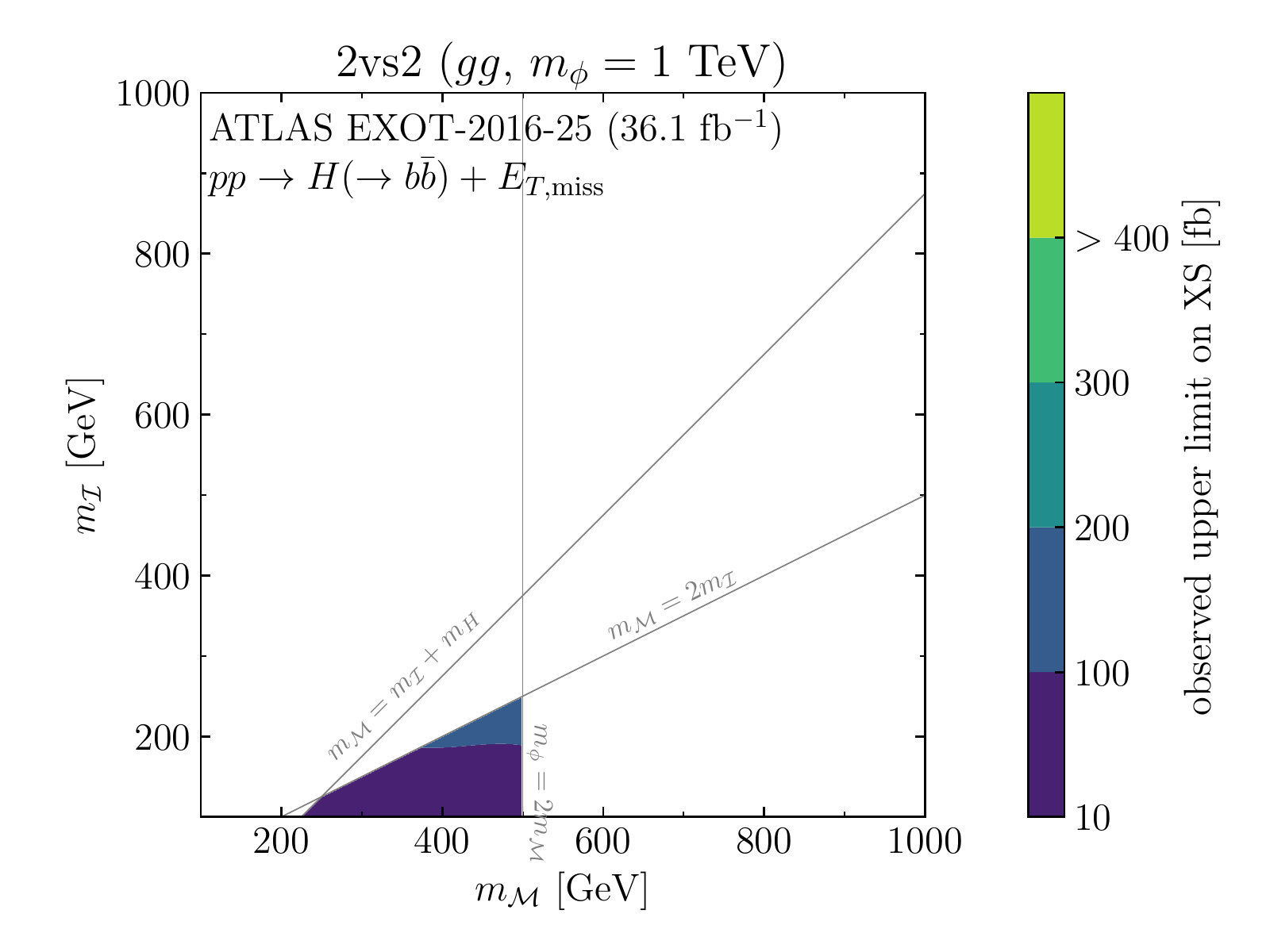}
\end{subfigure}
\caption{$95\%$ CL upper limits, derived by recasting the ATLAS mono-Higgs plus \ETmiss search for gluon-initiated production of \ccite{ATLAS:2017uis}, on the cross sections that are associated with the different simplified model topologies for this process. The results are shown in the $(m_\phi,m_\Inv)$ parameter plane for the 1vs1 topology (upper left panel), in the $(m_\Med,m_\Inv)$ parameter plane for the 2vs1 balanced topology (upper right panel), in the $(m_\Phi,m_\Med)$ parameter plane for the 2vs1 unbalanced topology (lower left panel), and in the $(m_\Med,m_\Inv)$ parameter plane for the 2vs2 topology (lower right panel). The kinematic constraints are depicted by gray lines.}
\label{fig:monoH+MET_example_upper_XS_limit}
\end{figure}

For each point in the parameter space of the simplified model topologies the acceptance $\times$ efficiency maps can be utilised to translate a given signal cross section into the corresponding number of signal events that would be expected to be observed in the considered experimental analysis. From this information, together with the number of expected SM background events and the number of the actually observed events in the experimental analysis (which are saved as part of the \texttt{MadAnalysis} analysis implementation~\cite{DVN/KAAYFM_2021}) we can directly obtain observed upper limits on the cross sections that are associated with the different simplified model topologies, following the procedure outlined in \ccite{Conte:2018vmg}. For models that can be mapped to the simplified model topologies, those cross section limits can easily be used for comparing the model predictions with the experimental limits. It is sufficient for this purpose to know the signal cross section for each of the simplified model topologies that is realised in the considered model as a function of the masses of the particles that correspond to $\Phi$, $\Med$ and $\Inv$. This direct application of the cross section limits works best for simple models where only one channel contributes to the signal. As we will discuss for the example of the MSSM below, for more complicated models where several channels contribute to the signal it can be advantageous to utilise the information contained in the acceptance $\times$ efficiency maps rather than just resorting to the cross section limits. However, in both kinds of applications the information about the signal cross section for the simplified model topologies is sufficient as input, i.e.\ no recasting or event generation is needed in order to apply the experimental results for testing the considered models.

The $95\%$ CL upper limits for the different simplified model topologies that we have obtained from recasting the ATLAS mono-Higgs plus \ETmiss search of \ccite{ATLAS:2017uis} are shown in \cref{fig:monoH+MET_example_upper_XS_limit}. For all four topologies, the strongest upper limit on the observed cross section is of $\mathcal{O}(10)$~fb within the considered parameter space. If the acceptance $\times$ efficiency value drops to zero (e.g.\ close to a kinematic edge), no upper limit on the observed cross section can be set. This is indicated by the upper bin of the colour coding, which represents the $\sigma_{95\%\text{ CL}}> 400$~fb region. The overall behaviour of the different topologies follows the one observed in \cref{fig:monoH+MET_example_eff_maps}. For the relationship between the acceptance $\times$ efficiency value and the observed upper limit on the cross section, we refer to \ccite{Conte:2018vmg}. As a consequence of this non-trivial relationship, the contours for the cross section limits displayed in \cref{fig:monoH+MET_example_upper_XS_limit} slightly differ from the patterns of the acceptance $\times$ efficiency maps shown in \cref{fig:monoH+MET_example_eff_maps}.

\begin{figure}\centering
\begin{subfigure}{.49\linewidth}\centering
\includegraphics[width=\textwidth]{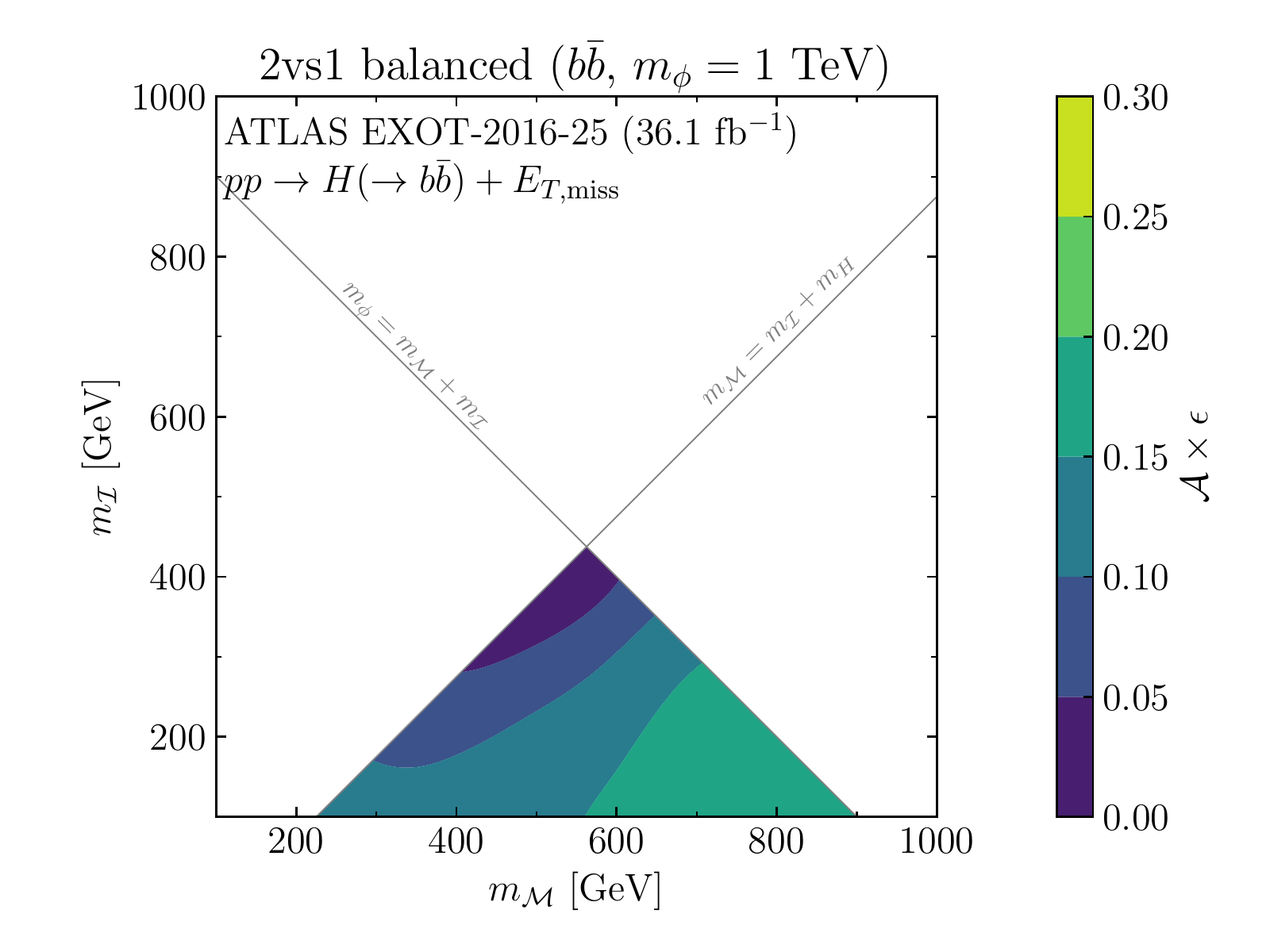}
\end{subfigure}
\begin{subfigure}{.49\linewidth}\centering
\includegraphics[width=\textwidth]{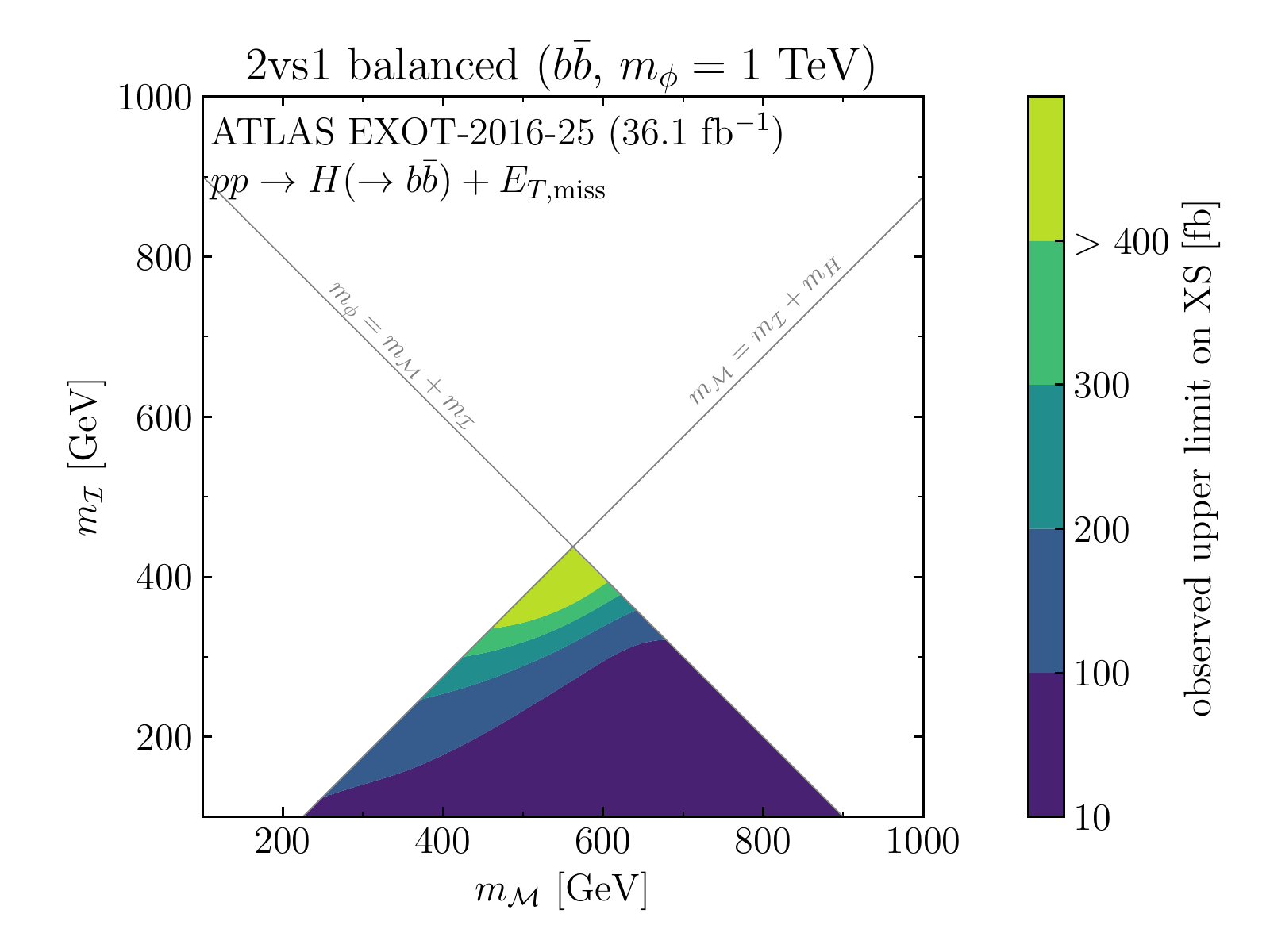}
\end{subfigure}
\caption{Acceptance $\times$ efficiency map (left) and observed upper limit on the cross section (right), derived by recasting the ATLAS mono-Higgs plus \ETmiss search for $b\bar b$-initiated production of \ccite{ATLAS:2017uis}, for the 2vs1 balanced topology.}
\label{fig:monoH+MET_example_bb}
\end{figure}

In addition to the results for gluon-initiated $\phi$ production, we show in \cref{fig:monoH+MET_example_bb} results for bottom-quark-associated $\phi$ production. Here, we discuss only the 2vs1 balanced topology, since we will make use of this result below. As a result of the additional $b$ jets in the final state, the obtained acceptance $\times$ efficiency values (shown in the left panel) are slightly lower than the values observed for the 2vs1 balanced topology for gluon-initiated $\phi$ production (see the upper right panel of \cref{fig:monoH+MET_example_eff_maps}), with a maximal value of $\sim 0.20$. Correspondingly, also the observed upper limit on the cross section (shown in the right panel) is slightly weaker in comparison to the case of gluon-initiated $\phi$ production (see the upper right panel of \cref{fig:monoH+MET_example_upper_XS_limit}).


\subsection{MSSM interpretation}

After demonstrating how our proposed simplified model framework can be utilised by the experimental collaborations in order to present their results, we now want to illustrate how the information provided in this way can be used by phenomenologists to constrain specific BSM models.

As an example, we consider the case of the MSSM and focus on the 2vs1 balanced topology, see the discussion in \cref{sec:H_ETmiss}. Specifically, we consider the decay of a heavy Higgs boson into two neutralinos. One of these neutralinos, $\tilde\chi_1$, is the lightest supersymmetric particle and therefore corresponds to the invisible particle in the simplified model framework. The second neutralino corresponds to the mediator particle. It decays to $\tilde\chi_1$ and a SM-like Higgs boson, which we denote by $h$ in this Section.

The MSSM  comprises not just a single heavy Higgs boson but two heavy neutral Higgs states, namely the \cp-even $H$ boson and the \cp-odd $A$ boson. They can be produced either via gluon fusion or bottom-associated Higgs production. We focus here on a scenario in which three of the four neutralinos, labelled by $\tilde\chi_{1,2,3}$ with $m_{\tilde \chi_1} \le m_{\tilde \chi_2} \le m_{\tilde \chi_3}$ are relatively light, whereas the fourth neutralino is much heavier and does not affect the Higgs decay processes. It should be noted that either $\tilde\chi_3$ or $\tilde\chi_2$ can be the mediator. Consequently, eight different sub-channels contribute to the mono-Higgs plus \ETmiss signature:\footnote{In principle, $\tilde\chi_3$ can also first decay into $\tilde\chi_2$ and a Higgs boson. This process would, however, correspond to a final state with two Higgs bosons, while we focus on a mono-Higgs final state here.}
\begin{itemize}
  \item $gg\to H \to \tilde\chi_2\tilde\chi_1 \to \tilde\chi_1\tilde\chi_1 h$,
  \item $gg\to H \to \tilde\chi_3\tilde\chi_1 \to \tilde\chi_1\tilde\chi_1 h$,
  \item $gg\to A \to \tilde\chi_2\tilde\chi_1 \to \tilde\chi_1\tilde\chi_1 h$,
  \item $gg\to A \to \tilde\chi_3\tilde\chi_1 \to \tilde\chi_1\tilde\chi_1 h$,
  \item $pp\to H b\bar b \to \tilde\chi_2\tilde\chi_1 b\bar b \to \tilde\chi_1\tilde\chi_1 h b\bar b$,
  \item $pp\to H b\bar b \to \tilde\chi_3\tilde\chi_1 b\bar b \to \tilde\chi_1\tilde\chi_1 h b\bar b$,
  \item $pp\to A b\bar b \to \tilde\chi_2\tilde\chi_1 b\bar b \to \tilde\chi_1\tilde\chi_1 h b\bar b$,
  \item $pp\to A b\bar b \to \tilde\chi_3\tilde\chi_1 b\bar b \to \tilde\chi_1\tilde\chi_1 h b\bar b$.
\end{itemize}
While we could compare the predicted cross-section values for each of these sub-channels with the corresponding upper limits presented in \cref{fig:monoH+MET_example_upper_XS_limit,fig:monoH+MET_example_bb}, a more stringent constraint can be obtained by combining all sub-channels. This is achieved by calculating the cross section for each sub-channel, multiplying with the corresponding acceptance $\times$ efficiency value (as provided in \cref{fig:monoH+MET_example_eff_maps,fig:monoH+MET_example_bb}), and then adding up the individual contributions in order to obtain
the result for the overall cross section $\times$ acceptance $\times$ efficiency,
\begin{align}
&\left(\sigma\times\mathcal{A}\times\epsilon\right)(pp \to (b\bar b)H,A \to (b\bar b)\tilde\chi_1\tilde\chi_1 h) = \nonumber\\
&= \sigma(gg\to H \to \tilde\chi_2\tilde\chi_1 \to \tilde\chi_1\tilde\chi_1 h) \cdot \left[(\mathcal{A}\times\epsilon)_{\text{2vs1-b.}}^{gg}(m_\phi=m_H,m_\Med=m_{\tilde\chi_2},m_\Inv=m_{\tilde\chi_1})\right] \nonumber\\
&\hspace{.4cm} + \sigma(gg\to H \to \tilde\chi_3\tilde\chi_1 \to \tilde\chi_1\tilde\chi_1 h) \cdot \left[(\mathcal{A}\times\epsilon)_{\text{2vs1-b.}}^{gg}(m_\phi=m_H,m_\Med=m_{\tilde\chi_3},m_\Inv=m_{\tilde\chi_1})\right] \nonumber\\
&\hspace{.4cm} + \sigma(gg\to A \to \tilde\chi_2\tilde\chi_1 \to \tilde\chi_1\tilde\chi_1 h) \cdot \left[(\mathcal{A}\times\epsilon)_{\text{2vs1-b.}}^{gg}(m_\phi=m_A,m_\Med=m_{\tilde\chi_2},m_\Inv=m_{\tilde\chi_1})\right] \nonumber\\
&\hspace{.4cm} + \sigma(gg\to A \to \tilde\chi_3\tilde\chi_1 \to \tilde\chi_1\tilde\chi_1 h) \cdot \left[(\mathcal{A}\times\epsilon)_{\text{2vs1-b.}}^{gg}(m_\phi=m_A,m_\Med=m_{\tilde\chi_3},m_\Inv=m_{\tilde\chi_1})\right] \nonumber\\
&\hspace{.4cm} + \sigma(pp\to b\bar b H \to b\bar b \tilde\chi_2\tilde\chi_1 \to b\bar b\tilde\chi_1\tilde\chi_1 h) \cdot \left[(\mathcal{A}\times\epsilon)_{\text{2vs1-b.}}^{b\bar b}(m_\phi=m_H,m_\Med=m_{\tilde\chi_2},m_\Inv=m_{\tilde\chi_1})\right] \nonumber\\
&\hspace{.4cm} + \sigma(pp\to b\bar b  H \to b\bar b \tilde\chi_3\tilde\chi_1 \to b\bar b \tilde\chi_1\tilde\chi_1 h) \cdot \left[(\mathcal{A}\times\epsilon)_{\text{2vs1-b.}}^{b\bar b}(m_\phi=m_H,m_\Med=m_{\tilde\chi_3},m_\Inv=m_{\tilde\chi_1})\right] \nonumber\\
&\hspace{.4cm} + \sigma(pp\to b\bar b A \to b\bar b \tilde\chi_2\tilde\chi_1 \to b\bar b \tilde\chi_1\tilde\chi_1 h) \cdot \left[(\mathcal{A}\times\epsilon)_{\text{2vs1-b.}}^{b\bar b}(m_\phi=m_A,m_\Med=m_{\tilde\chi_2},m_\Inv=m_{\tilde\chi_1})\right] \nonumber\\
&\hspace{.4cm} + \sigma(pp\to b\bar b A \to b\bar b \tilde\chi_3\tilde\chi_1 \to b\bar b \tilde\chi_1\tilde\chi_1 h) \cdot \left[(\mathcal{A}\times\epsilon)_{\text{2vs1-b.}}^{b\bar b}(m_\phi=m_A,m_\Med=m_{\tilde\chi_3},m_\Inv=m_{\tilde\chi_1})\right].
\end{align}
After multiplying with $\text{BR}(h\to b\bar b)$, this cross section $\times$ acceptance $\times$ efficiency value together with the expected background contribution can be directly compared to the number of experimentally observed events for the considered luminosity (following again the procedure outlined in \ccite{Conte:2018vmg}).

For our numerical analysis, we use \texttt{FeynHiggs 2.18.0}~\cite{Heinemeyer:1998yj,Heinemeyer:1998np,Hahn:2009zz,Degrassi:2002fi,Frank:2006yh,Hahn:2013ria,Bahl:2016brp,Bahl:2017aev,Bahl:2018qog} in order to obtain the predictions for the Higgs mass spectrum and the Higgs branching ratios. For calculating the Higgs production cross sections, we employ \texttt{SusHi 1.7.0}~\cite{Harlander:2012pb,Harlander:2016hcx,Harlander:2002wh,Harlander:2003ai,Aglietti:2004nj,,Bonciani:2010ms,Degrassi:2010eu,Degrassi:2011vq,Degrassi:2012vt,Harlander:2005rq,Chetyrkin:2000yt}. The neutralino branching ratios are obtained using \texttt{SUSY-HIT}~\cite{Djouadi:1997yw,Djouadi:2002ze,Muhlleitner:2003vg,Djouadi:2006bz}.

We set all squark soft SUSY-breaking parameters (as well as the gluino mass parameter $M_3$ and the Wino mass parameter $M_2$) equal to the common scale $M_\text{SUSY}$, which we fix to 2~TeV. All trilinear soft SUSY-breaking parameters are set to zero apart from the trilinear stop coupling, which we fix by setting the stop mixing parameter $X_t$ to 4~TeV. The ratio of the two vacuum expectation values, $\tan\beta$, is set to 10. The scale of the non-SM-like Higgs bosons is fixed to 1~TeV. All input parameters are assumed to be real implying that no mixing between the \cp-even $H$ and the \cp-odd $A$ bosons occurs.

\begin{figure}\centering
\includegraphics[width=.7\textwidth]{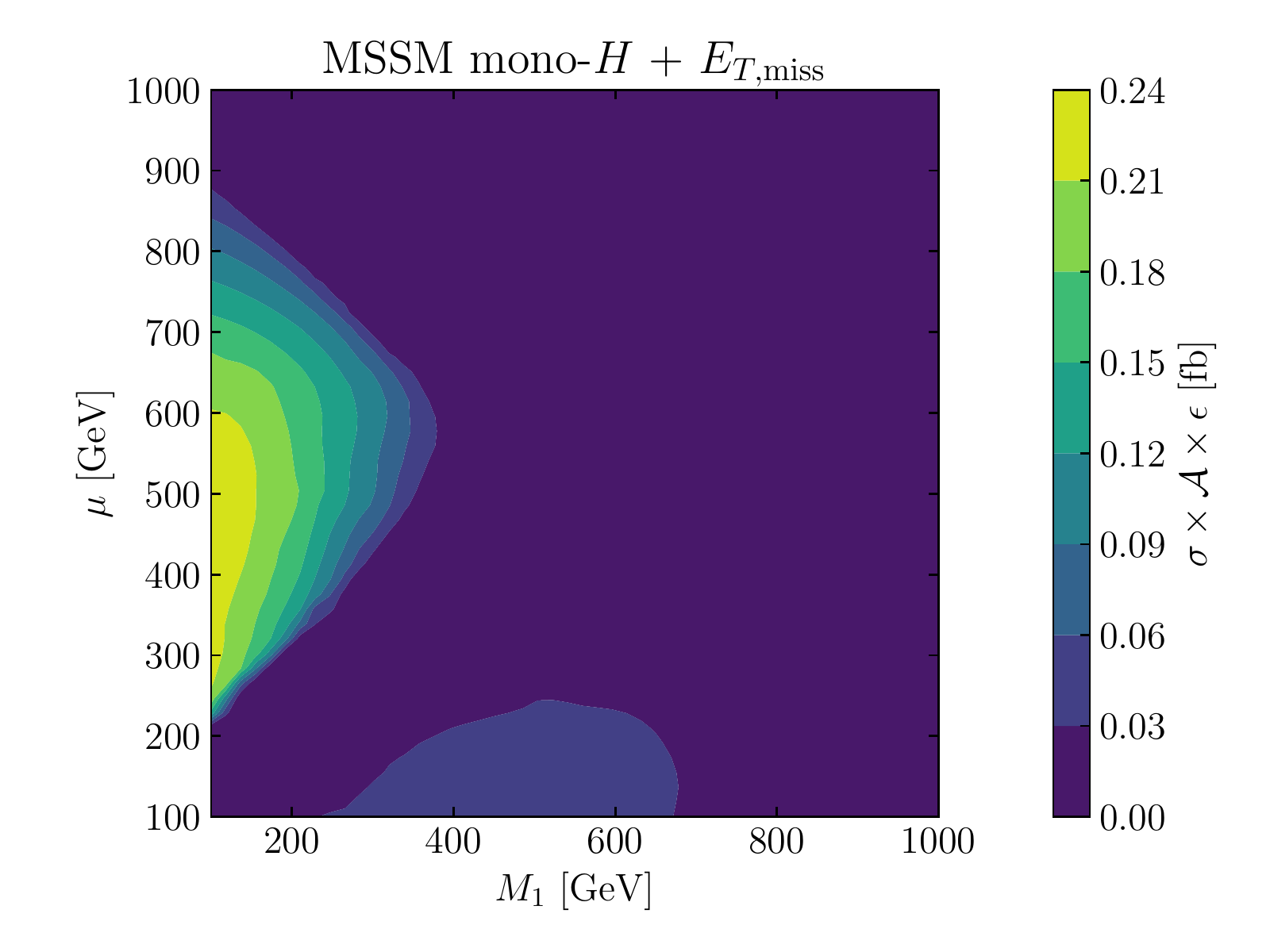} \caption{Predicted cross section $\times$ acceptance $\times$ efficiency for the mono-Higgs plus \ETmiss signature in the $(M_1,\mu)$ parameter plane of the MSSM according to the analysis carried out in \ccite{ATLAS:2017uis}.} \label{fig:monoH+MET_example_MSSM_prediction}
\end{figure}

The predicted values for the overall cross section $\times$ acceptance $\times$ efficiency in dependence of $M_1$ and the Higgsino mass parameter $\mu$ are shown in \cref{fig:monoH+MET_example_MSSM_prediction}. The highest values are reached for $250\gev \lesssim \mu \lesssim 850\gev$ and $M_1 \lesssim 400\gev$ with $\sigma\times\mathcal{A}\times\epsilon\lesssim 0.25\fb$. Even this maximal cross-section value is far smaller than the $68\%$ CL limit set in \ccite{ATLAS:2017uis} of $\sim 5\fb$. Even if we rescale this exclusion limit naively to $3000\invfb$ (as expected to be collected during the high-luminosity phase of the LHC), no part of the parameter plane as presented in \cref{fig:monoH+MET_example_MSSM_prediction} could be excluded on the basis of the mono-Higgs plus \ETmiss search at the HL-LHC.

Even though the recasting of the ATLAS mono-Higgs plus \ETmiss search of \ccite{ATLAS:2017uis} in the framework of the proposed simplified model topologies and its application to the production of heavy Higgs bosons in the MSSM has not resulted in an excluded region in the considered MSSM parameter plane, we regard this example as a useful illustration of our proposed framework. In particular we have demonstrated how the acceptance $\times$ efficiency maps as presented in \cref{sec:monoH+MET_example_eff_maps} can be used to constrain realistic BSM models (possibly involving multiple sub-channels) without the need for running any MC generation (for the desired situation where those acceptance $\times$ efficiency maps would be presented by the experimental collaborations as part of their analysis). Moreover, the result of our considered example motivates new analysis strategies that are optimized for e.g.\ the 2vs1 balanced topology (and not only for the 2vs1 unbalanced topology as done by most existing experimental searches).


\section{Conclusions}
\label{sec:conclusions}

Searches for additional Higgs bosons that are targeted at final states containing BSM particles have the potential to significantly enhance and complement the sensitivity of searches where the additional Higgs boson directly decays into SM particles. While Higgs searches involving BSM final states have been explored for some specific cases, a systematic investigation of BSM final states is lacking up to now. In this paper, we have proposed a simplified model framework in order to facilitate experimental analyses of dedicated searches for additional scalar resonances that decay into one or more BSM particles. We have pointed out that a simplified model interpretation can be very helpful in this context as a way to present the experimental results in a form that makes them easily applicable to a wide variety of possible models of BSM physics.

As a first step in the development of the simplified model framework, we have focused in this paper on signatures comprising a mono-$Z$ or a mono-Higgs plus missing transverse energy. Searches for a single SM particle recoiling against missing transverse energy are well-motivated in general by the quest to unravel the nature of dark matter and by the possibility that additional scalars of an extended Higgs sector could serve as a dark matter portal. While in the present paper we have restricted our discussion to this particular class of signatures, we stress that our general approach can easily be extended also to other final states.

Existing searches for mono-$Z$ and mono-Higgs plus missing transverse energy signatures concentrate on a specific decay topology in which the resonant particle decays into a mediator and a SM particle ($Z$ or Higgs boson) with the mediator particle decaying into two invisible particles. We have pointed out in our analysis that in many BSM models other decay topologies can be encountered.

As a first step, we have classified all possible decay topologies, taking into account only three-point interactions and considering decay chains which may contain a BSM mediator particle and invisible particles in the final state. Incorporating all different possibilities for the spin of the mediator and the invisible particles we encountered five distinct topologies: the 1vs1 topology, for which the resonant scalar directly decays to an invisible particle and a SM particle; the 2vs1 balanced topology, for which the resonant particle decays to an invisible particle and a mediator particle, which then decays to the SM particle and an invisible particle; the 2vs1 unbalanced topology, for which the resonant particle decays into a SM particle and a mediator particle, which then decays into two invisible particles; the 2vs2 topology, for which the initial resonance decays to two mediator particles, of which one decays to two invisible particles and the other one to a SM and an invisible particle; as well as the ISR topology, for which the SM particle is radiated off the initial state while the scalar resonance directly decays to two invisible particles.

As the next step, we investigated the kinematic distribution of the transverse momentum of the involved SM particle, which we regard as the most relevant experimental observable. We found that the spin character of the mediator and of the invisible particles has only a small impact on the kinematic distributions in most cases, indicating that a wide range of models can be covered by investigating a simplified model decay topology. Moreover, we found a similar behaviour for the 1vs1 and the 2vs1 unbalanced topologies. Both topologies feature a comparably sharp peak at large transverse momentum values, while the 2vs1 balanced, the 2vs2 and the ISR topologies feature a broad peak a lower transverse momentum values. Thus, the kinematic information obtained in an experimental analysis can be exploited for a possible discrimination between the different simplified model topologies.

These findings indicate that the proposed framework of decay topologies can serve as easy-to-use simplified models for the presentation of future experimental searches for mono-Higgs and mono-$Z$ plus missing transverse energy final states. Providing results not only for one specific decay topology but for the whole set of simplified models would make the results applicable to a wide variety of possible BSM models without the need to perform a tedious recasting based on MC generators.

As phenomenological applications, experimental results that are presented in terms of the simplified model topologies can be utilized for confronting predictions of different models with the experimental limits. This can either be carried out directly via the provided cross section limits, which should work best for relatively simple models, or via the acceptance $\times$ efficiency maps. The latter approach should be advantageous for more complicated models involving several channels. In both kinds of applications the information about the signal cross section for the simplified model topologies is sufficient as input, i.e.\ no recasting or event generation is needed in order to apply the experimental results for testing the considered models.

We have demonstrated how the proposed framework can be applied to experimental analyses using an existing ATLAS mono-Higgs plus missing energy search as an example. We have recasted this search to the various simplified model topologies, finding significant sensitivity to all of them. This clearly demonstrates the usefulness of our approach. As further step we have carried out a phenomenological analysis based on the experimental information that is expressed in terms of the simplified model approach. Considering the scenario where a heavy MSSM Higgs boson decays into a final state consisting of a mono-Higgs plus missing energy, a non-trivial signature composed of several sub-channels, we demonstrated that constraints on the parameter space of realistic models can be obtained in a straightforward way.

Consequently, the proposed framework is suitable to be implemented in codes which automatically test BSM models against experimental limits like \texttt{HiggsBounds}~\cite{Bechtle:2008jh,Bechtle:2013wla,Bechtle:2015pma,Bechtle:2020pkv,Bahl:2021yhk} or \texttt{SModelS}~\cite{Kraml:2013mwa,Ambrogi:2017neo,Dutta:2018ioj,Heisig:2018kfq,Ambrogi:2018ujg,Khosa:2020zar}. In this context, we also want to note that the proposed framework is applicable for search results that are given in the form of 95\% C.L.\ limits, but can also be employed for results comprising the full likelihood information.

In order to facilitate the use of our simplified model approach, we provide all necessary ingredients (i.e.\ model files) as ancillary material to this paper. While in the present paper our discussion focused on the mono-$Z$ and mono-Higgs final states, the proposed framework can directly be generalised to final states containing more than one $Z$ or Higgs boson. Also final states with $W$ bosons or leptons can be treated in the same manner. With this in mind, we hope that our approach can serve the community by improving the re-interpretability of future searches and as motivation to investigate so far unexplored decay topologies.


\section*{Acknowledgements}
\sloppy{
We thank Tim Stefaniak for collaboration in the early stages of this project, and we also we thank Danyer Perez Adan, Alexander Grohsjean and Christian Schwanenberger for useful discussions. We acknowledge support by the Deutsche Forschungsgemeinschaft (DFG, German Research Foundation) under Germany‘s Excellence Strategy -- EXC 2121 ``Quantum Universe'' – 390833306.  H.B.\ acknowledges support by the Alexander von Humboldt foundation.
}


\appendix


\section{\texorpdfstring{\texttt{UFO}}{UFO} model file}
\label{sec:model_file}

The \texttt{FeynRules} model file is called \texttt{simpBSM}; the corresponding \texttt{UFO} model file, \texttt{simpBSM\_UFO}. The model file contains seven particles in addition to the SM particles: \texttt{p0} (scalar resonance), \texttt{Ms} (scalar mediator), \texttt{Mv} (vector mediator), \texttt{Mf} (fermion mediator), \texttt{Is} (invisible scalar boson), \texttt{Iv} (invisible vector boson), and \texttt{If} (invisible fermion).\footnote{The model file also contains charged mediators as well as couplings to $W$ bosons. These are not needed for the work presented here but may be useful for future extensions.}

These particles are coupled to each other (and to SM particles) including all Lorentz-invariant dimension-four Lagrangian terms. By default, all the couplings added in addition to the SM couplings are set to the value one. These couplings are named by ``\texttt{g}'' followed by a short string denoting the involved particles (e.g.\ the coupling ``\texttt{gp0MsIs}'' controls the interaction between the scalar resonance, the scalar mediator, and the scalar invisible particle).

For scalar--fermion--fermion interactions, two couplings are involved. The coupling denoted by an additional ``\texttt{S}'' at the end controls the \cp-even part of the interaction; the coupling denoted by an additional ``\texttt{A}'' at the end the \cp-odd part (e.g.\ ``\texttt{gMsIfIfS}'' and ``\texttt{gMsIfIfA}''). Similarly, the coupling denoted by an addition ``\texttt{L}'' at the end controls the left-handed part of a fermion--fermion--vector interaction; and the coupling denoted by an addition ``\texttt{R}'' at the end controls the corresponding right-handed part.

The \texttt{FeynRules} and \texttt{UFO} model files are distributed as ancillary material for this paper.


\section{Event generation}
\label{sec:event_generation}

We use \texttt{MadGraph5\_aMC@NLO~2.8.2}~\cite{Alwall:2014hca} with \texttt{Pythia~8.244}~\cite{Sjostrand:2007gs} as parton shower. The \texttt{MSTW2008LO}~\cite{Martin:2009iq} parton-distribution set is used evaluated with \texttt{LHAPDF~6.3.0}~\cite{Whalley:2005nh}. We simulate the detector response using \texttt{Delphes~3.4.2}~\cite{deFavereau:2013fsa} with the ATLAS-LHC configuration card as provided by \texttt{Delphes}. The event analysis is performed using \texttt{MadAnalysis~5} (version~\texttt{1.9})~\cite{Dumont:2014tja,Conte:2012fm,Conte:2014zja,Conte:2018vmg}. We employ the four-flavour scheme.

\begin{table}\centering
\begin{tabular}{|c|c|c|}
\hline
topology                      & spin realization & \texttt{MG} decay syntax                          \\
\hline
\multirow{2}*{1vs1}            & $\Iv$           & \texttt{p0 > Z Iv}                                \\
                               & $\Is$           & \texttt{p0 > Z Is}                                \\
\hline
\multirow{5}*{2vs1 balanced}   & $\Mf\If$        & \texttt{p0 > Mf If\textasciitilde{}, (Mf > Z If)} \\
                               & $\Ms\Is$        & \texttt{p0 > Ms Is, (Ms > Z Is)}                  \\
                               & $\Mv\Iv$        & \texttt{p0 > Mv Iv, (Mv > Z Iv)}                  \\
                               & $\Ms\Iv$        & \texttt{p0 > Ms Iv, (Ms > Z Iv)}                  \\
                               & $\Mv\Is$        & \texttt{p0 > Mv Is, (Mv > Z Is)}                  \\
\hline
\multirow{4}*{2vs1 unbalanced} & $\Ms\Is$        & \texttt{p0 > Ms Z, (Ms > Is Is)}                  \\
                               & $\Ms\If$        & \texttt{p0 > Ms Z, (Ms > If If\textasciitilde{})} \\
                               & $\Ms\Iv$        & \texttt{p0 > Ms Z, (Ms > Iv Iv)}                  \\
                               & $\Mv\If$        & \texttt{p0 > Mv Z, (Mv > If If\textasciitilde{})} \\
\hline
\multirow{2}*{2vs2}            & $\Ms\Is$        & \texttt{p0 > Ms Ms, (Ms > Is Is, Ms > Z Is)}      \\
                               & $\Ms\Iv$        & \texttt{p0 > Ms Ms, (Ms > Iv Iv, Ms > Z Iv)}      \\
\hline
\multirow{2}*{ISR1}            & $\Iv$           & \texttt{p0 > Iv Iv}                               \\
                               & $\Is$           & \texttt{p0 > Is Is}                               \\
\hline
\multirow{2}*{ISR2}            & $\Ms\Is$        & \texttt{p0 > Ms Is, Ms > Is Is}                   \\
                               & $\Ms\Iv$        & \texttt{p0 > Ms Iv, Ms > Iv Iv}                   \\
\hline
\multirow{2}*{ISR3}            & $\Ms\Is$        & \texttt{p0 > Ms Ms, Ms > Is Is}                   \\
                               & $\Ms\Iv$        & \texttt{p0 > Ms Ms, Ms > Iv Iv}                   \\
\hline
\end{tabular}
\caption{List of mono-$Z$ topologies with \texttt{MadGraph} syntax for event generation. The ``ISR1'' topology corresponds to the left topology of \cref{fig:monoZ_topologies_ISR}; the ``ISR2'' topology to the middle topology of \cref{fig:monoZ_topologies_ISR}; and the ``ISR3'' topology to the right topology of \cref{fig:monoZ_topologies_ISR}.}
\label{tab:Z_MGsyntax}
\end{table}

\begin{table}\centering
\begin{tabular}{|c|c|c|}
\hline
topology                      & spin realization & \texttt{MG} decay syntax                          \\
\hline
\multirow{2}*{1vs1}            & $\Iv$           & \texttt{p0 > H Iv}                                \\
                               & $\Is$           & \texttt{p0 > H Is}                                \\
\hline
\multirow{5}*{2vs1 balanced}   & $\Mf\If$        & \texttt{p0 > Mf If\textasciitilde{}, (Mf > H If)} \\
                               & $\Ms\Is$        & \texttt{p0 > Ms Is, (Ms > H Is)}                  \\
                               & $\Mv\Iv$        & \texttt{p0 > Mv Iv, (Mv > H Iv)}                  \\
                               & $\Ms\Iv$        & \texttt{p0 > Ms Iv, (Ms > H Iv)}                  \\
                               & $\Mv\Is$        & \texttt{p0 > Mv Is, (Mv > H Is)}                  \\
\hline
\multirow{4}*{2vs1 unbalanced} & $\Ms\Is$        & \texttt{p0 > Ms H, (Ms > Is Is)}                  \\
                               & $\Ms\If$        & \texttt{p0 > Ms H, (Ms > If If\textasciitilde{})} \\
                               & $\Ms\Iv$        & \texttt{p0 > Ms H, (Ms > Iv Iv)}                  \\
                               & $\Mv\If$        & \texttt{p0 > Mv H, (Mv > If If\textasciitilde{})} \\
\hline
\multirow{2}*{2vs2}            & $\Ms\Is$        & \texttt{p0 > Ms Ms, (Ms > Is Is, Ms > H Is)}      \\
                               & $\Ms\Iv$        & \texttt{p0 > Ms Ms, (Ms > Iv Iv, Ms > H Iv)}      \\
\hline
\end{tabular}
\caption{List of mono-Higgs topologies with \texttt{MadGraph} syntax for event generation.}
\label{tab:H_MGsyntax}
\end{table}

Using this setup, we generate MC event samples employing the \texttt{UFO} model file introduced in \cref{sec:model_file}. Events for the scalar resonance production via gluon fusion are generated using the \texttt{MadGraph} syntax
\begin{verbatim}
  generate g g > p0 (MG_decay_syntax)
\end{verbatim}
where the ``\texttt{MadGraph} decay syntax'' for the various topologies can be found in  \cref{tab:Z_MGsyntax,tab:H_MGsyntax}. For the ISR topologies, the syntax
\begin{verbatim}
  generate g g > Z p0 (MG_decay_syntax)
\end{verbatim}
should be used instead.

The syntax for bottom-associated production reads
\begin{verbatim}
  generate p p > b b~ p0 PGG=0 QED=0, (MG_decay_syntax)
\end{verbatim}
and
\begin{verbatim}
  generate g g > b b~ Z p0 PGG=0, (MG_decay_syntax)
\end{verbatim}
for the ISR topologies. Here, the syntax ``\texttt{PGG=0}'' excludes the scalar production via gluon fusion.


\clearpage
\printbibliography

\end{document}